\documentclass{aa}

\usepackage{graphicx}

\usepackage{natbib}
\bibpunct{(}{)}{;}{a}{}{,} % to follow the A&A style

\usepackage{amsmath,amssymb}
\usepackage{url}
\usepackage{ulem}

\usepackage{txfonts}

\begin{document}

\title{Cold gas in hot star clusters: the wind from the red supergiant W26 in Westerlund 1}
\author{Jonathan Mackey\inst{1,2} \and 
        Norberto Castro\inst{2} \and
        Luca Fossati\inst{2} \and
        Norbert Langer\inst{2}
}
\offprints{mackey@ph1.uni-koeln.de}
\institute{
  I.\ Physikalisches Institut, Universit\"at zu K\"oln, Z\"ulpicher Stra\ss{}e 77, 50937 K\"oln, Germany
  \and
  Argelander-Institut f\"ur Astronomie, Auf dem H\"ugel 71, 53121 Bonn, Germany
}

\date{Received 23 March 2015 / Accepted 27 August 2015}

\abstract{
% Context
The massive red supergiant W26 in Westerlund 1 is one of a growing number of red supergiants shown to have winds that are ionized from the outside in.
The fate of this dense wind material is important for models of second generation star formation in massive star clusters.
% Aims
Mackey et al.~(2014) showed that external photoionization can stall the wind of red supergiants and accumulate mass in a dense static shell.
We use spherically symmetric radiation-hydrodynamic simulations of an externally photoionized wind to predict the brightness distribution of H$\alpha$ and [N\,II] emission arising from photoionized winds both with and without a dense shell.
% Methods
We analyse spectra of the H$\alpha$ and [N\,II] emission lines in the
circumstellar environment around W26
and compare them with simulations
to investigate whether W26 has a wind that is confined by external photoionization.
% Results
Simulations of slow winds that are decelerated into a dense shell show strongly limb-brightened line emission, with line radial velocities that are independent of the wind speed.
Faster winds ($\gtrsim22\ \mathrm{km}\,\mathrm{s}^{-1}$) do not form a dense shell, have less limb-brightening, and the line radial velocity is a good tracer of the wind speed.
The brightness of the [N\,II] and H$\alpha$ lines as a function of distance from W26 agrees reasonably well with observations when only the line flux is considered.
The radial velocity of the simulated winds disagrees with observations, however: the
brightest observed emission is blueshifted by $\approx25\ \mathrm{km}\,\mathrm{s}^{-1}$ relative to the radial velocity of the star, whereas a spherically symmetric wind has the brightest emission at zero radial velocity because of limb brightening.
% Conclusions
Our results show that the bright nebula surrounding W26 must be asymmetric, and we suggest that it is
confined by external ram pressure from the extreme wind of the nearby supergiant W9.
We obtain a lower limit on the nitrogen abundance within the nebula of 2.35 times solar.
The line ratio strongly favours photoionization over shock ionization,
and so even if the observed nebula is pressure confined there should still be an ionization front and a photoionization-confined shell closer to the star that is not resolved by the current observations, which could be tested with better spectral resolution and spatial coverage.
}

\authorrunning{Mackey et al.}
\titlerunning{Cold gas in hot star clusters: the wind of W26}
\maketitle

%%% ------------------------------------------------------------
%%% ------------------------------------------------------------
\section{Introduction}
\label{sec:intro}
%%% ------------------------------------------------------------
%%% ------------------------------------------------------------

Most stars form in clusters \citep{LadLad03} and, in particular, massive stars are predominantly found in clusters and associations \citep{Gie87}.
Furthermore, many O stars not currently in a cluster have been identified as runaway stars \citep[e.g.][]{GvaWeiKroEA12}.
Feedback from massive stars in clusters is important for the enrichment and evolution of the interstellar medium (ISM) of gas-rich galaxies \citep{KruBatArcEA14}, and it is a crucial ingredient in models of multiple stellar populations in globular clusters \citep[e.g.][]{GraCarBra12}.
Globular clusters formed early in cosmic time, during the epoch of first galaxy formation, and their multiple populations \citep[e.g.][]{PioBedAndEA07} demonstrate that somehow gas was able to accumulate, cool, and become gravitationally unstable within the star cluster \citep{DErVesDAnEA08, PalWunTen14}.
It is not at all clear how this happens; a number of theories and their shortcomings are discussed in depth in \citet{BasCabSal15}.
It is fair to say that there is no consensus on how globular clusters formed.
Because of this, researchers have begun to look at present-day massive star clusters to try to gain some insight \citep[e.g.][]{BasCabDavEA13, BasStr14}, with the result that little evidence of cold gas and dust has been found in the intracluster medium of nearby massive star clusters.

Models for multiple populations that involve mass loss from massive stars (as opposed to asymptotic giant branch stars, e.g.\ \citealt{DErVesDAnEA08}) all concentrate on winds from hot stars \citep[e.g.][]{CanRagRod00, WunTenPalEA08, DeMPolLanEA09, KraChaDecEA13}.
The difficulty these models face is how to cool down the hot shocked wind to temperatures low enough to allow star formation \citep{PalWunTen14}.
Surprisingly, these models do not explicitly consider red supergiants, even though Galactic stars with initial mass $M\lesssim30\ \mathrm{M}_{\odot}$ lose more mass as red supergiants than they do as main sequence stars \citep{ChiMae86}.
Their winds are already cold
\citep[temperature $\approx100-500$ K; e.g.][]{SchMarHorEA09,LeBMatGerEA12},
neutral, dusty, and dense, and so if this wind material can remain in a star cluster then it is the obvious candidate to host second generation star formation.
Furthermore, evolutionary models of very metal-poor stars with initial masses above 100 $\mathrm{M}_{\odot}$ show that they may evolve to red supergiants already on the main sequence \citep{MarChiKud03, SzeLanYoo15, KohLanDeKEA15, YooDieLan12}.
For this reason we consider it important to study mass loss from red supergiants in star clusters, to constrain whether their wind material can remain cold and dense in such a harsh environment.

A growing number of red supergiants are found to be surrounded by nebulae with emission lines from ionized gas, showing that their winds are photoionized by external radiation fields.
This was first proposed for NML Cyg \citep{MorJur83}, then for IRS 7 in the Galactic Centre \citep{YusMor91}, and more recently W26 \citep{WriWesDreEA14} and IRC\,$-$10414 \citep{GvaMenKniEA14, MeyGvaLanEA14}.
WOH G64 in the Large Magellanic Cloud also shows similar nebular emission lines \citep{LevMasPleEA09}.
\citet{MacMohGvaEA14} proposed that Betelgeuse's wind is also externally photoionized, to explain the presence of a static neutral shell around the star \citep{LeBMatGerEA12}.
We expect on this basis that many red supergiants have photoionized winds.
\citet{MacMohGvaEA14} showed that the winds of red supergiants could be decelerated and accumulated into a dense shell, if the wind is exposed to an external ionizing radiation field from all sides.
This could have important consequences for the fate of the mass lost during the red supergiant phase, and also for any subsequent evolutionary phases and the observational characteristics of the star's eventual supernova explosion \citep{MacMohGvaEA14}.
If this mechanism is effective, then the mass lost by red supergiants could remain within the parent star cluster for much longer than if the wind is freely expanding, increasing the likelihood that secondary star formation could happen.

\object{Westerlund 1} \citep{Wes61,Wes87} is probably the most massive young star cluster in our Galaxy \citep{BraClaStoEA08}, and its age ($4\pm1$ Myr) is such that it contains a population of extreme supergiant stars \citep{ClaNegCroEA05}.
Only at this age do star clusters contain the hottest (Wolf-Rayet stars) and coldest (red supergiants) stages of massive star evolution simultaneously.
The red supergiant W26 (RA: 16 47 05.403; DEC: -45 50 36.76;
SIMBAD identifier: \object{Cl* Westerlund 1 W 26})
is surrounded by a bright radio nebula \citep{ClaFenWatEA98, DouClaNegEA10}, also seen in mid infrared \citep{ClaFenWatEA98} and in narrow-band optical emission \citep{WriWesDreEA14}.
The wind of W26 is so bright in typical tracers of photoionized gas (H$\alpha$ and [N\,II] forbidden-line emission) that a number of authors have suggested that it could be externally photoionized by radiation from the O-type and Wolf-Rayet (WR) stars in the cluster \citep{DouClaNegEA10, WriWesDreEA14, MacMohGvaEA14}.
The main alternative explanation for the ionized nebula around W26 is that it is hydrodynamically confined by the pressure of the intracluster gas.
This could be ram pressure from stellar winds or a global cluster wind \citep{DouClaNegEA10}, or thermal pressure from thermalised shocked wind bubble and recent supernovae that must have occurred in Westerlund 1 \citep{MunLawClaEA06, KavNorMeu11}.

External photoionization and hydrodynamic confinement must both be present, because the combined winds of the evolved stars will create a high pressure environment \citep{MunLawClaEA06} and there are many hot stars in the cluster that emit ionizing photons, e.g., 24 WR stars \citep{CroHadClaEA06}.
The challenge then is rather to understand which process is producing the ionized nebula in W26, and by consequence in the winds of red supergiants in other star clusters.
The radius at which a stellar wind becomes ionized is well defined for a specified external radiation field \citep{MorJur83}, and the hydrodynamic confinement radius is set by the balance between the ram pressure of the stellar wind and the thermal/ram pressure of the environment.
Unfortunately there are no good estimates of the mass-loss rate or wind velocity of W26, and we do not know the physical distance separating it from the other stars (such as the supergiant sgB[e] star, W9) that are nearby in projection.

To make progress, we take the existing spectroscopic data on W26 and its circumstellar environment, and compare this to the predictions of numerical simulations of an externally photoionized wind, following the method of \citet{MacMohGvaEA14}.
The main observables are the radial velocity of the emitting gas, the intensity of this emission, and the line ratio, as a function of position along the slit in the H$\alpha$ and [N\,II] lines (in this paper we focus entirely on the brighter [N\,II] line at 6584\,\AA).
As shown below, these data already provide contraints that are difficult to satisfy with a spherically symmetric photoionized wind.

The observational data, reduction methods and results are described in Sect.~\ref{sec:observations}.
The results of numerical simulations are described and compared with observations in Sect.~\ref{sec:simulations}.
We discuss our results in Sect.~\ref{sec:discussion} and conclude in Sect.~\ref{sec:conclusions}.

%%% ------------------------------------------------------------
%%% ------------------------------------------------------------
\section{Dataset and analysis processing} 
\label{sec:observations}
%%% ------------------------------------------------------------
%%% ------------------------------------------------------------

\begin{figure}
\centering
\includegraphics[width=0.9\hsize]{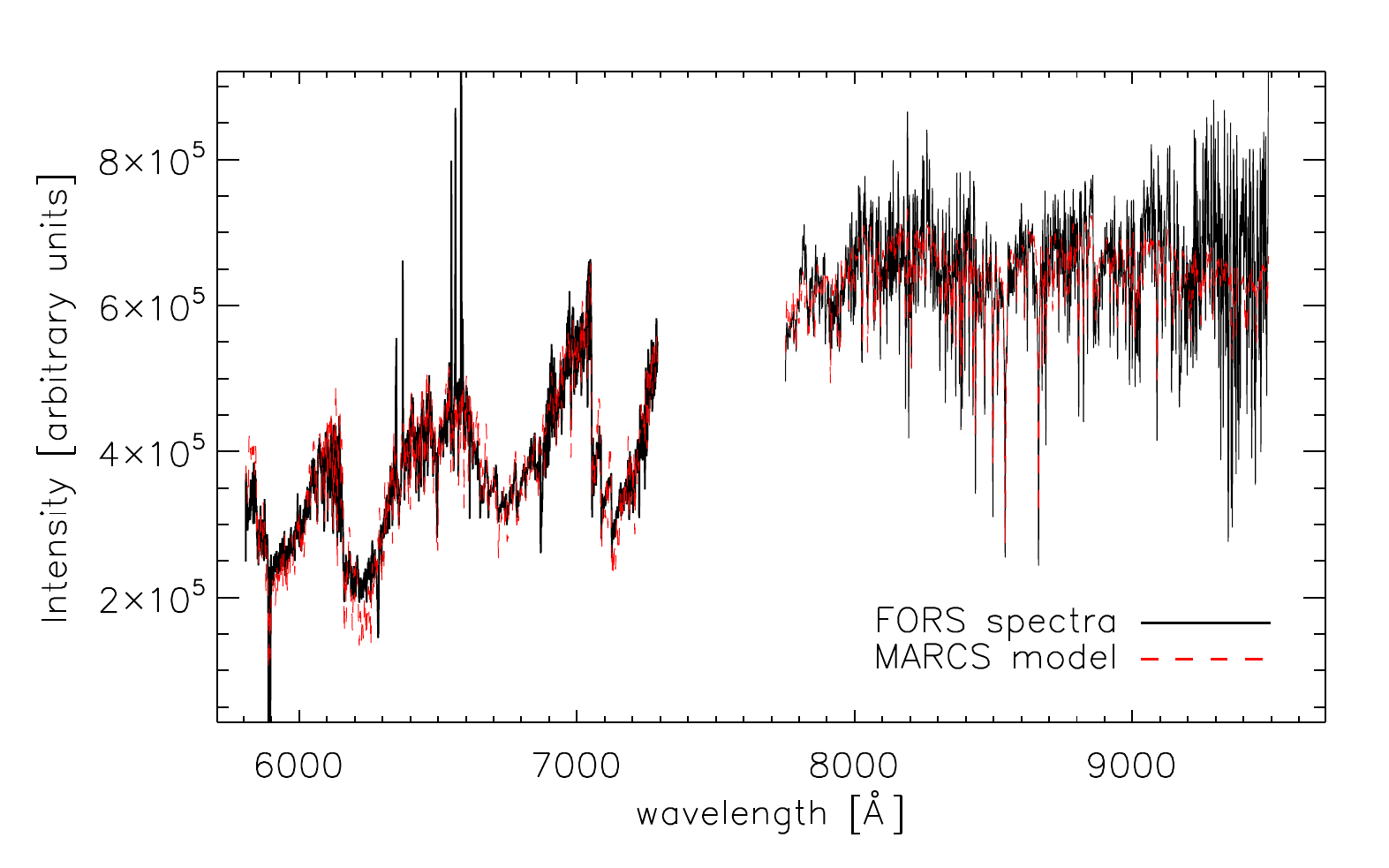}
\caption{
  Comparison between observed FORS spectra for W26 (black solid line) and MARCS model fluxes (dashed red line) for a $T_\mathrm{eff}=3600$ K supergiant.
  The wavelength gap where no spectrum was obtained arises from a gap between the two grisms.
}
\label{fig:w26sed}
\end{figure}

\begin{figure}
\centering
\includegraphics[width=0.9\hsize]{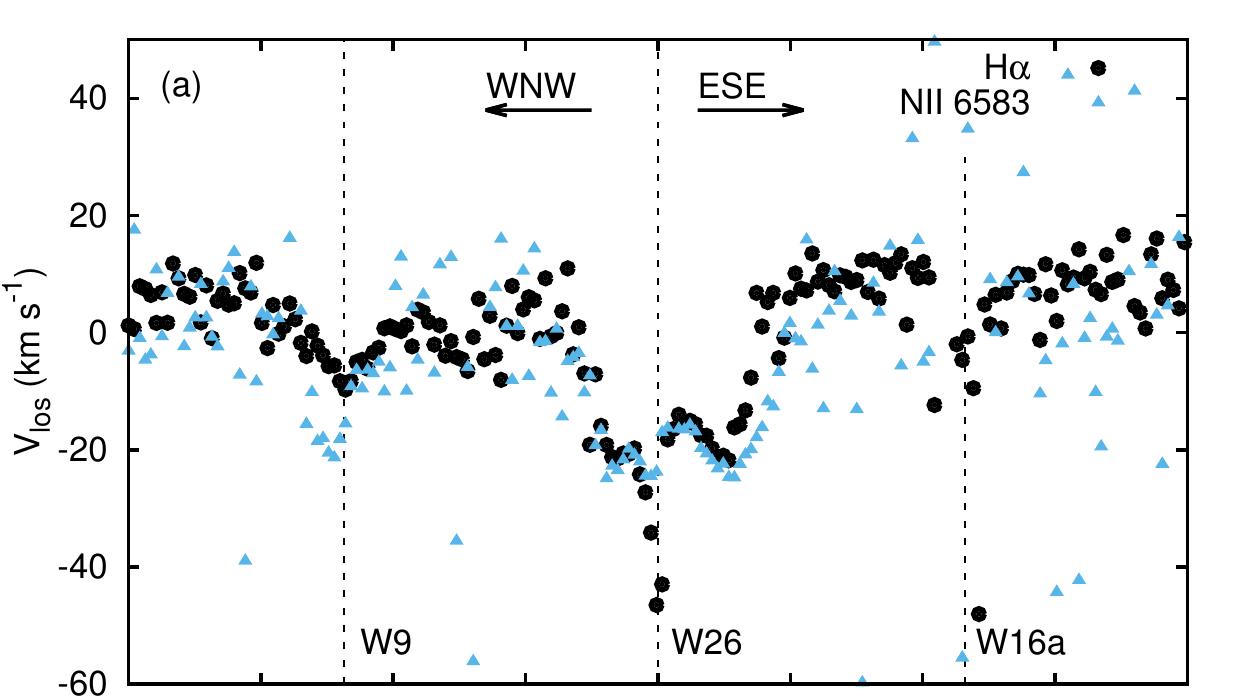}
\includegraphics[width=0.9\hsize]{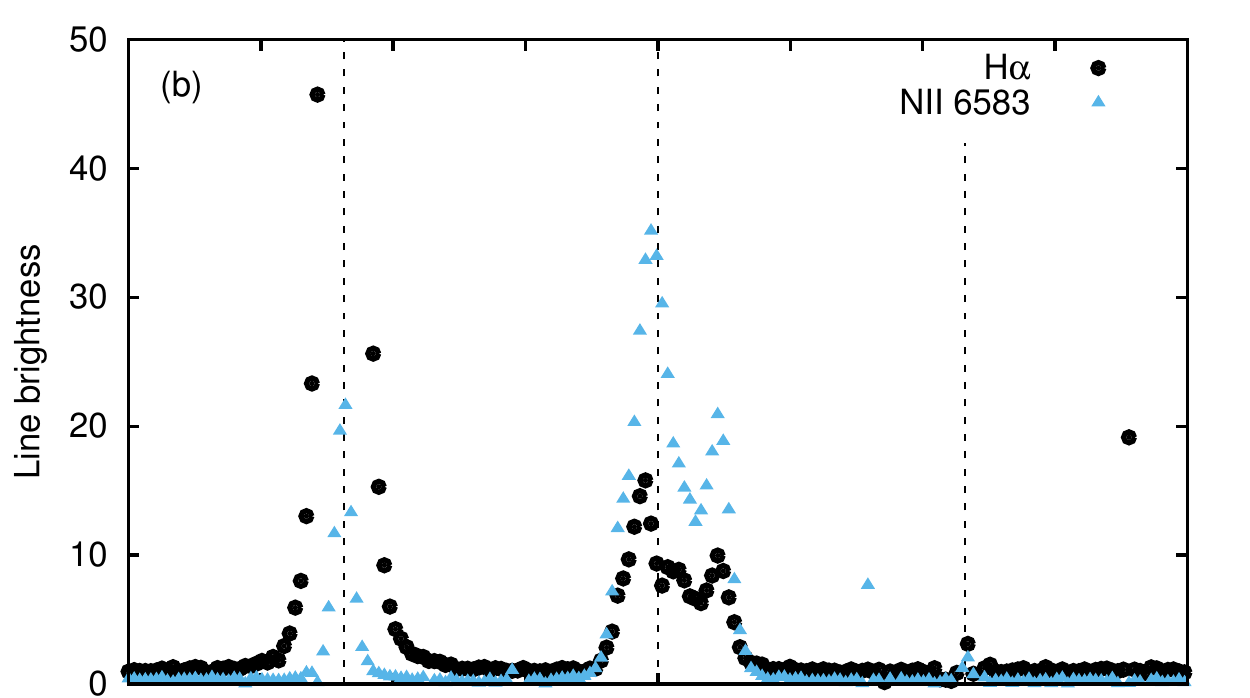}
\includegraphics[width=0.9\hsize]{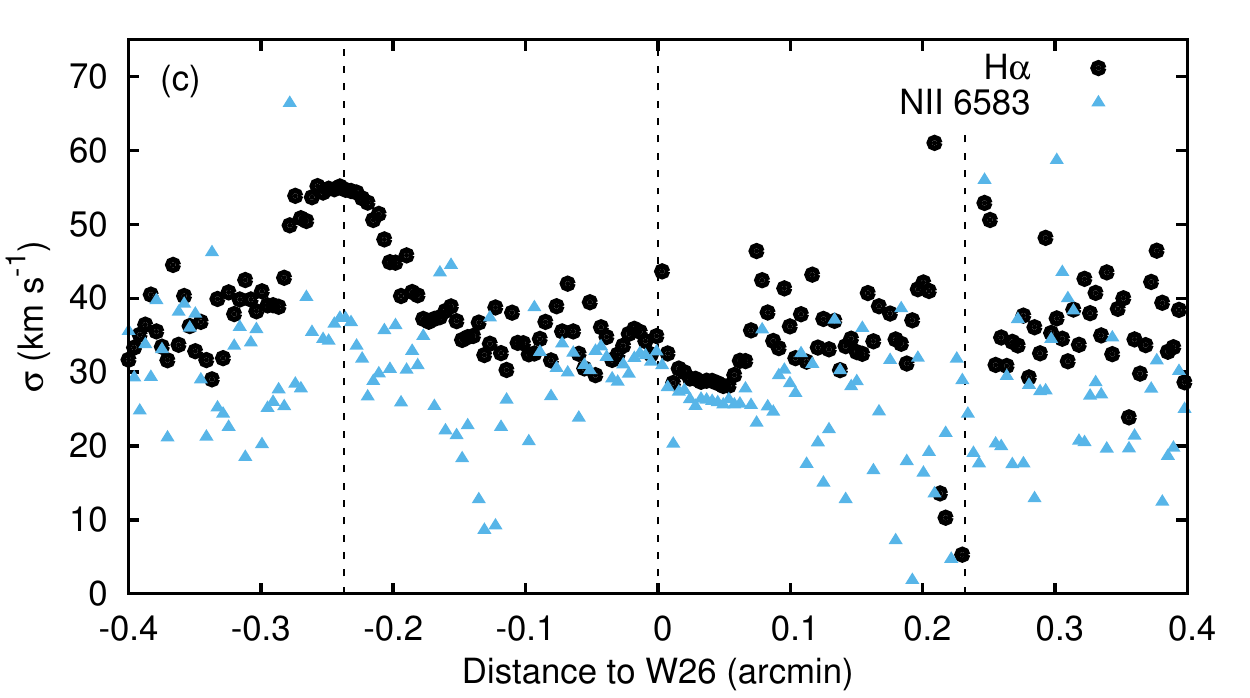}
\caption{
  Spectral observations of W26 and neighbouring stars.
  \textbf{(a)} radial velocity of the H$\alpha$ (black dots) and [N\,II] (blue triangles) lines along the slit in the W26 co-moving reference system ($\sim-32\,$km\,s$^{-1}$), with position measured relative to W26.
  The directions along the slit are also indicated: WNW for Westnorthwest and ESE for Eastsoutheast.
  \textbf{(b)} Line brightness (intensity) in the two spectral lines along the slit (arbitrary units).
  \textbf{(c)} Fitted Gaussian linewidth, $\sigma$ in the two spectral lines along the slit. 
  The three stars included in the slit, and no further away than 0.4 arcmin from W26, are labelled according to \citep{NegClaRit10}.
}
\label{fig:w26}
\end{figure}

W26 was observed with the FOcal Reducer and low dispersion Spectrograph (FORS\,2) \citep{AppRup92, AppFriFueEA98} attached to the Cassegrain focus of the ESO Very Large Telescope of the Paranal Observatory, Chile. The observations were carried out on the 16th of April changed 2011 under the ESO program ID 087.D-0673(A). The data were obtained in long-slit mode using the 1200R and 1028Z grisms and, a slit size of $0.3$ arcseconds to obtain the highest resolution available. This instrumental configuration offers a wavelength range between  $\sim5700-9300\,\AA$ and a resolving power of $R\sim7000$ \citep{NegClaRit10, ClaRitNeg13}. 
The slit location and orientation is shown in fig.~(2) of \citet{WriWesDreEA14}, where the data were first presented.

The data reduction  was performed employing IRAF\footnote{Image Reduction and Analysis Facility (IRAF -- {\tt http://iraf.noao.edu/}) is distributed by the National Optical Astronomy Observatory, which is operated by the Association of Universities for Research in Astronomy (AURA) under cooperative agreement with the National Science Foundation.} \citep{Tod93} and IDL routines following the standard steps and tasks for long-slit spectral data reduction (consisting of bias subtraction, flat-fielding, spectral extraction with and without background subtraction, and wavelength calibration) and analysis.

%%% ------------------------------------------------------------
\subsection{W26 Stellar parameters}
%%% ------------------------------------------------------------

We used the background-subtracted, wavelength-calibrated FORS2 spectra of W26 to roughly estimate the effective temperature ($T_{\mathrm{eff}}$).
We did this by comparing the observed spectrum with synthetic fluxes calculated with {\sc marcs} stellar atmosphere models\footnote{\tt http://marcs.astro.uu.se/} \citep{GusEdvEriEA08} with spherical symmetry and solar metallicity \citep{GreSau98} spanning between 3300\,K and 4500\,K in $T_{\mathrm{eff}}$, in steps of 100\,K, and $-$0.5 and $+$1.0 in $\log g$, in steps of 0.5\,dex.

Given the very low resolution of the FORS2 spectrum, hence the heavy line blending, there was no continuum available that could be used to properly normalise the observed spectrum.
For this reason, we normalised the observed spectrum using each model flux at a time as a reference before determining the total deviation between observed and model spectrum.
We obtained a best fitting effective temperature of $T_\mathrm{eff}=3600$\,K, which corresponds to a spectral type of about M2.5 \citep{VanLanThoEA99}, in agreement with previous spectral type determinations \citep{ClaRitNeg10}; a comparison of the model with the FORS spectrum is shown in Fig.~\ref{fig:w26sed}.
The formal uncertainty in $T_\mathrm{eff}$ is 200\,K, but systematic uncertainties in the modelling are likely to be larger.
It is important to recall here that W26 is a known variable star with a spectral type varying between M2 and M5 \citep{ClaRitNeg10}.

%%% ------------------------------------------------------------
\subsection{Radial velocity and intensity distribution maps}
%%% ------------------------------------------------------------

The kinematic study presented here is based on the the radial velocity distributions of H$\alpha$ $6562.79\AA$ and [N\,II] $6583.45\AA$ along the spatial direction of the FORS\,2 long-slit spectra.
This analysis was done with the 1200R grism data since these two transitions can only be reached in its wavelength coverage.
To explore the entire spatial information available in the spectra, each row of the science images was sliced and individually calibrated by the equivalent row in the calibration arc-lamp spectrum.
The targeted lines were fitted by a Gaussian function per row, identifying the position of the emission peaks and their amplitudes with respect to the spectral continuum.
The velocities are presented in the heliocentric reference system.
Previous studies  reported the possible presence of small instrumental flexures that might lead to shifts in the wavelength calibration as large as 5\,$\mathrm{km\,s}^{-1}$ \citep{BagFosKocEA13}. We corrected for such a possible instrumental bias by further calibrating the spectrum using two telluric sky emission lines at $5889.951$ and $6300.304$\,\AA.

Fig.~\ref{fig:w26} displays the radial velocity ($v_\mathrm{los}$, panel a) and line brightness (b)
spatial distributions  of H$\alpha$  and [N\,II] along the FORS\,2 slit  up to a maximum distance of 0.4 arcminutes from W26.
The radial velocity was shifted to the W26 radial velocity co-moving frame, $\approx-32\,$km\,s$^{-1}$ (Negueruela, private communication),
obtained using data from the AAOmega spectrograph on the Anglo-Australian Telescope, and reduced using the techniques described in \citet{MarNegGonEA14}, with a precision of $2-4\,$km\,s$^{-1}$.
We derived uncertainties of $\approx4$ and $7\,$km\,s$^{-1}$ for  H$\alpha$  and [N\,II] radial velocities, respectively.

We also measured the linewidths of the H$\alpha$ and [N\,II] lines along the slit, shown in panel (c) of Fig.~\ref{fig:w26}, and from the Gaussian fits we obtained $\sigma\approx 30\ \mathrm{km}\,\mathrm{s}^{-1}$ for both lines.
For H$\alpha$ most of the measurements are in the range 30-40 $\mathrm{km}\,\mathrm{s}^{-1}$, whereas for [N\,II] they are mostly in the range 20-40 $\mathrm{km}\,\mathrm{s}^{-1}$.
This is consistent with the instrumental resolution of FORS\,2.
The only exception is near W9, where the very bright H$\alpha$ line has a width $\sigma\approx55\ \mathrm{km}\,\mathrm{s}^{-1}$.
There is an apparent trend that $\sigma$ decreases from left to right across the nebula around W26, or with increasing distance from W9.
It is unclear if this is significant because we are so close to the instrumental resolution; higher resolution spectra could confirm this.
Such a trend might be expected if the wind from W9 were directly colliding with the wind of W26 (see section \ref{sec:discussion}).

For photoionized gas at $T\approx10^4$ K, H$\alpha$ has a thermal linewidth $\sigma\approx10\ \mathrm{km}\,\mathrm{s}^{-1}$, and the [N\,II] line is significantly narrower.
We therefore cannot hope to resolve any substructure in the nebular emission, for example blue- and redshifted components of the wind of W26.
This means that the only information we can reliably obtain from the spectral lines are the radial velocity of the peak and the line brightness.
It should be noted, however, that the observed line could be a blend of multiple unresolved lines with differing radial velocities.
Furthermore, the line brightness is uncalibrated which makes absolute flux/luminosity measurements impossible, and so we are limited to comparing the relative line brightness at different locations along the slit.
Spectroscopy with significantly higher spectral resolution, able to resolve multiple line components in photoionized gas, would provide much stronger constraints on models.

W9 has an enormous H$\alpha$ luminosity \citep{Wes87} with a measured equivalent width $W=-520\pm30 \ \AA$ \citep{ClaNegCroEA05}, which they point out is unprecedented for an early type, emission line star \citep[or $W=-640\pm40 \ \AA$;][]{ClaRitNeg13}.
The H$\alpha$ line is much brighter than the [N\,II] emission lines.
The diffuse emission in the ISM far away from W9 and W26 also has stronger H$\alpha$ than [N\,II] emission.
In contrast, the nebula around W26 has [N\,II] emission significantly brighter than H$\alpha$, as noted by \citet{WriWesDreEA14}.
This is not generally obtained in shock models with solar abundances, which predict intensity ratios [N\,II]/H$\alpha<1$ \citep[e.g.][]{Ray79}, and so can be a signature of enhanced nitrogen abundance.
Stellar evolution models predict that the nitrogen abundance in red supergiant winds should be enhanced by a factor of 3-5 \citep{BroDeMCanEA11}, and this has been measured in circumstellar nebulae around a number of evolved massive stars including the runaway red supergiant IRC\,$-$10414 \citep[][and references therein]{GvaMenKniEA14,MeyGvaLanEA14}.
Alternatively, ionization fronts can show enhanced [N\,II] emission under certain conditions ([N\,II]/H$\alpha\sim1-3$), even for solar abundances \citep{HenArtWilEA05}.

The three data points closest to W26 have significantly larger uncertainty than the surrounding points, because the continuum flux from W26 itself fills in the line at radial velocities near zero.
This biases the radial velocity measurement of the line to the blue, and so these three points blueshifted by 30-50 $\mathrm{km}\,\mathrm{s}^{-1}$ are probably spurious (the line intensities in these points are similarly more uncertain than neigbouring points).
Aside from these points, the radial velocity of the nebula around W26 is almost uniform at $\approx-20\ \mathrm{km}\,\mathrm{s}^{-1}$.

In the low-density limit, assuming that all N is in the form of N$^+$ and that H is fully ionized, the emissivity ratio of [N\,II] to H$\alpha$ can be expressed as \citep{Dop73}
\begin{equation}
  \frac{j_\mathrm{[N\,II]}}{j_{\mathrm{H\alpha}}} = 
  0.116 \sqrt{T} \exp \left(\frac{-21855}{T}\right) \frac{f(\mathrm{N})}{6.8\times10^{-5}} \,
  \label{eqn:em_ratio}
\end{equation}
where we assume $T\gg350$ K, and where $f(\mathrm{N})$ is the number fraction of N (relative to H) for which $6.8\times10^{-5}$ is the solar photosphere abundance \citep{AspGreSauEA09}.
In our models we assume $f(\mathrm{N})=2\times10^{-4}$, appropriate for nitrogen-enriched winds of rotating red supergiants \citep{BroDeMCanEA11}, leading to a pre-factor of $0.341$ instead of $0.116$.
For typical H\,\textsc{ii} region temperatures (7000-10\,000 K) the ratio is between 0.5 and 1.5 for solar abundance, but always $>1$ for the enhanced N abundance.

The low density limit applies (neglecting collisional de-excitation) because in our simulations the electron number density, $n_\mathrm{e}$,
is always less than the critical density for the $6584\AA$ line of [N\,II], which is $n_\mathrm{e,crit}=8.6\times10^4$ cm$^{-3}$ \citep[][table 3.11]{Ost89}.
For reference \citet{DouClaNegEA10} estimate $n_\mathrm{e}\approx5\times10^3$ cm$^{-3}$ from radio observations, if one assumes a constant density in the emitting region.
Comparing this estimate to the critical density, we expect a reduction in [N\,II] emission of $<10$\% because of collisional de-excitation.
Taking into account the possibility of doubly ionized N and of collisional de-excitation, any constraints we obtain on the nitrogen abundance are lower limits.

%%% ------------------------------------------------------------
%%% ------------------------------------------------------------
\section{Simulations of an externally photoionized wind} \label{sec:simulations}
%%% ------------------------------------------------------------
%%% ------------------------------------------------------------

\citet{MacMohGvaEA14} showed that external photoionization of the winds of red supergiants can have strong hydrodynamical effects, in particular decelerating the wind into a dense and massive shell that could contain up to 35\% of the mass lost.
This completely changes the density and velocity structure of the wind, and hence its emission properties.
The wind structure was calculated by \citet{MacMohGvaEA14} using analytic and numerical methods, in both cases assuming an isothermal equation of state (with temperatures of $T=10^2$ and $10^4$ K in the neutral and ionized parts of the wind, respectively).
This allowed them to predict 21 cm emission from the neutral shell around Betelgeuse, but this is a special case because it is so close to Earth that the wind can be spatially resolved at 21 cm.

Winds around other red supergiants have been detected (and spatially resolved) in optical nebular lines \citep{WriWesDreEA14, GvaMenKniEA14}.
We wish to model these lines but they are very senitive to temperature (see Eq.~\ref{eqn:em_ratio}), so it is important to calculate the gas temperature self-consistently with heating and cooling functions, particularly near the ionization front where  most of the emission is concentrated \citep[e.g.][]{HenArtWilEA05}.
The method we use to do this is described below.

Our models provide the first predictions for nebular emission from ionization fronts in red supergiant winds, including both radiation-hydrodynamics and postprocessing with radiative transfer.
The wind properties of red supergiants are often poorly constrained from both theory and observations \citep[see e.g.~the review by][]{MauJos11}.
Optical spectroscopy of nebular lines offers a new avenue to measure the wind velocity and density (and hence mass-loss rate, $\dot{M}$), through line intensities and radial velocities.
This is of crucial importance for constraining the evolution of evolved stars.

%%%%%%%%%%%%%%%%%%%%%%%%%%%%%%%%%%%%%%%%%%%%%%%%%%%%%%%%%%%%%%%%%%%%%
\subsection{Input physics for simulations}
%%%%%%%%%%%%%%%%%%%%%%%%%%%%%%%%%%%%%%%%%%%%%%%%%%%%%%%%%%%%%%%%%%%%%

We use the \textsc{pion} code \citep{Mac12} in spherical symmetry with a wind expanding from the origin ($r=0$) and an ionizing radiation (extreme ultraviolet, EUV, with $h\nu>13.6$ eV) source at radius $r=\infty$.
The method largely follows previous calculations of photoionization-confined shells around red supergiants \citep{MacMohGvaEA14},
except that we use non-equilibrium heating and cooling of the gas \citep{HenArtDeCEA09}, non-equilibrium ionization of hydrogen, and multi-frequency EUV radiation.
This calculates gas temperature more realistically than in \citet{MacMohGvaEA14}.
The implementation is almost exactly as described in \citet{MacGvaMohEA15}, where the scheme was used to describe expanding H\,\textsc{ii} regions around O stars; we refer the reader to section 2 of this work for details.
The rate equation of hydrogen ionization is solved including photoionization, radiative recombination, and collisional (electron impact) ionization.
We make the approximation that helium is singly ionized whenever hydrogen is, so the electron fraction is 1.1 times the H$^+$ fraction (for helium number fraction $n_\mathrm{He}/n_\mathrm{H}=0.1$).
We consider a blackbody radiation spectrum with radiation temperature $T_\mathrm{r}=4\times10^4$ K, in the energy range 13.6-54.4 eV, and also far-ultraviolet (FUV) heating radiation following \citet{HenArtDeCEA09} with equal photon flux in FUV and EUV photons.
This algorithm for thermal and ionization evolution captures the strong cooling in the neutral gas to tens or hundreds of Kelvin and the photoheating to $T\approx7000-10\,000$ K in photoionized gas, including spectral hardening (and extra heating) near the ionization front \citep[e.g.][]{HenArtWilEA05}, which is important because [N\,II] and H$\alpha$ have different temperature scaling (Eq.~\ref{eqn:em_ratio}).

%%%%%%%%%%%%%%%%%%%%%%%%%%%%%%%%%%%%%%%%%%%%%%%%%%%%%%%%%%%%%%%%%%%%%
\subsection{Wind properties for W26}
%%%%%%%%%%%%%%%%%%%%%%%%%%%%%%%%%%%%%%%%%%%%%%%%%%%%%%%%%%%%%%%%%%%%%

There is considerable uncertainty in the wind properties of W26 \citep[e.g.][]{WriWesDreEA14}, so we have run simulations for two different mass-loss rates, $\dot{M}=2\times10^{-5}$ and $10^{-4}\ \mathrm{M}_{\odot}\,\mathrm{yr}^{-1}$, and four different wind speeds, $v_\infty=15,\,20,\,25,$ and $30\ \mathrm{km}\,\mathrm{s}^{-1}$.
According to the \citet{MauJos11} reformulation of the mass-loss rate of \citet{NieDeJ90}, we can estimate
\begin{equation}
\dot{M}=5.6\times10^{-6}
\left(\frac{L}{10^5\ \mathrm{L}_\odot}\right)^{1.64}
\left(\frac{T_\mathrm{eff}}{3500\ \mathrm{K}}\right)^{-1.61} \mathrm{M}_{\odot}\,\mathrm{yr}^{-1} \;.
\end{equation}
Using $T_\mathrm{eff}=3600$ K (estimated above) and $L=3.5\times10^5\ \mathrm{L}_\odot$ \citep{WriWesDreEA14} we obtain $\dot{M}=4.2\times10^{-5}\ \mathrm{M}_{\odot}\,\mathrm{yr}^{-1}$.
This is only a very approximate estimate, however, as individual red supergiants often have mass-loss rates a factor of 3 larger or smaller than empirical relations \citep{MauJos11}.
Furthermore, \citet{MauJos11} show that measured mass-loss rates for the same object often differ by a factor of a few depending on the technique used.

\citet{FokNakYunEA12} compare near- and mid-infrared photometry for W26 with models and estimate $\dot{M}=2.2\times10^{-5}\ \mathrm{M}_{\odot}\,\mathrm{yr}^{-1}$, although their model requires W26 to have a total luminosity ($10^{6.03}\ \mathrm{L}_\odot$) that is much larger than the above-quoted estimates.
\citet{DouClaNegEA10} study the nebula around W26 with radio observations, but did not obtain a measurement of the mass-loss rate.
They estimate that the nebular mass of W26 is comparable to that surrounding VY CMa \citep{SmiHumDavEA01}.
VY CMa has a time-averaged $\dot{M}\approx4\times10^{-5}\ \mathrm{M}_{\odot}\,\mathrm{yr}^{-1}$ \citep{DecHonDeKEA06, MauJos11} from gas emission measurements, but dust emission gives much larger estimates: $\dot{M}=(1.6-4)\times10^{-4}\ \mathrm{M}_{\odot}\,\mathrm{yr}^{-1}$ \citep{DanBesDegEA94, SmiHinRyd09, MauJos11}.
The discrepency may arise partly because the mass loss is variable.
On the other hand, \citet{DouClaNegEA10} estimate $\dot{M}=2\times10^{-5}\ \mathrm{M}_{\odot}\,\mathrm{yr}^{-1}$ for W237, another (presumably coeval) red supergiant in Westerlund 1, and then note that it has a similar nebular mass to W26.
These results show that the two mass-loss rates that we use are reasonable estimates for W26, although they are not hard lower or upper limits.

The distance from W26 of the edge of the H$\alpha$ emission is about 0.02--0.04 arcmin on 
the Westnorthwest side (hereafter WNW) and 0.04--0.06 arcmin on the Eastsoutheast side (hereafter ESE); see Fig.~\ref{fig:w26}.
We adjust the ionizing flux to set up simulations that have a photoionized nebula at a separation from W26 that is appropriate for each side.
Westerlund 1 contains a number of Wolf-Rayet and O stars with ionizing photon luminosities $Q_0\sim10^{49}$ s$^{-1}$, providing ionizing fluxes at W26 on the order of $F_\gamma\sim(10^{10}-10^{13})\ \mathrm{cm}^{-2}\,\mathrm{s}^{-1}$ (see section \ref{sec:disc:pion}), so the fluxes that we use are about what we expect for the environment of W26.

%%%%%%%%%%%%%%%%%%%%%%%%%%%%%%%%%%%%%%%%%%%%%%%%%%%%%%%%%%%%%%%%%%%%%
\subsection{Simulation parameters}
%%%%%%%%%%%%%%%%%%%%%%%%%%%%%%%%%%%%%%%%%%%%%%%%%%%%%%%%%%%%%%%%%%%%%

We have 16 simulations for two mass-loss rates, two positions for the ionization front, and four wind velocities.
The identifier (ID) of each simulation and the parameters used are listed in Table~\ref{tab:sims}.
In the main text we present only the simulations with $\dot{M}=2\times10^{-5}\ \mathrm{M}_{\odot}\,\mathrm{yr}^{-1}$ and ionizing fluxes chosen for the ESE side of the nebula (i.e.~only $v_\infty$ is varied).
We focus on the ESE side of the nebula because it extends further from the star and hence the data are less contaminated by the central star.
The results from the simulations with $\dot{M}=10^{-4}\ \mathrm{M}_{\odot}\,\mathrm{yr}^{-1}$ are very similar to those with $\dot{M}=2\times10^{-5}\ \mathrm{M}_{\odot}\,\mathrm{yr}^{-1}$ except that the wind in the former case is five times denser than the latter, and so the line emission is approximately 25 times brighter.
It is not possible to calibrate the data in absolute flux, and so we do not know the absolute line brightness and cannot constrain the wind density (but see Sect.~\ref{sec:discussion}).
For completeness, the other simulations are presented in Appendices \ref{app:ESEside}  and \ref{app:WNWside}.

All simulations use 10\,240 
uniformly spaced
grid zones from $r=9.5\times10^{15}$ cm ($\approx0.0031$ pc) to $6.172\times10^{17}$ cm (0.2 pc).
This is more than sufficient to resolve the structure of the flow and the internal structure of the shocked shell.
The simulations were run for 0.1 Myr, after which the star has lost $10\ \mathrm{M}_{\odot}$ for $\dot{M}=10^{-4}\ \mathrm{M}_{\odot}\,\mathrm{yr}^{-1}$ or $2\ \mathrm{M}_{\odot}$ for $\dot{M}=2\times10^{-5}\ \mathrm{M}_{\odot}\,\mathrm{yr}^{-1}$.
For $v_\infty\geq20\ \mathrm{km}\,\mathrm{s}^{-1}$ the simulations had reached a steady state at this time.
The shocked shell has not reached a steady state for simulations with $v_\infty=15\ \mathrm{km}\,\mathrm{s}^{-1}$ (it is still accumulating mass), but this does not affect the ionized gas line emission.
We checked this by running some simulations for 0.2 Myr and comparing the resulting emission properties.

\begin{table}
  \centering
  \caption{
    Parameters used for the 16 spherically symmetric radiation-hydrodynamics simulations.
    In the identifier (ID), M5/M4 refer to the mass-loss rates, V15/V20/V25/V30 refers to the wind velocity, and W/E denote WNW side or ESE side of W26 (i.e.\ when the ionization front is closer to or further from W26).
    The ionizing fluxes have been set to give ionization fronts that match the observations as closely as possible in terms of H$\alpha$ brightness as a function of distance from the star.
  }
  \begin{tabular}{ l c c c c}
    %\hline
    ID  &  $v_\infty\ (\mathrm{km}\,\mathrm{s}^{-1})$ & $\dot{M}\ (\mathrm{M}_{\odot}\,\mathrm{yr}^{-1})$ & $F_\gamma \ (\mathrm{cm}^{-2}\,\mathrm{s}^{-1})$ \\
    \hline
    M5V15W & 15 & $2\times10^{-5}$ & $7.87\times10^{10}$   \\
    M5V20W & 20 & $2\times10^{-5}$ & $1.29\times10^{11}$   \\
    M5V25W & 25 & $2\times10^{-5}$ & $8.29\times10^{10}$   \\
    M5V30W & 30 & $2\times10^{-5}$ & $5.83\times10^{10}$   \\
    \hline
    M4V15W & 15 & $1\times10^{-4}$ & $1.70\times10^{12}$   \\
    M4V20W & 20 & $1\times10^{-4}$ & $2.34\times10^{12}$   \\
    M4V25W & 25 & $1\times10^{-4}$ & $1.44\times10^{12}$   \\
    M4V30W & 30 & $1\times10^{-4}$ & $9.95\times10^{11}$   \\
    \hline
    M5V15E & 15 & $2\times10^{-5}$ & $1.70\times10^{10}$   \\
    M5V20E & 20 & $2\times10^{-5}$ & $2.78\times10^{10}$   \\
    M5V25E & 25 & $2\times10^{-5}$ & $1.79\times10^{10}$   \\
    M5V30E & 30 & $2\times10^{-5}$ & $1.26\times10^{10}$   \\
    \hline
    M4V15E & 15 & $1\times10^{-4}$ & $3.67\times10^{11}$   \\
    M4V20E & 20 & $1\times10^{-4}$ & $5.05\times10^{11}$   \\
    M4V25E & 25 & $1\times10^{-4}$ & $3.10\times10^{11}$   \\
    M4V30E & 30 & $1\times10^{-4}$ & $2.15\times10^{11}$   \\
  \end{tabular}
  \label{tab:sims}
\end{table}

%%%%%%%%%%%%%%%%%%%%%%%%%%%%%%%%%%%%%%%%%%%%%%%%%%%%%%%%%%%%%%%%%%%%%
\subsection{Structure of the photoionized wind}
%%%%%%%%%%%%%%%%%%%%%%%%%%%%%%%%%%%%%%%%%%%%%%%%%%%%%%%%%%%%%%%%%%%%%

\begin{figure}
\centering
\includegraphics[width=\hsize]{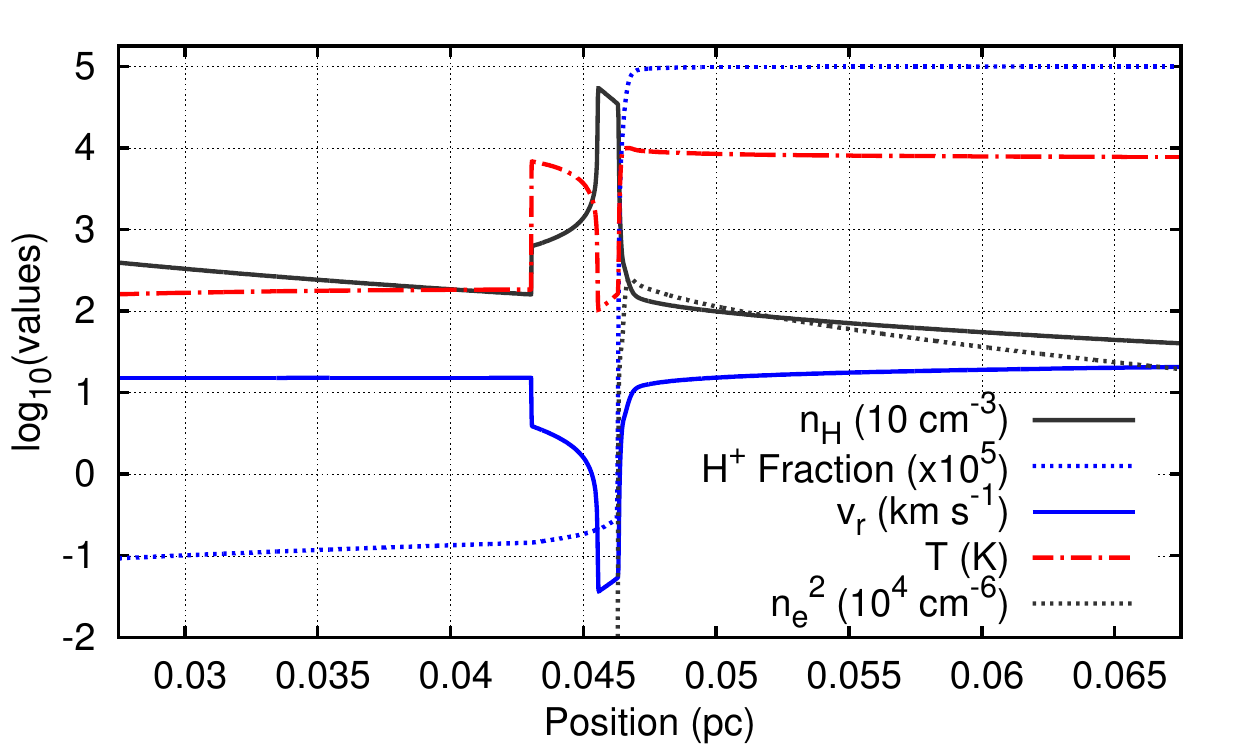}
\caption{
  Radial structure of the wind in simulation M5V15E: we show the H number density $n_\mathrm{H}$, H$^{+}$ fraction, gas radial velocity $v_\mathrm{r}$, and temperature $T$ on a logarithmic scale as a function of distance from the red supergiant.
  }
\label{fig:M5V15E_radial}
\end{figure}

The structure of the photoionized wind as a function of radius is shown in Fig.~\ref{fig:M5V15E_radial} for simulation M5V15E.
The ionization front is D-type \citep[\emph{dense};][]{Kah54} because the wind velocity satisfies $v_\infty \leq 2a_\mathrm{i}$, where $a_\mathrm{i}$ is the isothermal sound speed in photoionized gas.
It has the classical structures of (from right to left) the ionization front, a dense shocked shell, and a shock in the neutral gas.
\citet{MacMohGvaEA14} assumed isothermal neutral gas, and so the shock was very compressive, whereas here the shock is adiabatic with a factor of 4 increase in density.
Then there is a post-shock cooling region where the temperature decreases (from $r\approx0.043-0.0455$ pc), and density increases to compensate (velocity also decreases to conserve mass).
Finally there is a dense cooled shell bounded by the cooling layer and the ionization front.

\begin{figure}
\centering
\includegraphics[width=\hsize]{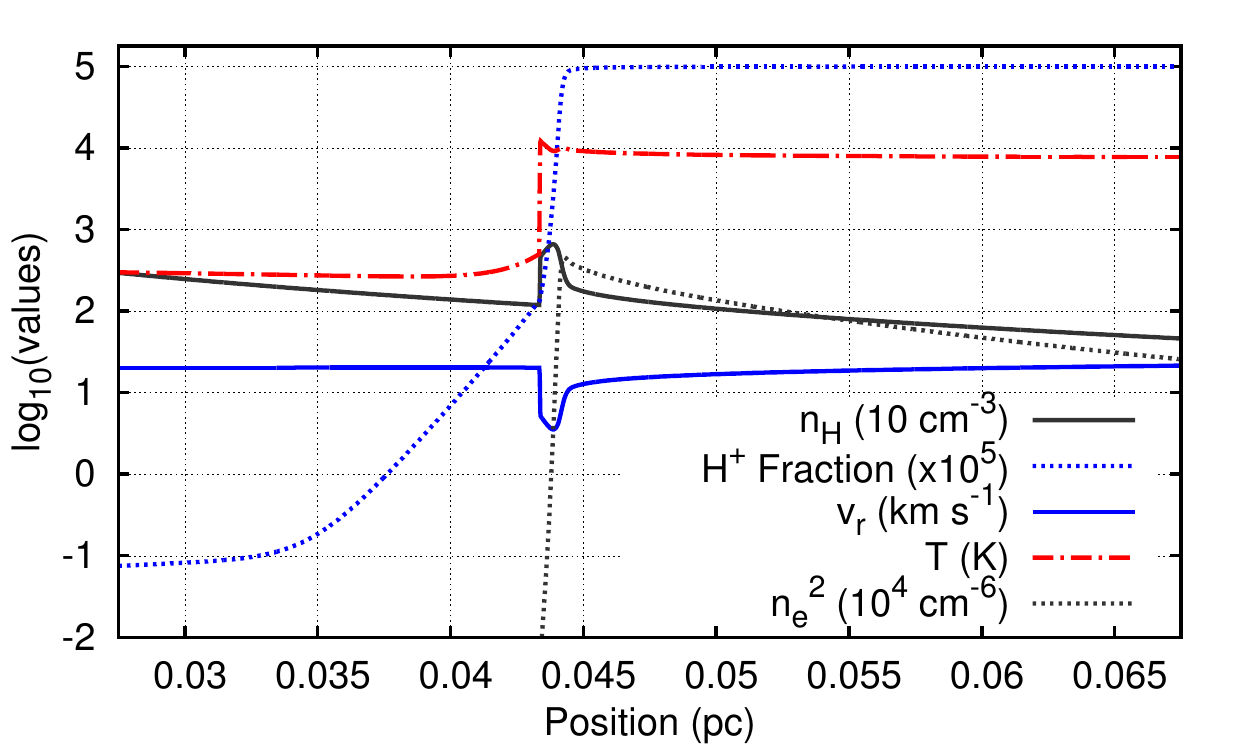}
\caption{
  As Fig.~\ref{fig:M5V15E_radial} but for simulation M5V20E.
  }
\label{fig:M5V20E_radial}
\end{figure}

\begin{figure}
\centering
\includegraphics[width=\hsize]{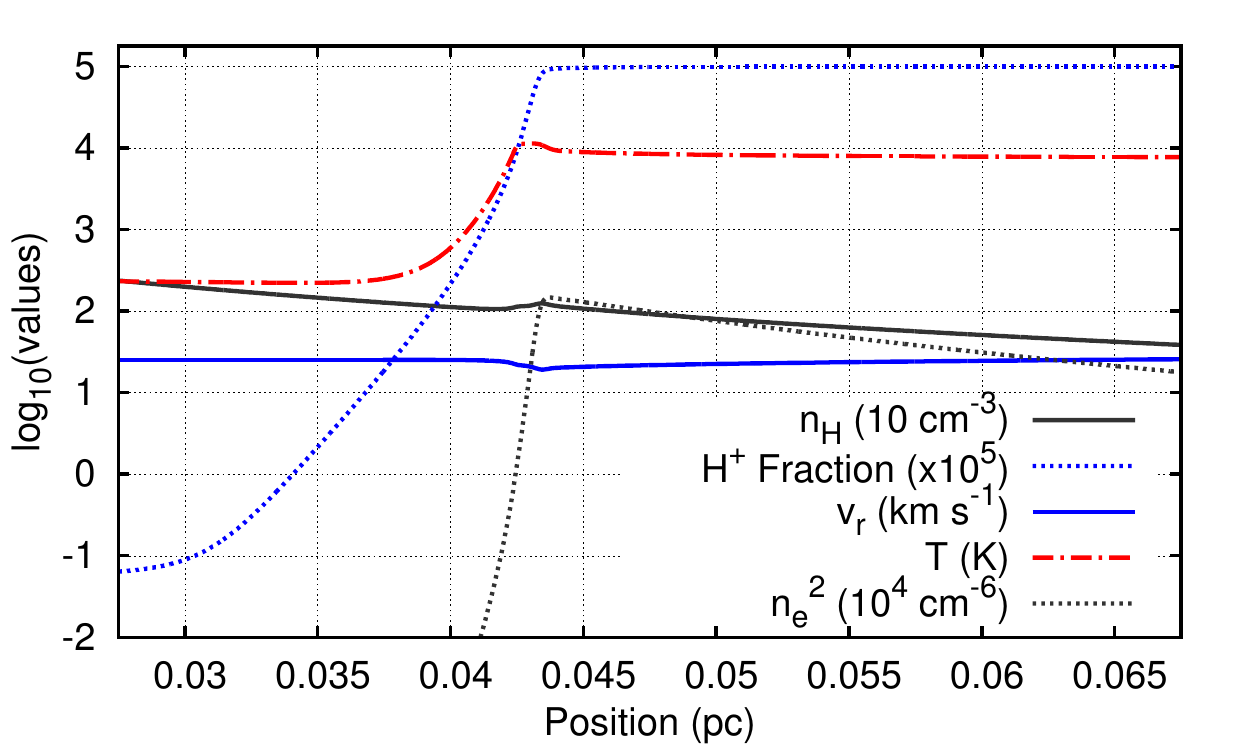}
\caption{
  As Fig.~\ref{fig:M5V15E_radial} but for simulation M5V25E.
  }
\label{fig:M5V25E_radial}
\end{figure}

Simulations with $v_\infty=20$ and $25\ \mathrm{km}\,\mathrm{s}^{-1}$ are shown in Figs.~\ref{fig:M5V20E_radial} and \ref{fig:M5V25E_radial}, respectively.
For $v_\infty=25\ \mathrm{km}\,\mathrm{s}^{-1}$ the wind is too fast to sustain a shocked region and the ionization front is R-type (\emph{rarefied}), characterised by weak density and velocity changes across the front \citep{Kah54}.
The temperature increases smoothly in the wind from a few hundred to about 10\,000 K.
The ionization front is also much broader because there is no dense shell to absorb all of the photons.
Simulation M5V20E (Fig.~\ref{fig:M5V20E_radial}) is a transition case on the boundary between D-type and R-type ionization fronts.
There is a shock in the wind, but the shock cannot propagate far upstream from the ionization front and so there is too little time for the post-shock gas to cool.
The separation between the shock and the ionization front is calculated in \citet{MacMohGvaEA14} and goes to zero as $v_\infty\rightarrow 2a_\mathrm{i}$.
In this case there is a shocked shell, but it cannot accumulate significant mass.

Figs.~\ref{fig:M5V15E_radial}--\ref{fig:M5V25E_radial} also plot $n_\mathrm{e}^2$, which determines (along with temperature) the emissivity of the nebular emission lines.
The peak value of $n_\mathrm{e}^2$ is always right at the ionization front, and drops by an order of magnitude between $r=0.044$ pc and $r=0.065$ pc.
We therefore expect most of the line emission to arise from a relatively thin shell of ionized gas around the ionization front.

%%%%%%%%%%%%%%%%%%%%%%%%%%%%%%%%%%%%%%%%%%%%%%%%%%%%%%%%%%%%%%%%%%%%%
\subsection{Properties of the H$\alpha$ and [N\,II] line emission}
%%%%%%%%%%%%%%%%%%%%%%%%%%%%%%%%%%%%%%%%%%%%%%%%%%%%%%%%%%%%%%%%%%%%%
We project the spherically symmetric wind onto the plane of the sky and trace rays through the wind for various impact parameters, $b$, the radius of closest approach for each ray to the red supergiant.
The emission and absorption coefficient for the H$\alpha$ and [N\,II] lines are calculated for every grid zone and tabulated.
Then each ray is split up into line segments corresponding to the grid zones, and the radiative transfer equation is solved analytically along each segment assuming a constant source function \citep[see e.g.][]{RybLig79}, first inwards to $r=b$ and then back outwards.
To calculate velocity information we obtain the line-of-sight radial velocity of the gas from the gas dynamics, and the emission is then thermally broadened about this value using the gas temperature.
We use 501 velocity bins between $-50.1$ and $50.1\ \mathrm{km}\,\mathrm{s}^{-1}$ to ensure that the lines are always well resolved in velocity.

The radial velocity of peak emission is then extracted for each pixel, and the line brightness is obtained by summing the emission at all velocities for each pixel.
The line ratio is the ratio of these line intensities.
We convert from impact parameter, $b$, to an angular distance from W26, $\theta$, by assuming a distance to Westerlund 1 of 3.55 kpc \citep{BraClaStoEA08}, consistent with the distance derived more recently from eclipsing binary stars ($3.7\pm0.6$ kpc) by \citet{KouBon12}.
We also spatially smooth the intensity maps to a resolution of 1 arcsec at the distance of Westerlund 1, for easier comparison with W26.

\begin{figure}
\centering
\includegraphics[width=0.9\hsize]{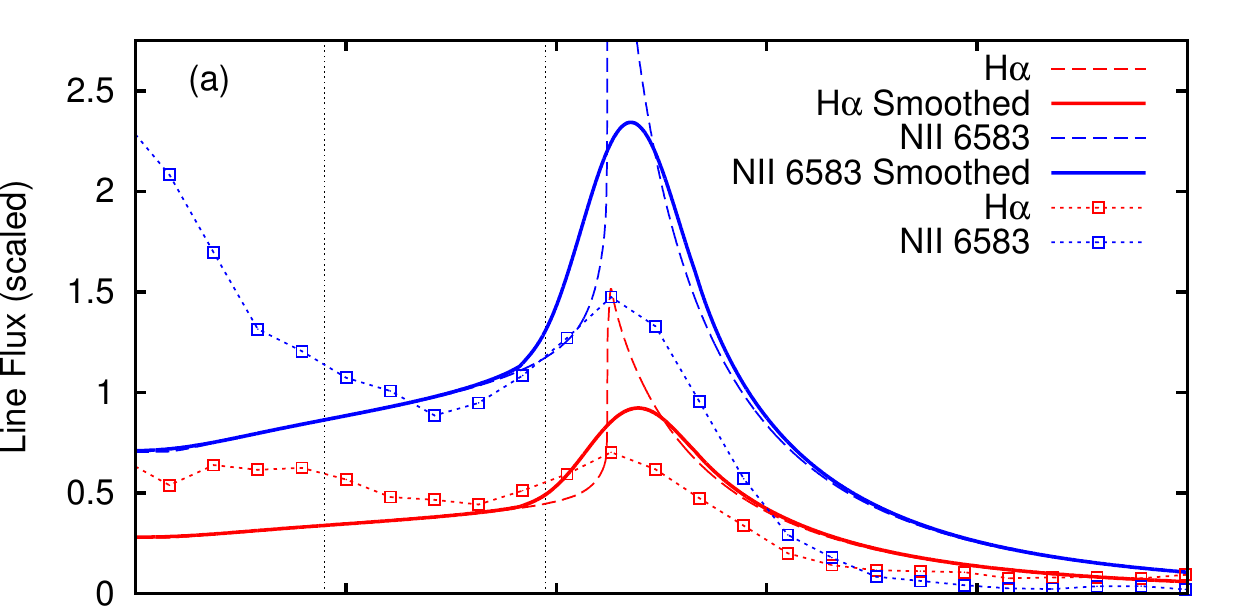}
\includegraphics[width=0.9\hsize]{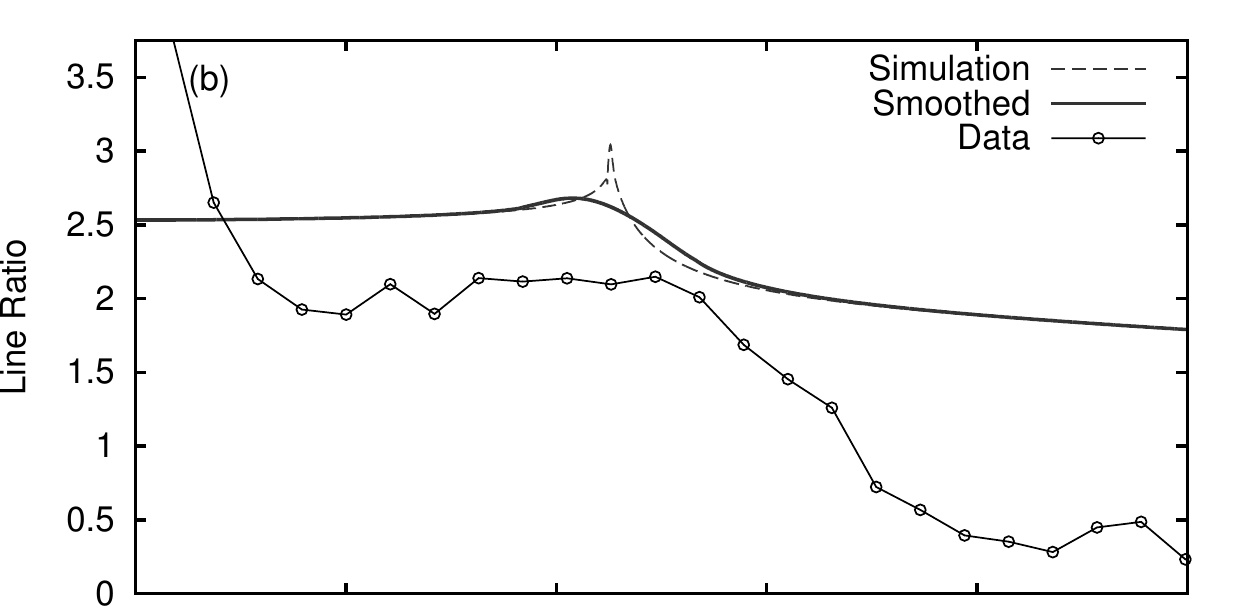}
\includegraphics[width=0.9\hsize]{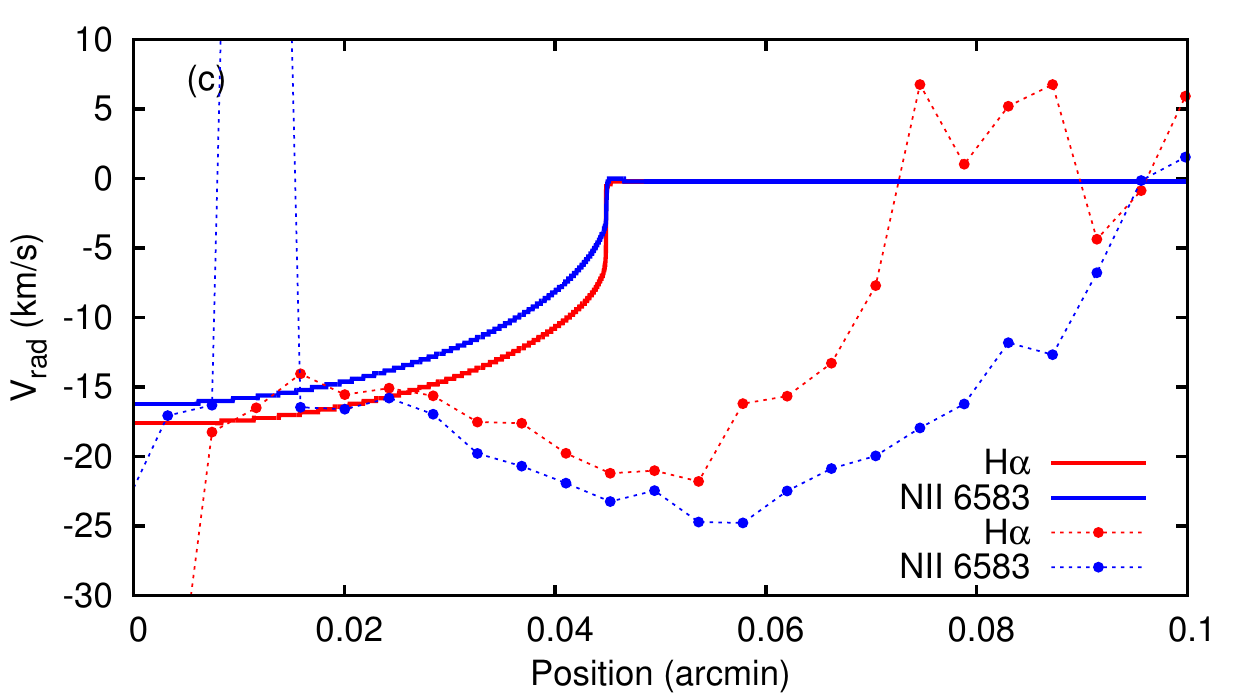}
\caption{
  The H$\alpha$ and [N\,II] spectral lines from the simulation M5V15E, with $\dot{M}=2\times10^{-5}\ \mathrm{M}_{\odot}\,\mathrm{yr}^{-1}$, $v_\infty=15\ \mathrm{km}\,\mathrm{s}^{-1}$, and $F_\gamma=1.70\times10^{10}\ \mathrm{cm}^{-2}\,\mathrm{s}^{-1}$, after 0.1 Myr of evolution.
  The dashed lines show the simulation,
  solid lines smoothed to 1 arcsec resolution at the distance of Westerlund 1,
  and the dotted lines with points show the observations to the ESE of W26.
  The panels show \textbf{(a)} line brightness, \textbf{(b)} line ratio [N\,II]/H$\alpha$, and \textbf{(c)} radial velocity of the peak emission, all as a function of distance from the star.
  Vertical dotted lines in panel (a) show the range of radii used to set the normalisation of the simulated and observed [N\,II] line brightness.
  }
\label{fig:M5V15E}
\end{figure}

The results are plotted in Fig.~\ref{fig:M5V15E} for simulation M5V15E, with the observational data overplotted, as a function of distance from W26 measured in arcminutes.
Line intensities for [N\,II] and H$\alpha$ are plotted in panel (a), the line ratio in panel (b), and radial velocity of the line peaks in panel (c).
The relative normalisation of the simulation to the data is chosen by setting the mean [N\,II] line intensity to unity between the two radii shown by vertical dotted lines in panel (a).
Note again that the simulation results assume nitrogen is enhanced in the stellar wind by a factor of 2.8 (see section \ref{sec:observations}).
The spectral data close to W26, within about 0.015 arcmin, are less reliable than at larger separations because of stellar continuum subtraction (see section \ref{sec:observations}).

This simulation has a D-type ionization front with a dense static shell, so the ionization front is much denser than in the higher velocity simulations.
This leads to a thinner region of peak emission because the newly ionized gas accelerates, and so its density (and hence emissivity) decreases more rapidly with radius than for a constant velocity wind.
This simulation therefore has the most limb-brightened emission, compared with simulations with faster winds.

\begin{figure}
\centering
\includegraphics[width=0.9\hsize]{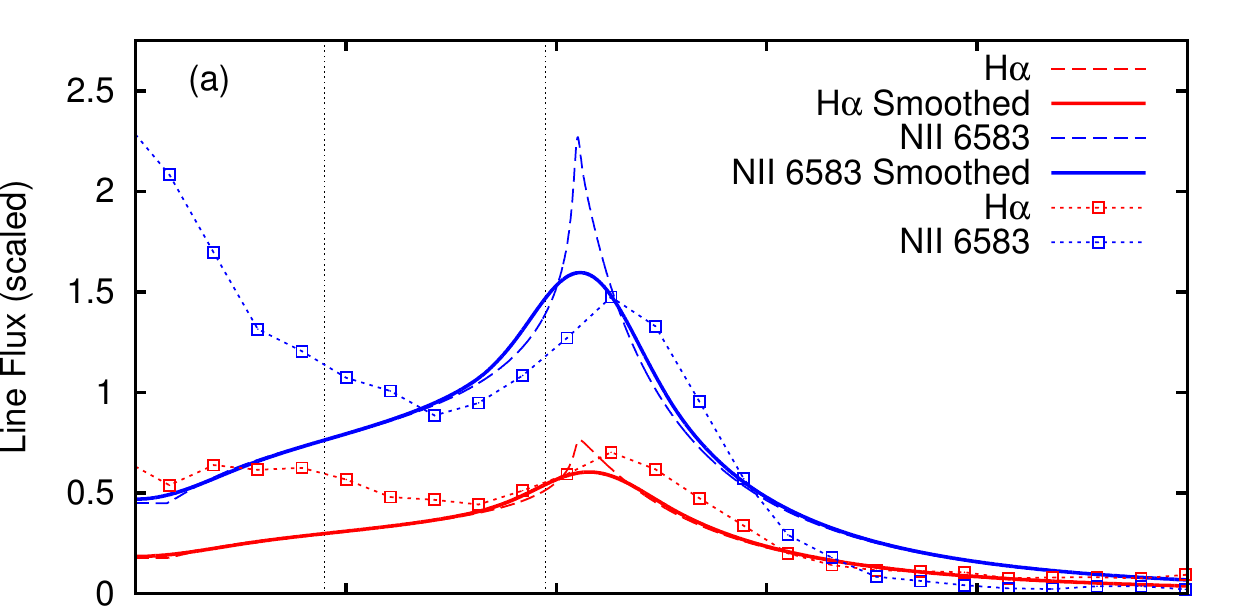}
\includegraphics[width=0.9\hsize]{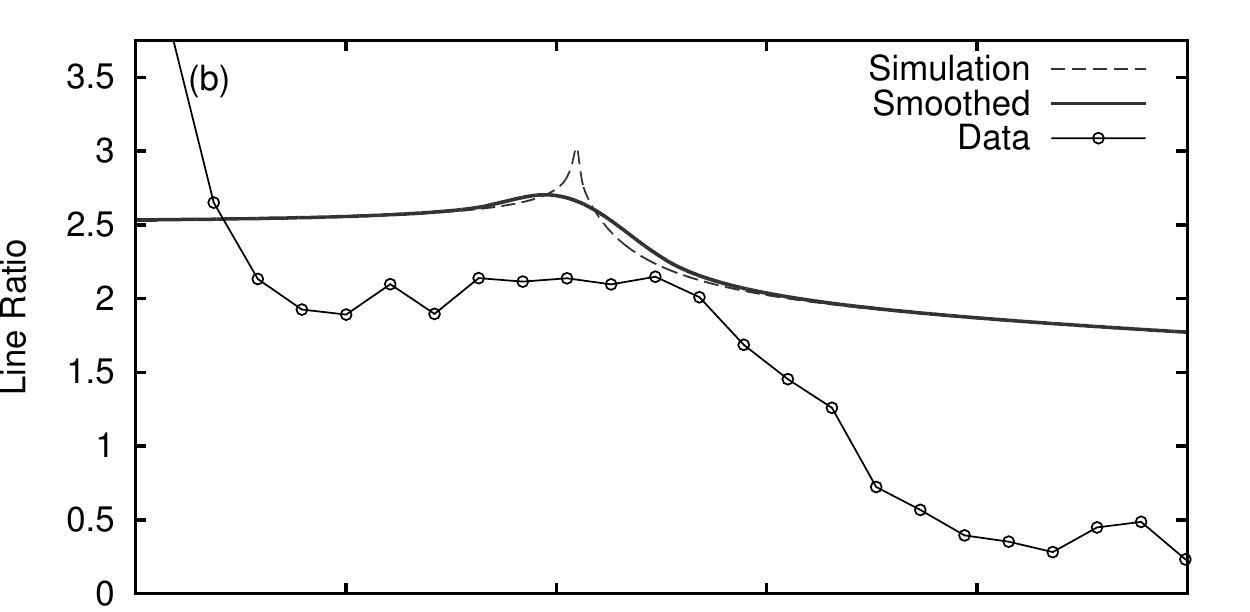}
\includegraphics[width=0.9\hsize]{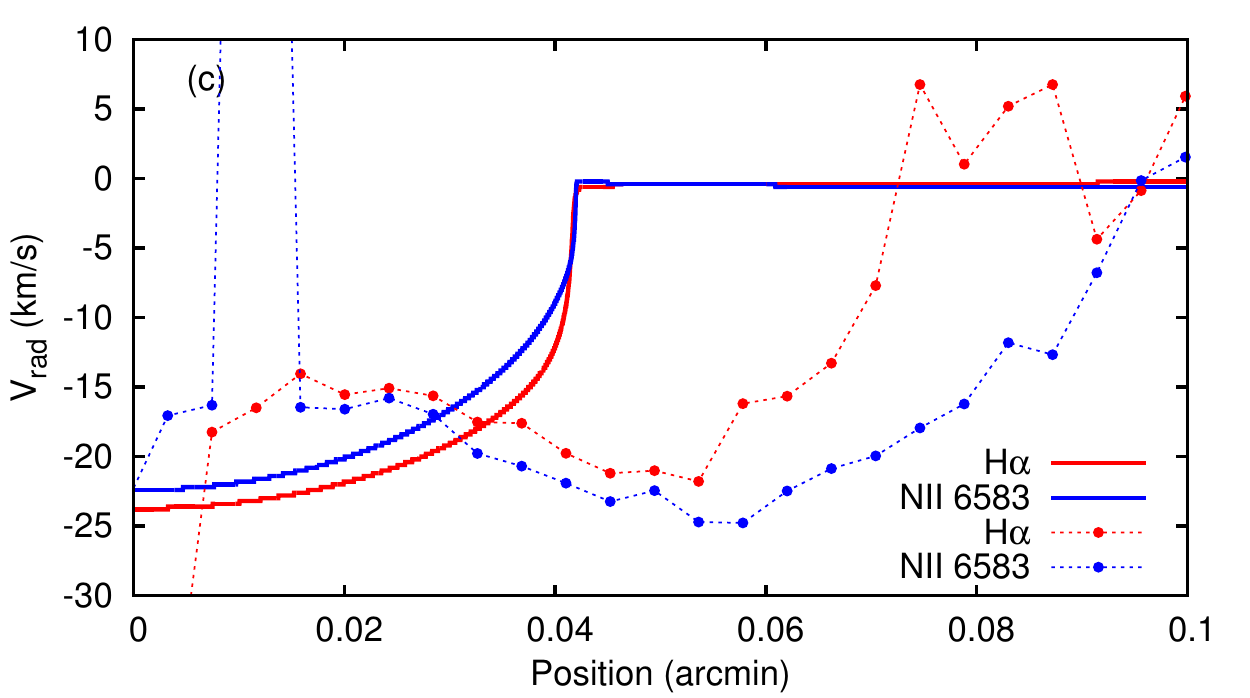}
\caption{
  The H$\alpha$ and [N\,II] spectral lines from the simulation M5V25E, again after 0.1 Myr of evolution.
  The lines and symbols are the same as in Fig.~\ref{fig:M5V15E}.
  }
\label{fig:M5V25E}
\end{figure}

The peak emission in panel (a) is at the same position as the observations by construction, because we constrained the ionizing flux to locate the ionization front at this radius.
The observed emission is much less peaked than the synthetic data for both lines, although this can partly be explained by the finite spatial resolution of the observations.
The degree of limb brightening constrains the thickness of the shell of brightly emitting gas, with a thinner shell being more limb-brightened.
The data thus appear to imply a thicker shell of brightly-emitting gas compared with what we find from this simulation.

The [N\,II] brightness decreases more rapidly with distance from W26 than the synthetic data, but 
this is less clear with the H$\alpha$ brightness because the ratio between maximum and minimum observed brightness is not as large as for [N\,II].
The observed diffuse H$\alpha$ emission at large radii is a constant component from Westerlund 1 and not related to the W26 nebula (see Fig.~\ref{fig:w26}), even though its brightness is similar to our predictions at separation $0.07-0.1$ arcmin.
Subtraction of this constant component complicates any estimate of the rate of decrease at these larger radii, but it may be that the observed H$\alpha$ also decreases more rapidly with radius than we predict.
The rate that the brightness decreases with distance from the peak constrains the density of the flow as a function of radius if the gas is isothermal (assuming spherical expansion).
The data then indicate that either nitrogen begins to be doubly ionized further from the star, or the emitting gas density decreases much more rapidly than our models predict.
This could arise if the wind of W26 is hydrodynamically confined by much hotter intracluster gas (see Sect.~\ref{sec:discussion}).

The brightness ratio, [N\,II]/H$\alpha$, is plotted in panel (b) and shows that within the bright part of the nebula the synthetic and observed data have a broadly constant line ratio that is $\approx2$ for the observations and $\approx2.5$ for the simulations.
If the wind were not enriched in nitrogen, the simulations would predict a line ratio of $\approx0.9$ in this region, inconsistent with the results.
The observations therefore indicate that nitrogen is enhanced by a factor of $\sim2-3$ in the nebula.
If we consider the nitrogen abundance to be a free parameter, we obtain good agreement between simulations and observations for $f(N)=1.60\times10^{-4}$, or 2.35 times the solar abundance \citep{AspGreSauEA09}.
This enhancement is strong evidence that the nebula contains gas processed by the CNO cycle, and hence that it consists of mass lost by W26, at least out to a distance of $\approx0.06$ arcmin.

The line ratio decreases in the simulations from the ionization front ($\theta\approx0.045$ arcmin) outwards, whereas the observed line ratio stays constant to the edge of the bright nebula and then decreases sharply from $\approx1.5$ to  $\lesssim0.5$ between $\theta=0.06$ and $\theta=0.08$ arcmin.
The synthetic data never decrease to such a small line ratio, so again we suggest for this low-emission gas that either nitrogen is more highly ionized or the gas is not enriched in nitrogen (i.e.\ it is not part of the wind of W26).
Whatever the explanation, clearly there is something happening in the region between $\theta=0.06$ and $\theta=0.08$ arcmin that is not seen in the synthetic data.

This is further shown by looking at the radial velocity of the line peak as a function of position, shown in panel (c).
The inner part of the nebula has radial velocity $v_\mathrm{rad}\approx-15\ \mathrm{km}\,\mathrm{s}^{-1}$, similar to the simulated lines.
The observations then show that the brightest emission at 0.035-0.055 arcmin is also the most blueshifted, whereas the simulations show that the radial velocity of the brightest emission is zero.
This is the biggest discrepency between our model and the data: both the line ratio and the radial velocity stay approximately constant throughout the bright nebula, while the simulations show a change at the brightest part of the nebula.
The decrease in observed line ratio is spatially correlated with the decrease in radial velocity.

Fig.~\ref{fig:M5V25E} shows the same plots for simulation M5V25E, with a faster wind such that the ionization front is R-type, with only weak density and velocity changes in the wind (see Fig.~\ref{fig:M5V25E_radial}).
The agreement of the line brightness with observations is better than for M5V15E, in that the line emission is not so strongly peaked at the ionization front, but the decrease in intensity at $\theta>0.05$ arcmin still has the wrong shape for [N\,II].
The line ratio also agrees well for $\theta\leq0.06$ arcmin.
We see the same discrepencies for this simulation as for M5V15E, however.
The simulation predicts that the line goes to zero radial velocity at the emission peak, whereas the data show the maximum radial velocity at the peak.
The same holds true for the line ratio, which in the observational data remains large as long as the radial velocity is significantly different from zero.

The situation is somehat different on the opposite side of W26, Westnorthwest (WNW) in the direction of the extreme sgB[e] star W9.
Here the nebula is confined closer to W26, possibly because of the proximity of W9.
The observed data shows no limb-brightened peak in the nebular emission, rather just a shoulder of emission between 0.02 and 0.03 arcmin.
The results of a comparison between simulation and observation are very similar to what we have already discussed for the ESE side of the nebula, and are shown in Appendix~\ref{app:WNWside}.

%%%%%%%%%%%%%%%%%%%%%%%%%%%%%%%%%%%%%%%%%%%%%%%%%%%%%%%%%%%%%%%%%%%%%
\subsubsection{Wind velocity as traced by H$\alpha$ and [N\,II] lines}
%%%%%%%%%%%%%%%%%%%%%%%%%%%%%%%%%%%%%%%%%%%%%%%%%%%%%%%%%%%%%%%%%%%%%

\begin{figure}
\centering
\includegraphics[width=0.9\hsize]{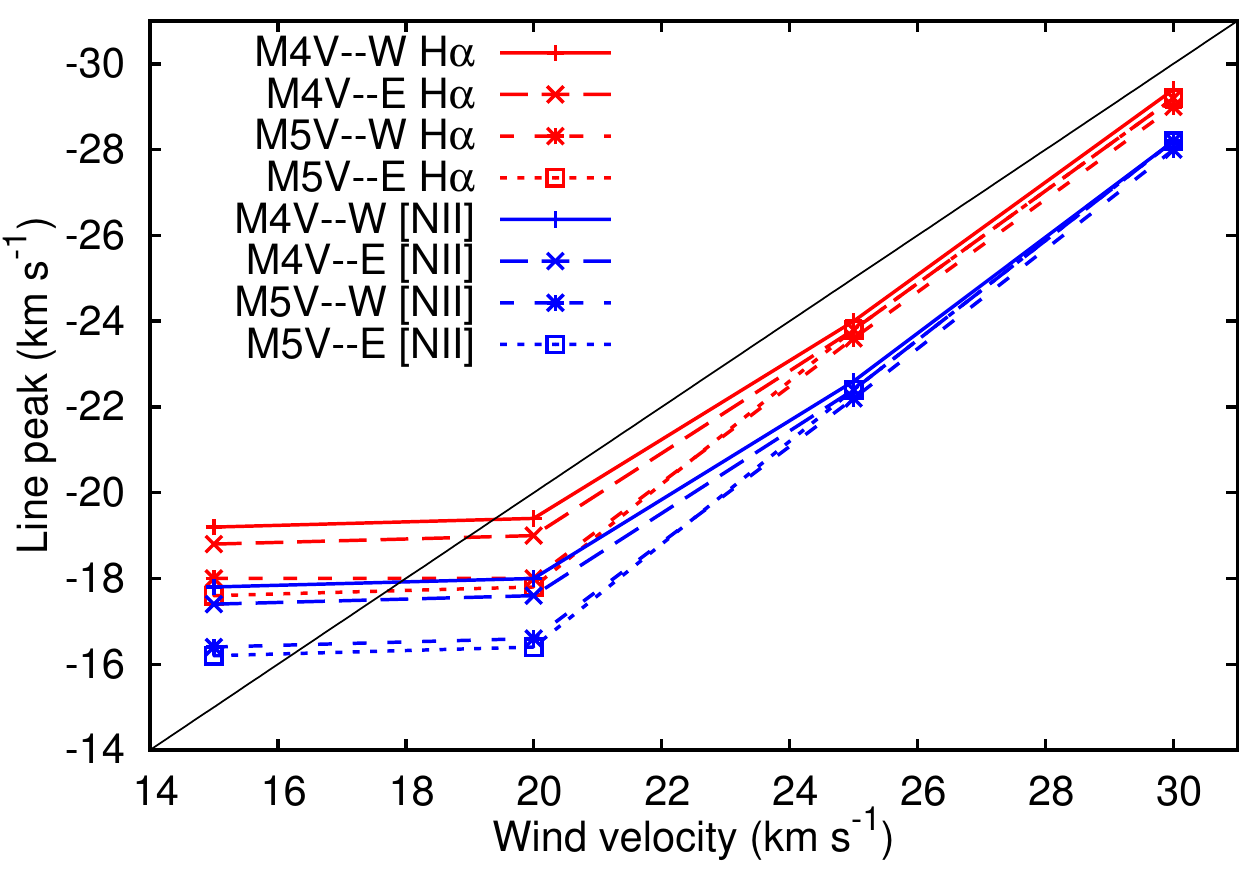}
\caption{
  Comparison between wind velocity, $v_\infty$, and the blueshift of the H$\alpha$ and [N\,II] spectral lines for lines of sight close to the star.
  The boundary between D-type and R-type ionization fronts is at $v_\infty\approx22\ \mathrm{km}\,\mathrm{s}^{-1}$.
  The red lines are for H$\alpha$ and the blue for [N\,II].
}
\label{fig:linevel}
\end{figure}

Fig.~\ref{fig:linevel} plots the blueshift of the two lines as a function of wind velocity for the four sets of simulations (two values of $\dot{M}$ and two sets of ionizing fluxes).
For slow winds with D-type ionization fronts ($v_\infty\lesssim22\ \mathrm{km}\,\mathrm{s}^{-1}$) the ionized gas is accelerating outwards from the outer edge of a static shell, and so the line velocity is independent of wind velocity.
In this velocity range the H$\alpha$ line is blueshifted by about 2\ km\,s$^{-1}$ more than the [N\,II] line for all simulations.
This is a temperature effect: Eq.~\ref{eqn:em_ratio} shows that the two lines will be produced from different depths in the ionization front \citep[see e.g.][]{HesScoSanEA96, HenArtWilEA05}.
We also find that the simulations with denser winds ($\dot{M}=10^{-4}\ \mathrm{M}_{\odot}\,\mathrm{yr}^{-1}$) are blueshifted by about 2\ km\,s$^{-1}$ compared with simulations that have less dense winds.
This is a density effect because within the ionization front, the heating rate scales linearly with density, whereas the cooling rate is quadratic for collisional coolants, implying that the temperature structure within the front should be density-dependent.
The ionization front thickness is inversely proportional to density, and so its internal structure is also expected to depend on $\dot{M}$.
We find that for a given temperature in the photoevaporation flow, the denser winds have an expansion velocity about 2\ km\,s$^{-1}$ larger than the less dense winds.

This result for D-type ionization fronts likely depends weakly on the temperature of the ionizing radiation field: harder radiation fields create hotter ionization fronts, faster photoevaporation flows, and hence the lines will probably be more blueshifted.
The metallicity of the wind is also important, because lower metallicity gas has a higher equilibrium temperature, which will also drive faster photoevaporation flows.

For faster winds ($v_\infty\gtrsim22\ \mathrm{km}\,\mathrm{s}^{-1}$) the ionization front becomes R-type, and this has very little effect on the wind dynamics except for a slight deceleration.
The ionized gas exiting the ionization front has a velocity given by \citep[][ch.~106]{MihMih84}
\begin{equation}
v_\mathrm{i} \approx v_\infty \left( 1+ \left[\frac{a_\mathrm{i}}{v_\infty}\right]^2 \right)^{-1}
\end{equation}
where the term $(a_\mathrm{i}/v_\infty)^2<0.25$ for R-type ionization fronts.
This explains why the line velocity is always less blueshifted than $v_\infty$, and also why both lines approach $v_\infty$ for larger $v_\infty$.
Again, [N\,II] is less blueshifted than H$\alpha$ because it forms deeper in the ionization front before thermal re-acceleration of the wind can take place.
We conclude that these lines are useful but imperfect tracers of the wind velocity for red supergiants, with their accuracy increasing as wind velocity increases.

%%%%%%%%%%%%%%%%%%%%%%%%%%%%%%%%%%%%%%%%%%%%%%%%%%%%%%%%%%%%%%%%%%%%%
\subsubsection{Limb brightening of the H$\alpha$ and [N\,II] lines}
%%%%%%%%%%%%%%%%%%%%%%%%%%%%%%%%%%%%%%%%%%%%%%%%%%%%%%%%%%%%%%%%%%%%%

\begin{figure}
\centering
\includegraphics[width=0.9\hsize]{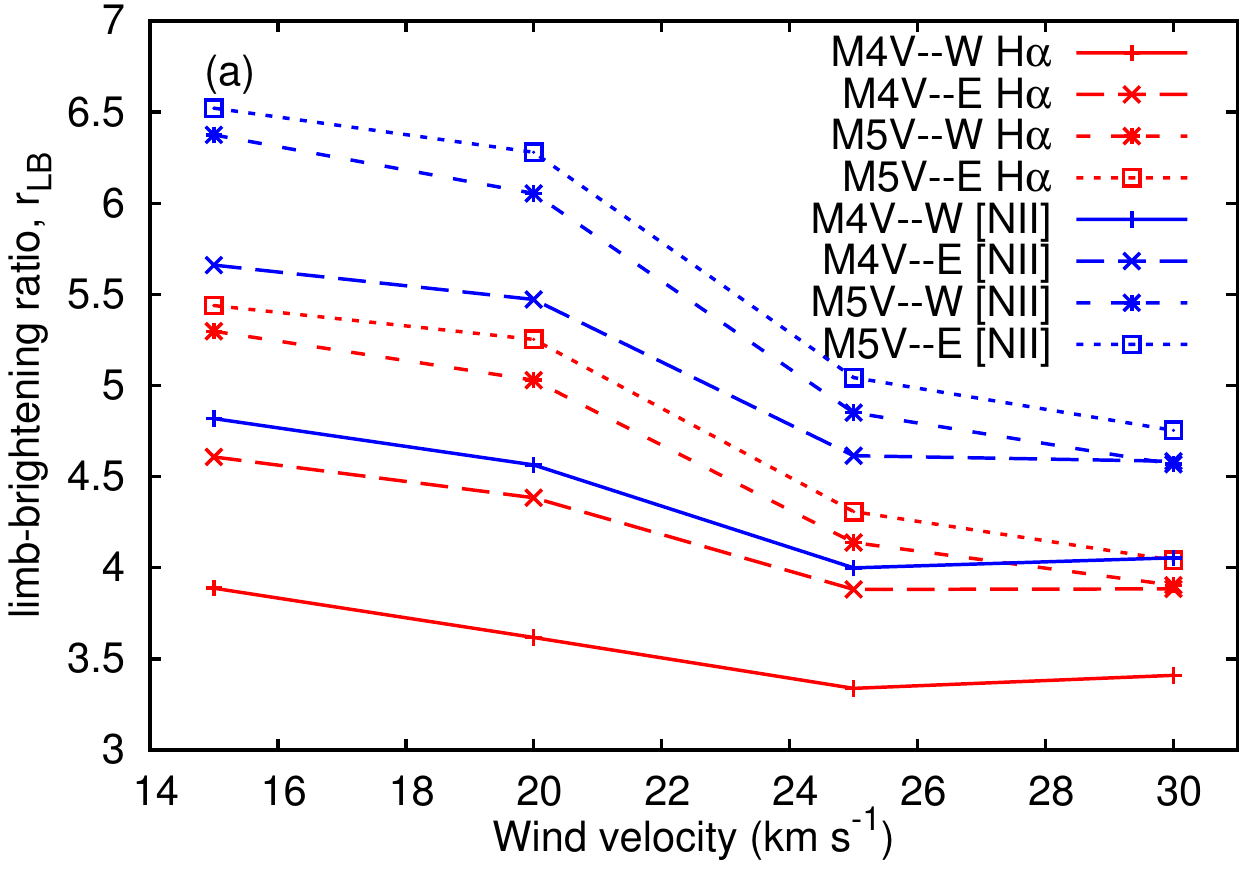}
\includegraphics[width=0.9\hsize]{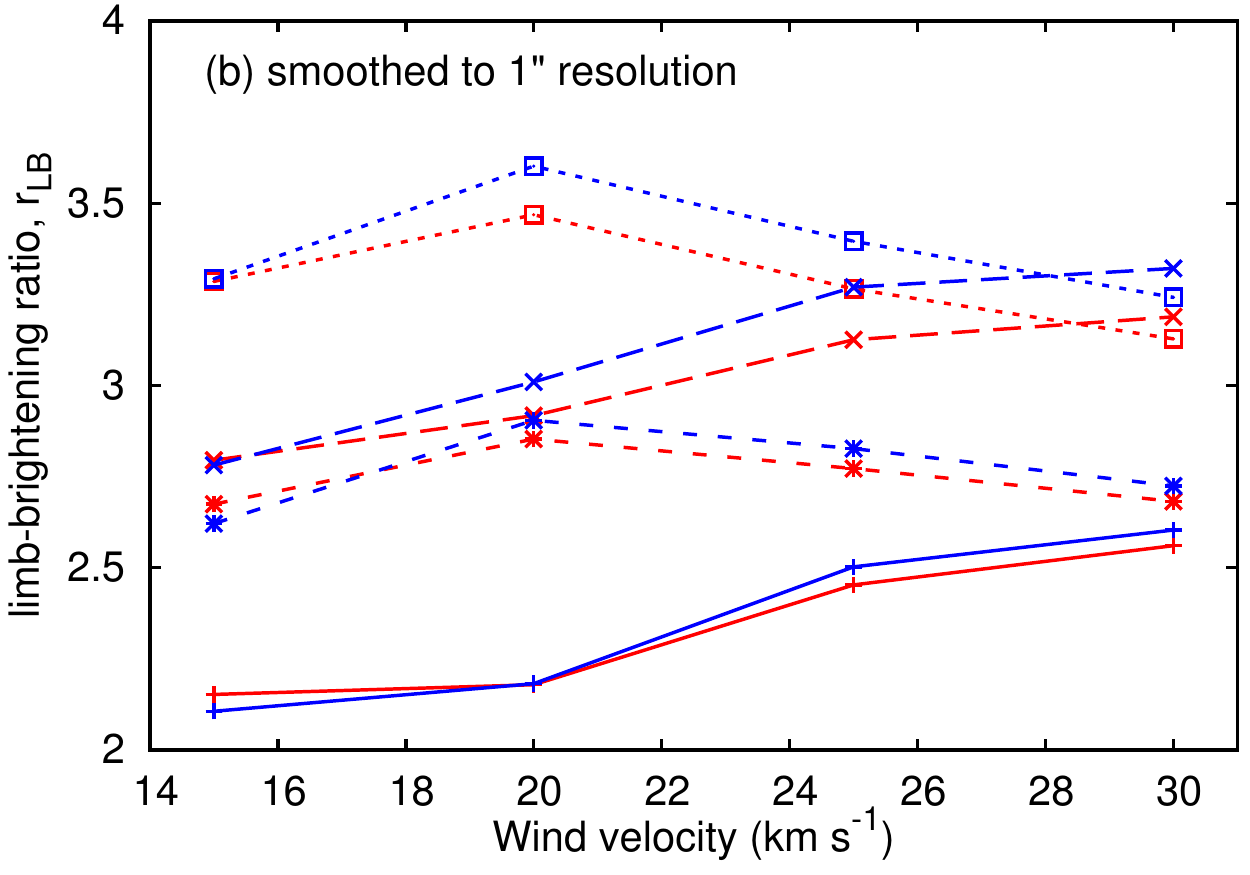}
\caption{
  Comparison between wind velocity, $v_\infty$, and the degree of limb brightening of the H$\alpha$ and [N\,II] spectral lines, defined as the ratio of the peak brightness to that at lines of sight near the star ($r=0$).
  The boundary between D-type and R-type ionization fronts is at $v_\infty\approx22\ \mathrm{km}\,\mathrm{s}^{-1}$.
  The red lines are for H$\alpha$ and the blue for [N\,II].
  Panel (a) shows the raw simulation data, and panel (b) when smoothed to a spatial resolution of 1 arcsec.
  The lines in panel (b) follow the legend in panel (a).
}
\label{fig:limbb}
\end{figure}

Figs.~\ref{fig:M5V15E} and \ref{fig:M5V25E} show that the slow wind simulations have more limb-brightened emission than faster winds.
In Fig.~\ref{fig:limbb}(a) this is quantified with plots of the ratio, $r_\mathrm{LB}$, of the peak to the central intensity for all 16 simulations and for both spectral lines.
This confirms the trend that the two slow-wind simulations ($v_\infty=15$ and 20 km\,s$^{-1}$) with D-type ionization fronts have stronger limb brightening than the two fast-wind simulations ($v_\infty=25$ and 30 km\,s$^{-1}$) with R-type ionization fronts.
The slow-wind simulations have $r_\mathrm{LB}\approx3.5-5.5$ for H$\alpha$ and $r_\mathrm{LB}\approx4.5-6.5$ for [N\,II].
The fast-wind simulations have $r_\mathrm{LB}\approx3.3-4.3$ for H$\alpha$ and $r_\mathrm{LB}\approx4-5.1$ for [N\,II].
This arises, as noted above, because the emitting region is thinner for the slow-wind simulations, and limb brightening is stronger the thinner the emitting shell.
[N\,II] emission arises from a thinner shell of emitting gas within the ionization front than H$\alpha$ emission, and so it has larger $r_\mathrm{LB}$ than H$\alpha$ for the same reason.

The simulations with denser winds have smaller $r_\mathrm{LB}$, which arises because of self-absorption.
Lines of sight towards the star have very little absorption (from the front side; the back side is invisible) because the wind density decreases outwards.
Lines of sight at the limb pass through much more material, however, and the opacity is significant for the simulations with the densest winds.

Fig.~\ref{fig:limbb}(b) shows, however, that the limb brightening is much decreased when the data are smoothed to a resolution of 1 arcsec, with most values of $r_\mathrm{LB}$ lying between 2 and 3.5.
We still see the trend that denser winds have less limb-brightening, but the difference between H$\alpha$ and [N\,II] is almost completely gone.
Also, the trend of decreasing $r_\mathrm{LB}$ with increasing $v_\infty$ is no longer present, and is even reversed for simulations with dense winds.
The smoothed values are still somewhat larger than the observed ratio for W26, $r_\mathrm{LB}\approx1.5$.

%%% ------------------------------------------------------------
%%% ------------------------------------------------------------
\section{Discussion}
\label{sec:discussion}
%%% ------------------------------------------------------------
%%% ------------------------------------------------------------

%%% ------------------------------------------------------------
\subsection{Constraining wind properties using photoionized winds}
%%% ------------------------------------------------------------

Photoionized red supergiant winds can be extremely bright, being very strongly irradiated by nearby stars.
The peak brightness of the simulated nebulae in H$\alpha$ (smoothed to a spatial resolution of 1 arcsec for the distance of Westerlund 1) ranges from, $I_\mathrm{max}\approx10^{-13}$ erg\,cm$^{-2}$\,s$^{-1}$\,arcsec$^{-2}$ for M5V30E to $I_\mathrm{max}\approx10^{-11}$ erg\,cm$^{-2}$\,s$^{-1}$\,arcsec$^{-2}$ for M4V20W.
This is, for example, $200-20\,000$ times brighter than the photoionized bow shock around the red supergiant IRC\,$-$10414 \citep{MeyGvaLanEA14}, and $\approx1-100$ times brighter than the H$\alpha$ emission from the brightest of the Eagle Nebula pillars \citep{HesScoSanEA96}.
The overall brightness scales approximately with the square of the electron density (hence the wind density).
This can potentially provide a good constraint on the mass-loss rate of red supergiants, in the same way that radio data gives the emission measure and hence electron density (with assumptions about geometry).

Using the brightness estimate from \citet{WriWesDreEA14} for the circumstellar nebula around W26, and assuming the same extinction as they find for the nearby triangular nebula (8.4 mag at H$\alpha$), the mean intrinsic brightness of the circumstellar nebula is $I(\mathrm{H}\alpha)\sim6\times10^{-12}$ erg\,cm$^{-2}$\,s$^{-1}$\,arcsec$^{-2}$.
This puts the nebula at the upper end of the emission range of our models, and may be evidence that the mass-loss rate is closer to our upper value than our lower value.
Our simulations have similar electron densities (few $\times10^3$ cm$^{-3}$, see Figs.~\ref{fig:M5V15E_radial}-\ref{fig:M5V25E_radial}) to the radio estimates from \citet{DouClaNegEA10}, so this agreement is not surprising.
This holds even though our model does not match the data very well, because the mass-loss rate is the primary parameter determining the gas density at a given distance from the star.

Winds that are slow enough to allow D-type ionization fronts ($\lesssim22\ \mathrm{km}\,\mathrm{s}^{-1}$ in these models) have blueshifted lines for which the velocity is determined by the ionization front and not by the wind.
We find line blueshifts of $16-19\ \mathrm{km}\,\mathrm{s}^{-1}$, although this depends somewhat on wind density (and probably also on the radiation temperature).
For faster winds an R-type ionization forms, and this has only weak dynamical effects on the wind.
In this regime, nebular lines become good tracers of the wind velocity, although the line velocity is always an underestimate.
This is potentially very useful for estimating wind terminal velocities because the best estimates (usually from masers) come from the wind acceleration region itself and require models for the turbulent velocities in the wind \citep[e.g.~for NML Cyg,][]{ZubLiLimEA04, NagTakOmoEA06}.
\citet{FokNakYunEA12} detected maser emission from the wind of W26, but were not able to measure the wind velocity with their data.
Advances in instrumental capabilities will soon enable measurements of the wind velocity using similar methods as for NML Cyg, so an independent constraint from nebular emission at larger radii is a useful check on these models.

%%% ------------------------------------------------------------
\subsection{Comparison with W26}
%%% ------------------------------------------------------------
We have shown that the there is some agreement between observations and the predictions of the spherically symmetric photoionized wind described in \citet{MacMohGvaEA14}, but that there are also important differences that seem difficult to reconcile with any spherically symmetric wind model.
The whole photoionized nebula seems to be significantly blueshifted wherever there is significant emission, favouring an interpretation where W26 is on the near side of Westerlund 1 and its wind is being swept towards us by some process (either radiative or hydrodynamical) associated with winds, radiation, or supernovae from the star cluster.

This seems more similar to the one-sided photoionization of the wind of NML Cyg \citep[by the Cyg OB2 association;][]{MorJur83} than to the diffuse ionizing radiation field that is invoked to explain the photoionization of Betelgeuse's wind \citep{MacMohGvaEA14}.
If the \citet{MacMohGvaEA14} model is correct, then Betelgeuse is irradiated by an isotropic ionizing radiation field, creating a spherically symmetric ionization front in the star's wind and producing a static shell.
Betelgeuse is in this relatively quiet environment because it is a runaway star that is now far from its place of birth.
NML Cyg is very different, being near a massive and young OB association, and so experiencing a very anisotropic radiation field.
W26 experiences an even more extreme environment because it is still within (at least in projection) the most massive star cluster in our Galaxy, that contains many WR and O-type stars.
It follows then, that we should not be surprised if W26 were to experience an asymmetric radiation field, and if a multi-dimensional extension of the \citet{MacMohGvaEA14} model were required to explain the data.
We discuss this and other alternative explanations in the following subsections.

%%% ------------------------------------------------------------
\subsection{Bow shock}
%%% ------------------------------------------------------------
If the nebula is a bow shock then it must be oriented away from us, i.e.,~W26 is embedded in an ISM with a bulk flow directed approximately towards us.
The nebula is blueshifted by about 25 $\mathrm{km}\,\mathrm{s}^{-1}$ relative to W26, which represents the velocity of the wind material that has been entrained in the bulk flow by the bow shock.
This radial velocity then measures the relative velocity between the star and the ISM.
The standoff distance, $R_\mathrm{SO}$, is then less than or equal to the smallest radial extent of the nebula, which we estimate to be 0.037 pc, and is given by
\begin{equation}
R_\mathrm{SO} = \sqrt{\frac{\dot{M}v_\infty}{4\pi\rho_0 v_0^2}} \;,
\end{equation}
where $\rho_0$ is the ISM density and $v_0=25\ \mathrm{km}\,\mathrm{s}^{-1}$ the flow velocity of the ISM past W26.
The wind is photoionized, so we assume this accelerates the ionized part of the wind to at least $v_\infty=35\ \mathrm{km}\,\mathrm{s}^{-1}$ \citep{MeyGvaLanEA14}.
We then use our upper and lower estimates of $\dot{M}$ for W26 to estimate the value of $\rho_0$ that is required to produce the observed nebula.
We find $\rho_0=2.15\times10^{-20}\ \mathrm{g\,cm}^{-3}$ (or $n_0=10^4\ \mathrm{cm}^{-3}$, where $n_0$ is the number density of H and He nuclei) for $\dot{M}=10^{-4}\ \mathrm{M}_{\odot}\,\mathrm{yr}^{-1}$, or 5 times smaller for $\dot{M}=2\times10^{-5}\ \mathrm{M}_{\odot}\,\mathrm{yr}^{-1}$.
This seems an unreasonably large ISM density for the intracluster medium of Westerlund 1 (X-ray emission suggests $n_0\approx0.5\ \mathrm{cm}^{-3}$; see below), arguing against a straightforward bow shock interpretation for the nebula.
For reference to the following discussion, the ram/thermal pressure required to confine the wind to $R_\mathrm{SO}$ is $p=1.3\times 10^{-7}$ dyne\,cm$^{-2}$ if $\dot{M}=10^{-4}\ \mathrm{M}_{\odot}\,\mathrm{yr}^{-1}$, or $p=2.7\times 10^{-8}$ dyne\,cm$^{-2}$ if $\dot{M}=2\times10^{-5}\ \mathrm{M}_{\odot}\,\mathrm{yr}^{-1}$, both calculated for $v_\infty=35\ \mathrm{km}\,\mathrm{s}^{-1}$.

To further test this model, we relax the assumption that the shocked wind of W26 is accelerated to the full velocity of the bulk flow, which might be possible if there is little mixing at the contact discontinuity.
This then allows a much faster bulk flow with a lower density, but to decrease the flow density to $n_0\sim1\ \mathrm{cm}^{-3}$ we require a gas velocity, $v_0\gtrsim10^3\ \mathrm{km}\,\mathrm{s}^{-1}$.
This sort of gas flow in the intracluster medium is only likely in the immediate vicinity of an extreme driver of momentum, e.g.\ a massive star wind or a supernova.
In comparison to other massive star clusters, \citet{PovBenWhiEA08} estimate ram pressures of $\rho_0v_0^2\sim(0.2-5)\times10^{-8}$ dyne\,cm$^{-2}$ for bow shocks around O stars in the M17 and RCW 49 massive-star-forming regions.
This shows that the nebula around W26 is probably even more extreme than any of the bow shocks studied by \citet{PovBenWhiEA08}, if the bow shock interpretation is correct, suggesting again that perhaps we need an extreme source of momentum very nearby, rather than a cluster-scale gas outflow.

%%% ------------------------------------------------------------
\subsection{Thermal pressure confinement}
%%% ------------------------------------------------------------
Thermal pressure confinement is a promising mechanism, because X-ray observations have revealed a hot, diffuse, intracluster medium \citep{KavNorMeu11}.
This is produced by the shocked wind of the WR and O stars in Westerlund 1, by the numerous supernovae that have already occurred, or a combination of both.
These authors estimate the H number density and temperature of the X-ray emitting plasma to be $n_\mathrm{H}=0.5\ \mathrm{cm}^{-3}$ and $kT=3.7$ keV ($T=4.3\times10^7$ K), respectively.
For a fully ionized plasma containing $1/11$ helium by number, this gives a thermal pressure of $p_0=2.2n_\mathrm{H}kT = 6.5\times10^{-9}$ dyne\,cm$^{-2}$.
Balancing this against the ram pressure of the wind from W26 gives the radius of the wind termination shock, $R_\mathrm{ts}$, of 
\begin{equation}
R_\mathrm{ts} = 0.17\ \mathrm{pc} \sqrt{\frac{\dot{M}}{10^{-4}\ \mathrm{M}_{\odot}\,\mathrm{yr}^{-1}}}
\sqrt{\frac{v_\infty}{35\ \mathrm{km}\,\mathrm{s}^{-1}}} \;.
\end{equation}
For our two mass-loss rates we obtain $R_\mathrm{ts} =0.17$ pc ($\dot{M}=10^{-4}\ \mathrm{M}_{\odot}\,\mathrm{yr}^{-1}$) and $R_\mathrm{ts} =0.075$ pc ($\dot{M}=2\times10^{-5}\ \mathrm{M}_{\odot}\,\mathrm{yr}^{-1}$), both of which are larger than the observed extent of the nebula (0.037 pc).
For the lower mass-loss rate, however, $R_\mathrm{ts}$ is not much larger than the observed nebula, and a factor of 4 increase in the external pressure would bring the numbers into agreement
(or a $4\times$ decrease in $\dot{M}$ to $5\times10^{-6}\ \mathrm{M}_{\odot}\,\mathrm{yr}^{-1}$, although this is unlikely for such a massive red supergiant).
Some asymmetric pressure is still required, for example a subsonic bulk flow \citep[e.g.][]{MacGvaMohEA15}, to create the asymmetric nebula, so purely thermal-pressure confinement is not a viable explanation.

%%% ------------------------------------------------------------
\subsection{Wind-wind interaction}
%%% ------------------------------------------------------------
The star W9 is 0.23 arcmin from W26 in projection, or 0.24 pc at the distance of Westerlund 1.
It is the strongest radio emitter in Westerlund 1 and is surrounded by a dense wind with $\dot{M}\approx3.3\times10^{-4}\ \mathrm{M}_{\odot}\,\mathrm{yr}^{-1}$ and $v_\infty \sim200\ \mathrm{km}\,\mathrm{s}^{-1}$ \citep[][model 3]{DouClaNegEA10}.
If the winds of the two stars collide, then the wind of W26 will be shocked and decelerated at a distance, $R_\mathrm{wc}$, from W26 that is given by
\begin{equation}
R_\mathrm{wc} = R_0 \sqrt{c} \,\frac{1-\sqrt{c}}{1-c} \;, \label{eqn:windcoll}
\end{equation}
where $R_0$ is the separation of the two stars and $c$ is the ratio of the wind momenta:
\begin{equation}
c= \frac{\left(\dot{M} v_\infty\right)_\mathrm{W26}}{\left(\dot{M} v_\infty\right)_\mathrm{W9}} \;.
\end{equation}
For the W9 wind properties above, and using $\dot{M}=10^{-4}\ \mathrm{M}_{\odot}\,\mathrm{yr}^{-1}$ and $v_\infty=35\ \mathrm{km}\,\mathrm{s}^{-1}$ for W26, we find $R_\mathrm{wc}=0.19R_0$.
For W26 with $\dot{M}=2\times10^{-5}\ \mathrm{M}_{\odot}\,\mathrm{yr}^{-1}$, this reduces to $R_\mathrm{wc}=0.093R_0$.
The observed radius of the nebula is 0.037 pc, implying $R_0=0.20$ pc for the larger $\dot{M}$ estimate or $R_0=0.40$ pc for the smaller $\dot{M}$.
This is again comparable to the observed separation between the two stars, and so should be considered as a plausible alternative scenario to a simple photoionized wind.

Considering the pressure of the X-ray plasma quoted above ($6.5\times10^{-9}$ dyne\,cm$^{-2}$), the wind of W9 would have a termination shock at $R_\mathrm{ts} =0.73$ pc.
The freely expanding wind region would have a diameter of 1.4 arcmin at the distance of Westerlund 1, covering a large fraction of the core of Westerlund 1.
This means that W26 and many other stars in the cluster may well be embedded in the freely-expanding wind of W9, and could be experiencing a wind-wind collision.

%%% ------------------------------------------------------------
\subsection{Photoionized wind}
\label{sec:disc:pion}
%%% ------------------------------------------------------------
We have shown that a spherically symmetric model for a photoionized wind has severe problems explaining the observations, but perhaps an asymmetric model could work.
This was proposed by \citet{MorJur83} for NML Cyg, to explain the arc-shaped H\,\textsc{ii} region on one side of the star.
In this model there is one dominant radiation source so that the wind is photoionized mainly from one direction and the ionization front has a shape similar to a bow shock \citep[for mathematical details, see][]{MorJur83}.
If the ionization front is D-type then the wind can be decelerated into a one-sided shell, and the asymmetry of the shell leads to non-radial flows away from the radiation source because of the rocket effect \citep{OorSpi55}.
This requires multi-dimensional simulations to model quantitatively, beyond the scope of this paper, but we intend to investigate this in more detail in future work.

The EUV fluxes in Table~\ref{tab:sims} show what is required to ionize the wind of W26, and simple estimates show that nearby stars can provide these fluxes.
The Wolf-Rayet star WR77o (WR B) \citep{CroHadClaEA06} is about 0.5 pc from W26 (projected distance).
According to \citet{Cro07}, it should have an ionizing photon luminosity, $Q_0=10^{49.4}$ s$^{-1}$.
For a distance of 0.5 pc, the EUV flux at W26 is $F_\gamma=8.4\times10^{11}\ \mathrm{cm}^{-2}\,\mathrm{s}^{-1}$, comparable to the largest fluxes used in the spherically symmetric simulations.
Its wind, by contrast, is probably not strong enough to shape the nebula, because \citet{Cro07} estimate $\dot{M}=1.6\times10^{-5}\ \mathrm{M}_{\odot}\,\mathrm{yr}^{-1}$ and $v_\infty=1300\ \mathrm{km}\,\mathrm{s}^{-1}$, too weak to confine W26's wind according to Equation~(\ref{eqn:windcoll}).

Much closer to W26, the O9\,Iab star W25 has an ionizing photon luminosity of $Q_0=(10^{48.8}-10^{49.0})$ s$^{-1}$ \citep{MarSchHil05}, and is only 0.09 pc from W26 in projection.
This corresponds to an ionizing flux of $F_\gamma=(6.4-10.0)\times10^{12}\ \mathrm{cm}^{-2}\,\mathrm{s}^{-1}$, and so could ionize the wind of W26 even deeper than the observed bright nebula, if the physical separation of the two stars is comparable to the projected separation.
Another candidate is W9, discussed above in the context of wind-wind collision, but we do not have any reliable estimates of its ionizing photon luminosity.

%%% ------------------------------------------------------------
\subsection{Hot companion star?}
%%% ------------------------------------------------------------
Westerlund 1 was observed with \textit{Chandra} in X-rays and analysis of point sources was presented by \citet{SkiSimZheEA06}.
W9 was clearly detected, which they argue is because W9 is a colliding-wind binary, but W26 (referred to as Ara A) was not detected although it was observed.
If W26 was a binary system with a neutron star (or black hole) companion that provided the ionizing photons to produce the nebula, then it would also be bright in X-rays, and this is not observed.
W26 is also too young to host a white dwarf companion, so we can exclude that a compact companion is ionizing the wind.
It is also not a photometric variable star \citep{Bon07} and, although it has spectral variations, these are consistent with other massive red supergiants and not indicative of duplicity \citep{ClaRitNeg10}.

%%% ------------------------------------------------------------
\subsection{Outlook}
%%% ------------------------------------------------------------
We see that there are many radiation sources that should photoionize the wind of W26 to at least the radius of the observed nebula, and potentially much closer to the star,
although it is not yet clear if one-sided photoionization could produce the velocity signatures of the observed nebula.
We have also shown that the measured X-ray plasma pressure could contribute to confining the wind.
The large temperature difference between the photoionized red supergiant wind and the X-ray plasma of the intracluster medium may also set up a strong thermal conduction front at the contact discontinuity between wind and ISM, which has been shown to strongly modify the properties of O star bow shocks \citep{ComKap98, MeyMacLanEA14}.
We suggest, however, that
a wind-wind collision between the winds of W9 and W26 
is the most obvious scenario that could produce
the observed nebula.
Naively, we expect that a bow shock model (possibly from a wind-wind collision with the wind of W9) is the only one that can explain why \emph{all} of the bright emission is blueshifted.

It would be very useful to have higher resolution spectra, to reduce the uncertainties in the radial velocity of the spectral lines, and to search for substructure in the lines (e.g.~redshifted absorption).
Better spatial coverage of the nebula is also required.
In particular spectra of the triangular nebula discovered by \citet{WriWesDreEA14} and the emission connecting it with the W26 nebula could reveal velocity gradients with distance from W26.
This would show whether the gas is being accelerated away from W26, as expected for gas that is entrained by a wind.
This would also provide more conclusive evidence for whether W26 is on the near or far side of Westerlund 1: if on the near side then the gas should be more blueshifted the further it is from W26.

%%% ------------------------------------------------------------
%%% ------------------------------------------------------------
\section{Conclusions}
\label{sec:conclusions}
%%% ------------------------------------------------------------
%%% ------------------------------------------------------------
We have presented the first predictions for nebular emission from photoionized winds of red supergiants, using spherically symmetric, radiation-hydrodymamical simulations and postprocessing radiative transfer.
Although the results have been compared only to W26, they also have application to photoionized winds around other red supergiants.
The simulations improve on the previous work of \citep{MacMohGvaEA14} in that the gas temperature is calculated self-consistently from mechanical and radiative heating/cooling, allowing us to make realistic predictions of temperature-sensitive nebular emission from ionized gas.

We produce position-velocity datasets for H$\alpha$ and [N\,II] spectral lines from the simulations, and study the properties of this emission.
The peak brightness of the simulated nebulae is $10^{-13}-10^{-11}$ erg\,cm$^{-2}$\,s$^{-1}$\,arcsec$^{-2}$, for mass-loss rates, $\dot{M}=(0.2-1)\times10^{-4}\ \mathrm{M}_{\odot}\,\mathrm{yr}^{-1}$, and wind velocities, $v_\infty=(15-30)$ km\,s$^{-1}$.
Simulations of slow winds that are decelerated into a dense shell show strongly limb-brightened line emission, and radial velocity of spectral lines is independent of the wind speed.
Faster winds ($\gtrsim22\ \mathrm{km}\,\mathrm{s}^{-1}$) that cannot form a dense shell have less limb-brightening and the line radial velocity is a good tracer of the wind speed.
All of the simulations have the most blue-shifted emission at the position of the star, and the radial velocity of the lines goes to zero at the region of brightest emission (the limb), as is typical for emission from an expanding shell.

If nitrogen is not enriched in the wind, then the predicted line ratio of [N\,II] to H$\alpha$ is about 0.9.
Nitrogen is enhanced in red supergiant winds of a factor of about 3 \citep[cf.][]{BroDeMCanEA11}, and for this we obtain line ratios of about $2.5-2.8$.

We have compared spectra of the H$\alpha$ and [N\,II] emission lines from the nebula around the red supergiant W26 in Westerlund 1 \citep{WriWesDreEA14}
with our simulations
to investigate the origin and physical properties of the nebula.
We also obtain a temperature for the star $T_\mathrm{eff}=3600$ K from the same spectra.
The brightness distribution of the emission and the line ratio of [N\,II] to H$\alpha$ 
  show similarities to
the photoionization model, but there are discrepencies for the radial velocity of the emission.
All of the observed bright nebular emission is blueshifted with respect to W26 by 15-25 km\,s$^{-1}$, and the brightest emission is the most blueshifted, neither of which can be reproduced by a spherically symmetric model of a photoionized wind.
From the observed line ratio ($\gtrsim2.25$), we can conclude that the nebula is photoionized, however, and that it is enriched in nitrogen by a factor of about 2.35.
This shows that the nebula is produced by the wind of W26, although the discrepencies with our model suggest that the outer edge of the nebula may be hydrodynamically confined rather than freely expanding.

We propose that a wind-wind collision scenario,
where the wind of W26 is swept towards the observer by the stronger wind of the nearby star W9,
may match the data better.
Our results suggest that the circumstellar medium may have more structure even closer to the star than the observed nebula, because it does not appear as though we have detected the ionization front and its associated shell in the current observations.
Multidimensional models and/or simulations, and spectra with higher resolution and better spatial coverage are necessary to progress further in understanding the circumstellar nebula around W26.
Further study of this nebula will contribute to our understanding of what happens to the mass lost by red supergiants in star clusters.
It is important to know whether it is rapidly heated and expelled from the cluster, or whether it can remain within the cluster for an extended period of time, remain cool and dense, and potentially be available for secondary star formation in the cluster.

%%% ------------------------------------------------------------
%%% ------------------------------------------------------------
\begin{acknowledgements}
This project was supported by the the Deutsche Forschungsgemeinschaft priority program 1573, Physics of the Interstellar Medium.
LF acknowledges financial support from the Alexander von Humboldt foundation.
The authors gratefully acknowledge the computing time granted by the John von Neumann Institute for Computing (NIC) and provided on the supercomputer JUROPA at J\"ulich Supercomputing Centre (JSC).
We thank I.\ Negueruela for radial velocity data for W26 and A.\ Whitworth for useful discussions on star formation within massive star clusters.
Based on observations made with ESO Telescopes at the La Silla Paranal Observatory under programme ID\,087.D-0673(A).
We are grateful to the referee for comments that significantly improved the presentation of our work.
\end{acknowledgements}
%%% ------------------------------------------------------------
%%% ------------------------------------------------------------

\bibliographystyle{aa}
\bibliography{../../../../../documentation_misc/bibtex/refs}

\appendix

%%% ------------------------------------------------------------
%%% ------------------------------------------------------------
\section{Results for the ESE side of the nebula} \label{app:ESEside}
\begin{figure}
\centering
\includegraphics[width=0.9\hsize]{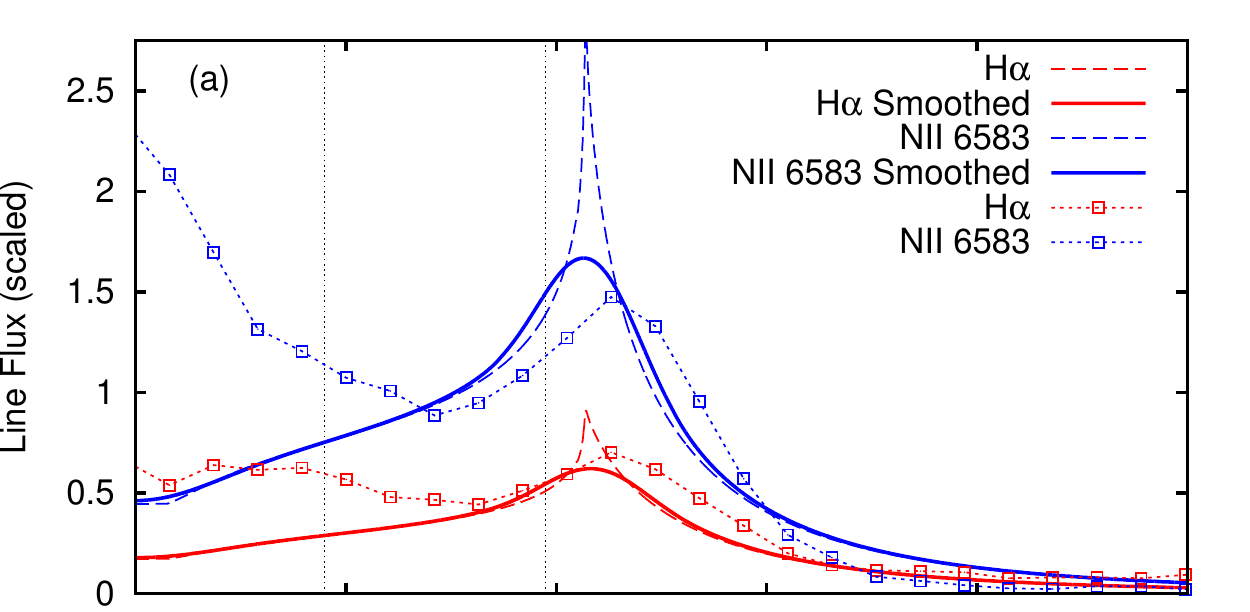}
\includegraphics[width=0.9\hsize]{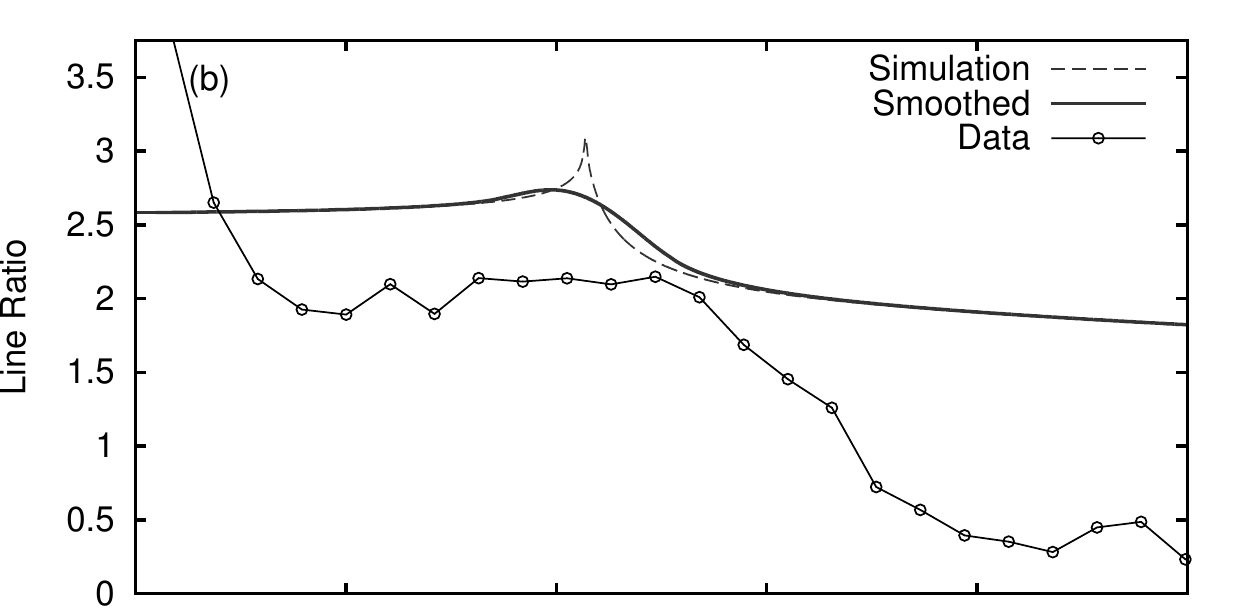}
\includegraphics[width=0.9\hsize]{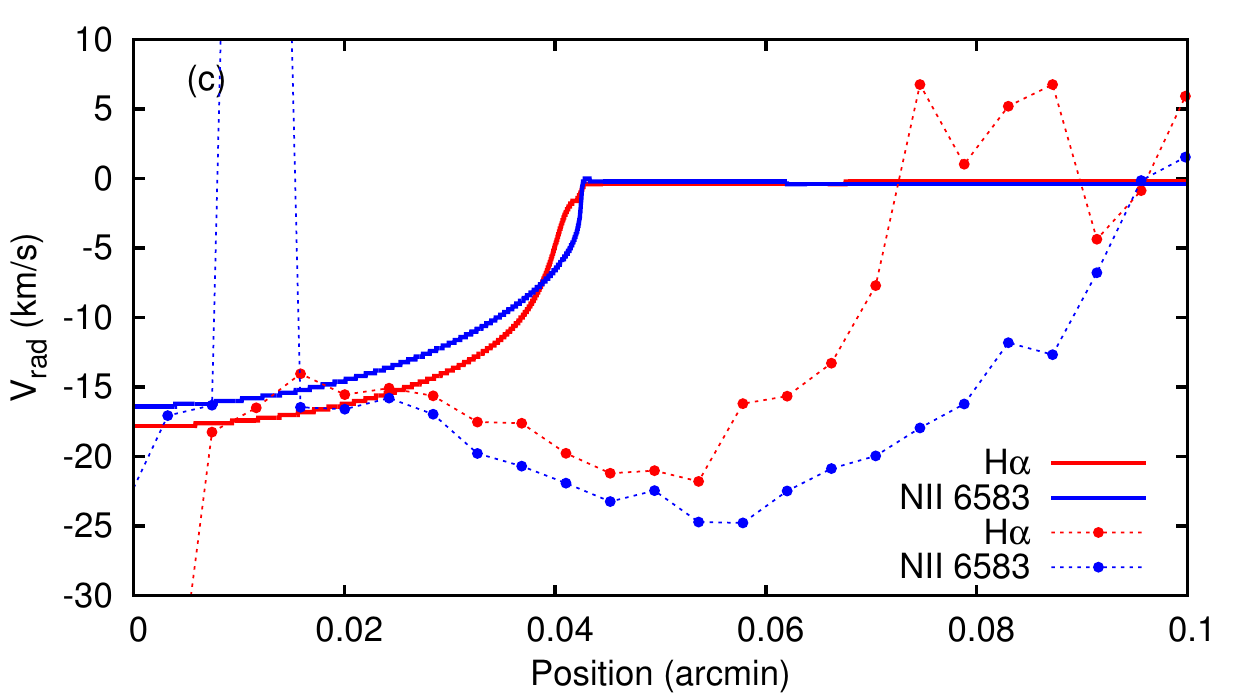}
\caption{
  The H$\alpha$ and [N\,II] spectral lines from the simulation M5V20E, with $\dot{M}=2\times10^{-5}\ \mathrm{M}_{\odot}\,\mathrm{yr}^{-1}$, $v_\infty=20\ \mathrm{km}\,\mathrm{s}^{-1}$, and $F_\gamma=2.78\times10^{10}\ \mathrm{cm}^{-2}\,\mathrm{s}^{-1}$, after 0.1 Myr of evolution.
  The solid lines show the simulation, and the dotted lines with points show the observations to the ESE of W26.
  The panels show \textbf{(a)} line brightness, \textbf{(b)} line ratio [N\,II]/H$\alpha$, and \textbf{(c)} radial velocity of the peak emission, all as a function of distance from the star.
  }
\label{fig:M5V20E}
\end{figure}

\begin{figure}
\centering
\includegraphics[width=0.9\hsize]{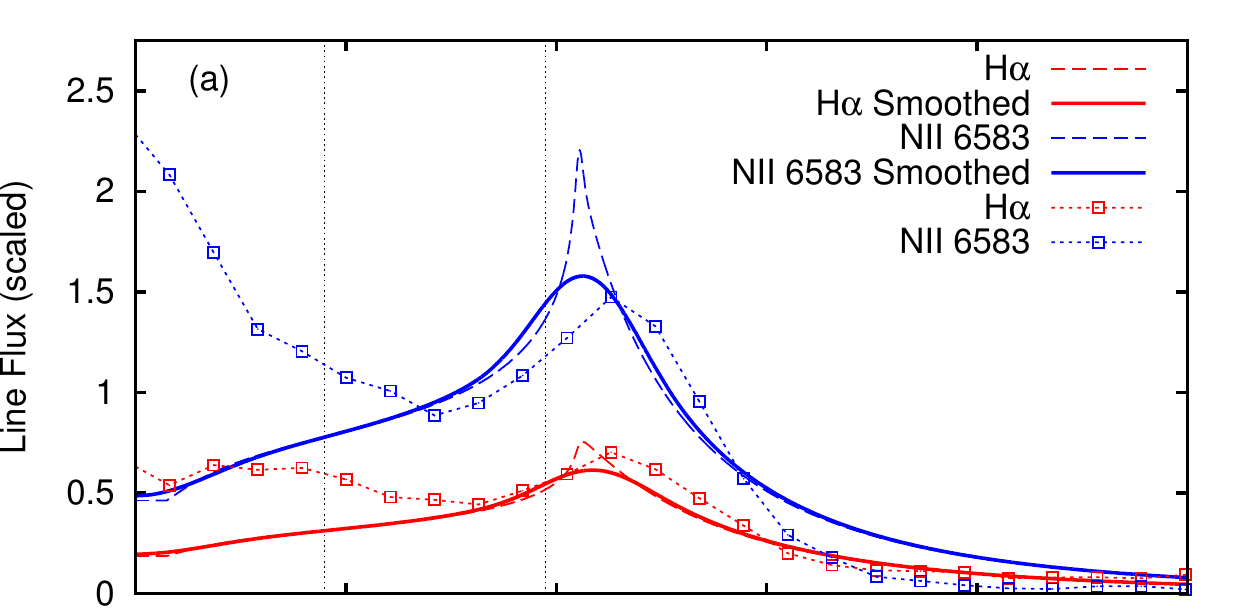}
\includegraphics[width=0.9\hsize]{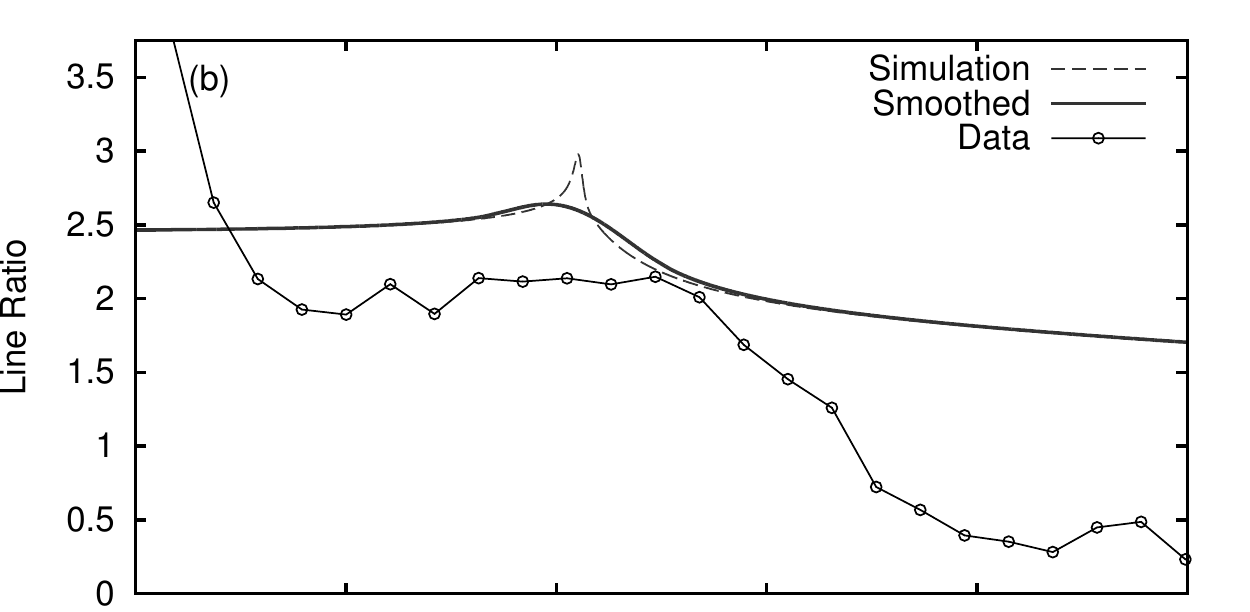}
\includegraphics[width=0.9\hsize]{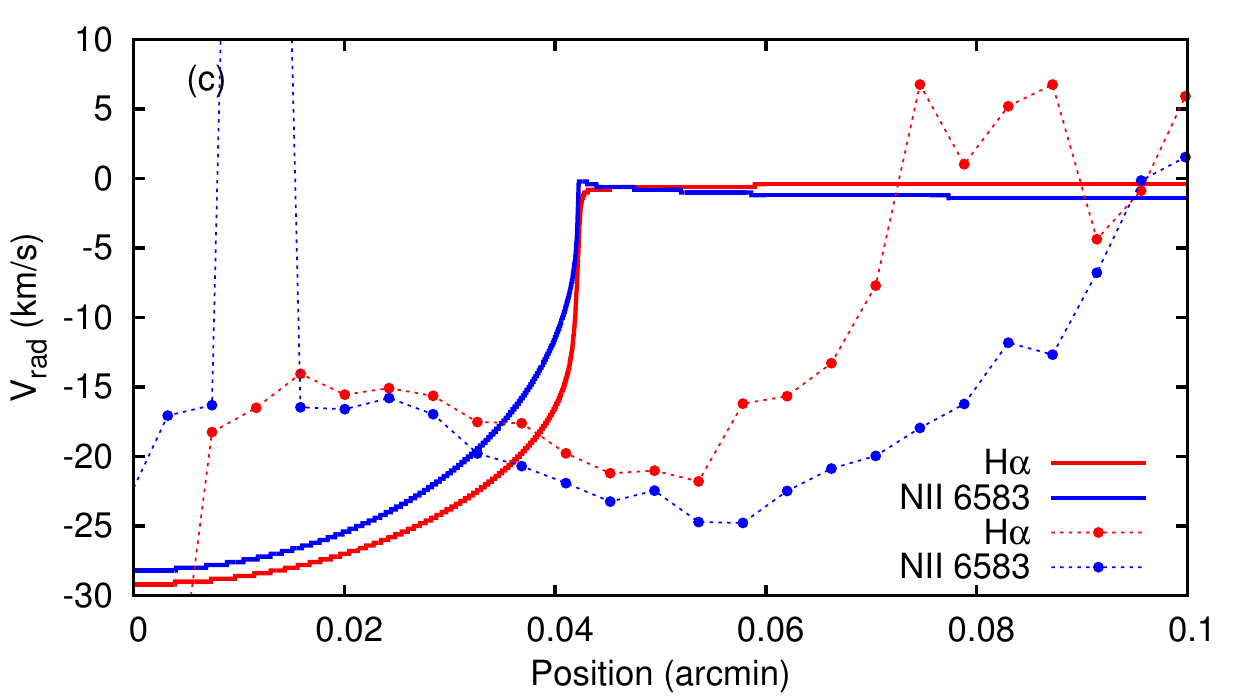}
\caption{
  The H$\alpha$ and [N\,II] spectral lines from the simulation M5V30E, with $\dot{M}=2\times10^{-5}\ \mathrm{M}_{\odot}\,\mathrm{yr}^{-1}$, $v_\infty=30\ \mathrm{km}\,\mathrm{s}^{-1}$, and $F_\gamma=1.26\times10^{10}\ \mathrm{cm}^{-2}\,\mathrm{s}^{-1}$, after 0.1 Myr of evolution.
  The lines and symbols are the same as in Fig.~\ref{fig:M5V20E}.
  }
\label{fig:M5V30E}
\end{figure}

\begin{figure}
\centering
\includegraphics[width=0.9\hsize]{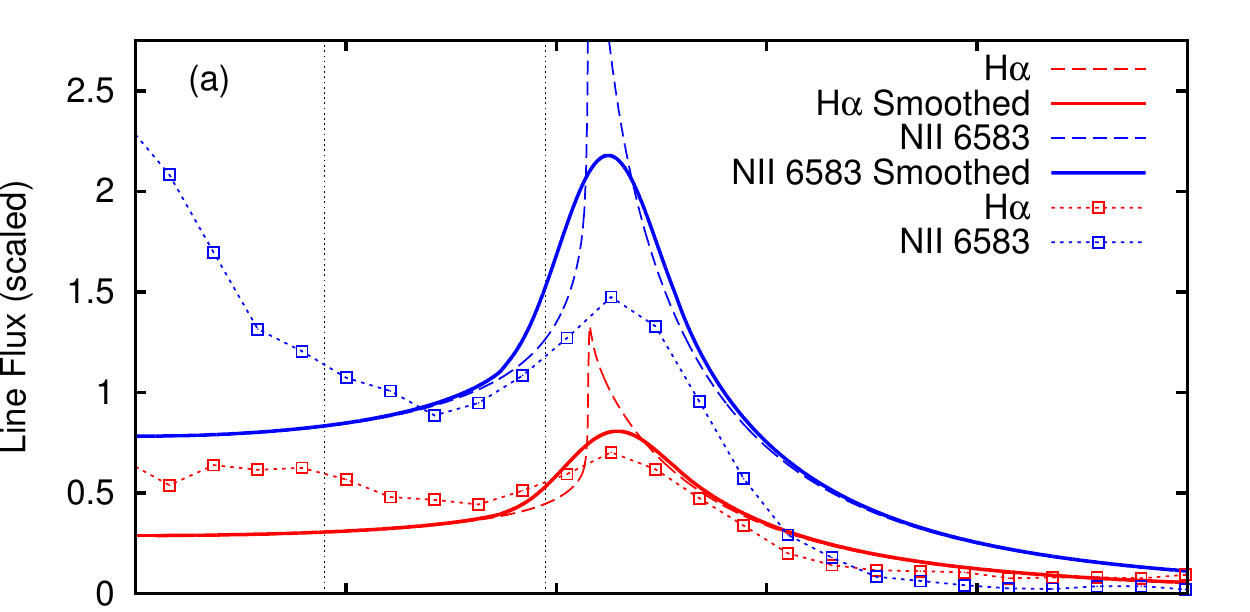}
\includegraphics[width=0.9\hsize]{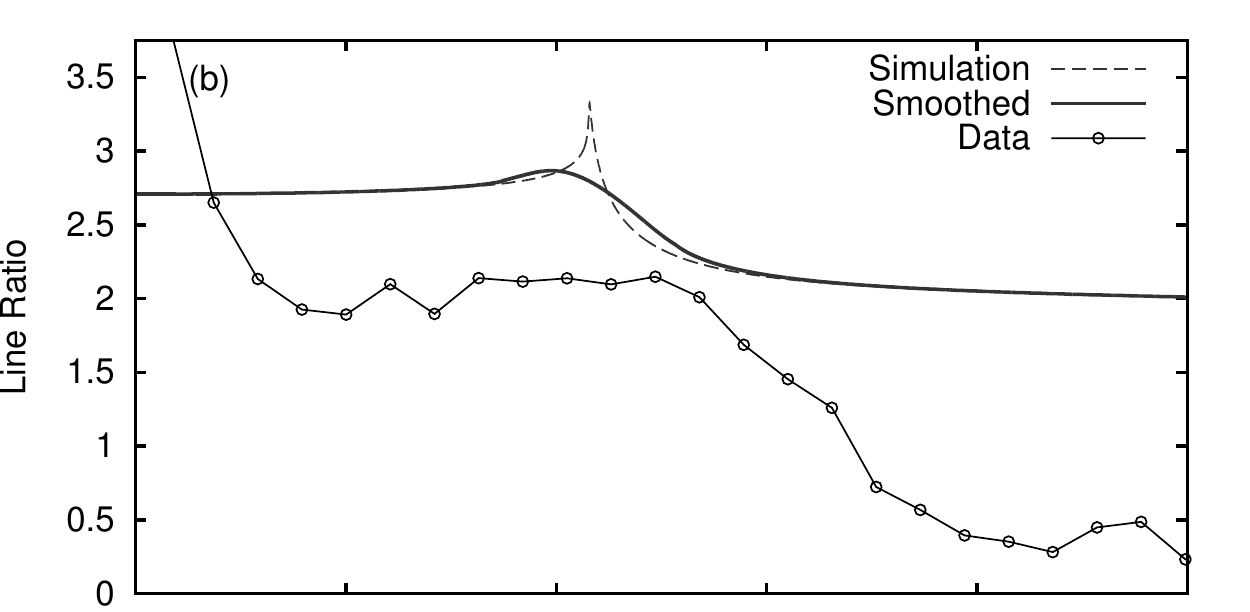}
\includegraphics[width=0.9\hsize]{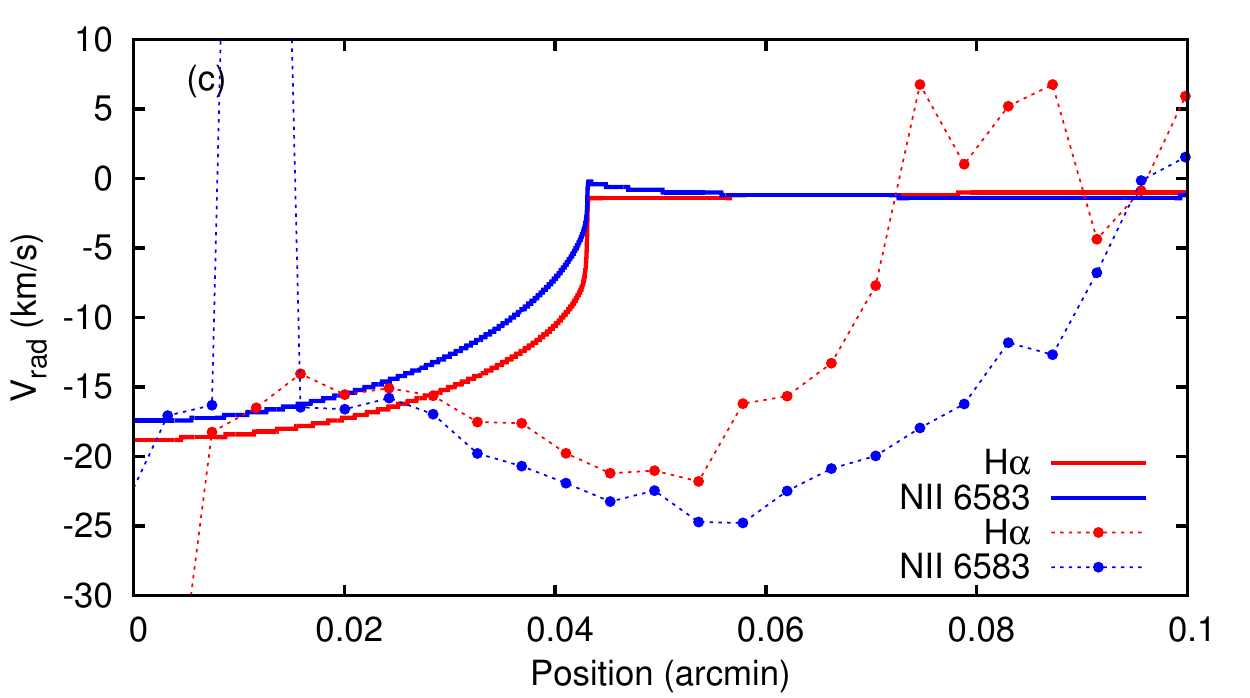}
\caption{
  The H$\alpha$ and [N\,II] spectral lines from the simulation M4V15E, with $\dot{M}=10^{-4}\ \mathrm{M}_{\odot}\,\mathrm{yr}^{-1}$, $v_\infty=15\ \mathrm{km}\,\mathrm{s}^{-1}$, and $F_\gamma=3.67\times10^{11}\ \mathrm{cm}^{-2}\,\mathrm{s}^{-1}$, after 0.1 Myr of evolution.
  The lines and symbols are the same as in Fig.~\ref{fig:M5V20E}.
  }
\label{fig:M4V15E}
\end{figure}

\begin{figure}
\centering
\includegraphics[width=0.9\hsize]{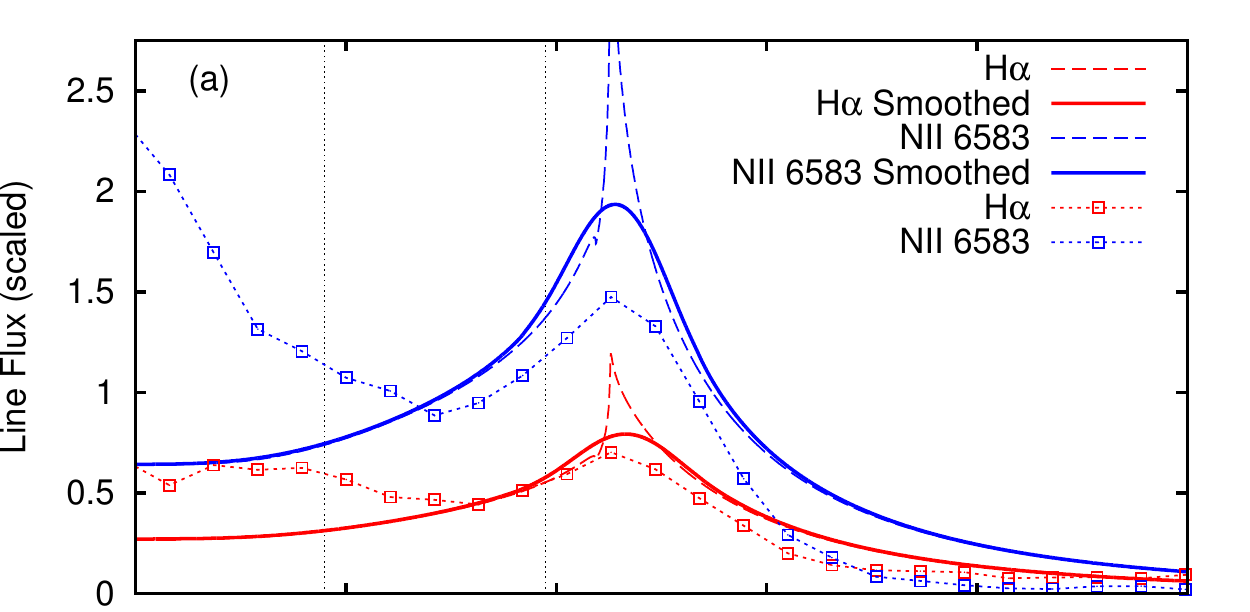}
\includegraphics[width=0.9\hsize]{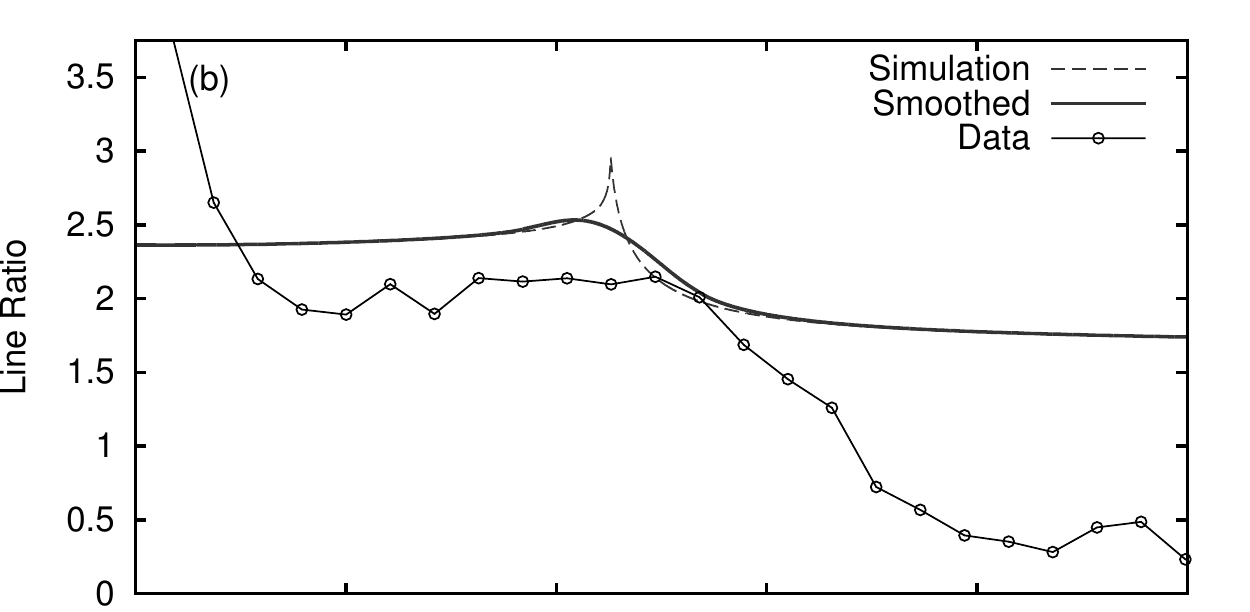}
\includegraphics[width=0.9\hsize]{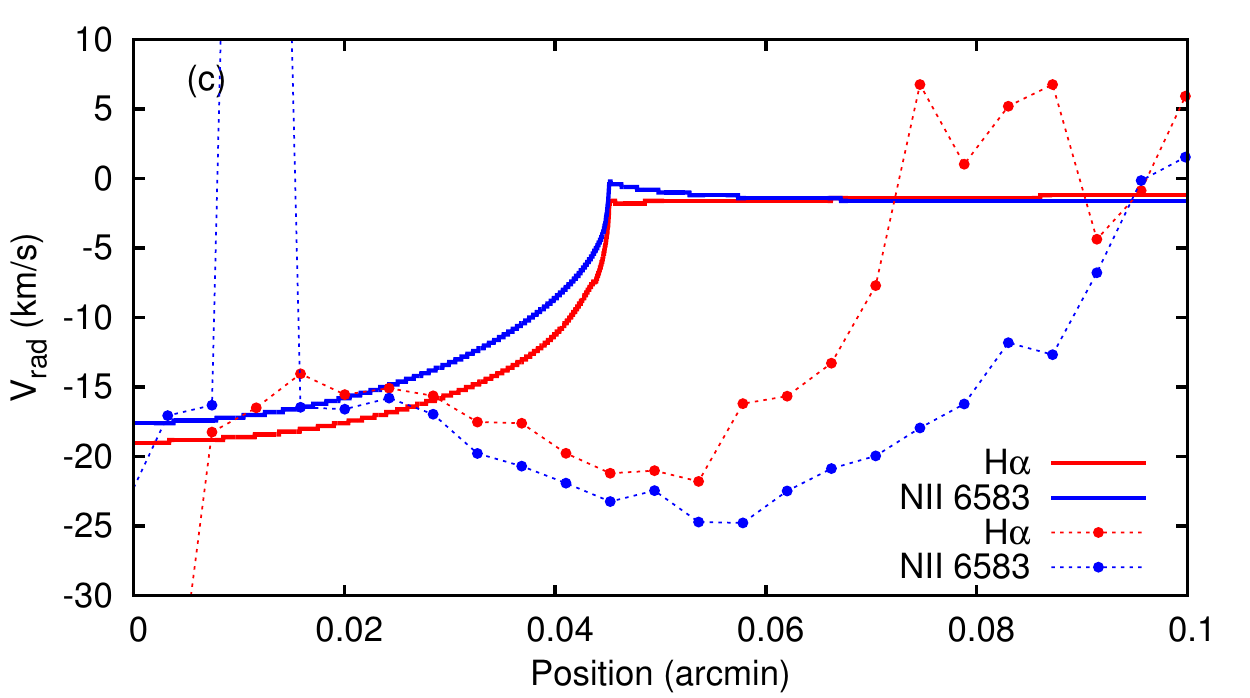}
\caption{
  The H$\alpha$ and [N\,II] spectral lines from the simulation M4V20E, with $\dot{M}=10^{-4}\ \mathrm{M}_{\odot}\,\mathrm{yr}^{-1}$, $v_\infty=20\ \mathrm{km}\,\mathrm{s}^{-1}$, and $F_\gamma=5.05\times10^{11}\ \mathrm{cm}^{-2}\,\mathrm{s}^{-1}$, after 0.1 Myr of evolution.
  The lines and symbols are the same as in Fig.~\ref{fig:M5V20E}.
  }
\label{fig:M4V20E}
\end{figure}

\begin{figure}
\centering
\includegraphics[width=0.9\hsize]{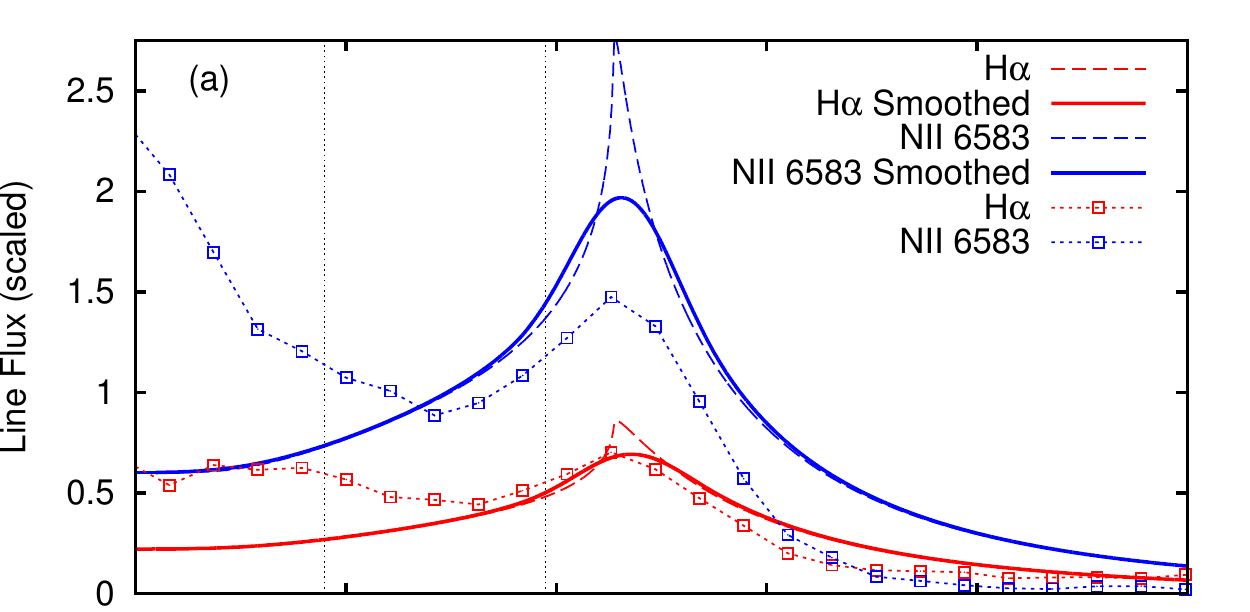}
\includegraphics[width=0.9\hsize]{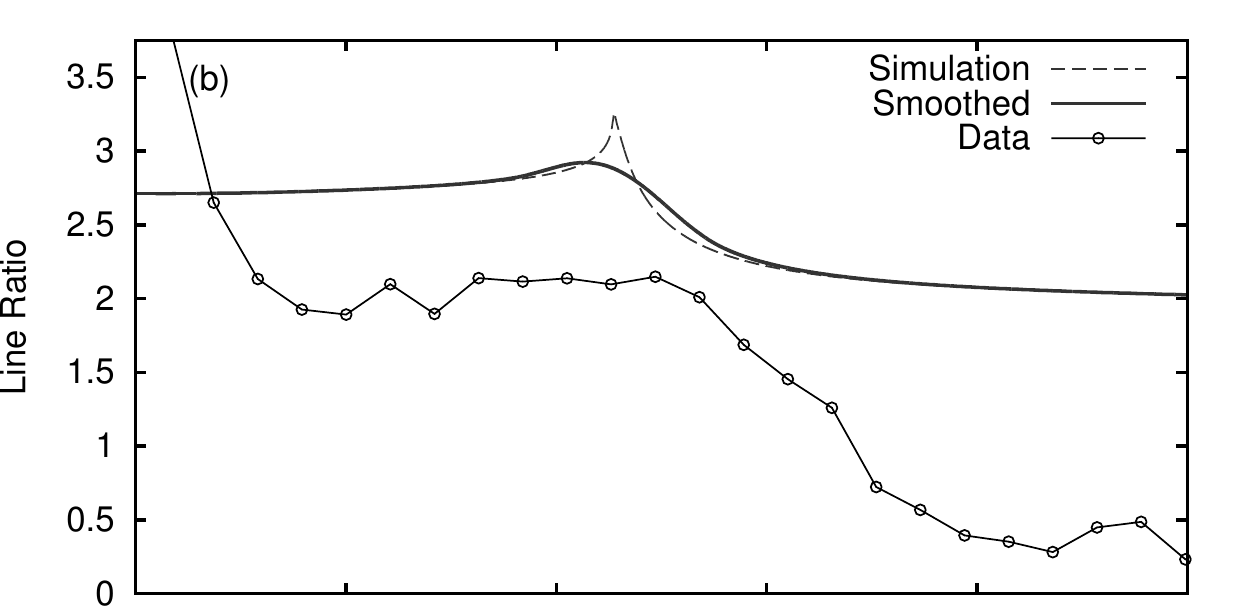}
\includegraphics[width=0.9\hsize]{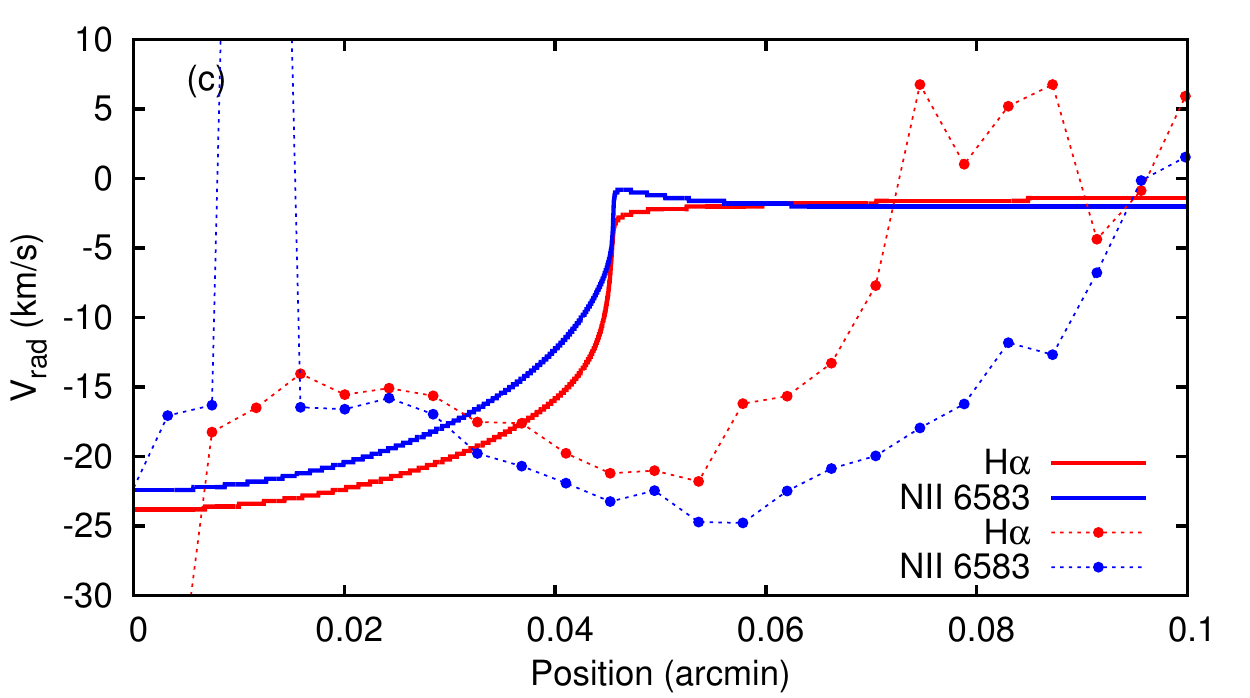}
\caption{
  The H$\alpha$ and [N\,II] spectral lines from the simulation M4V25E, with $\dot{M}=10^{-4}\ \mathrm{M}_{\odot}\,\mathrm{yr}^{-1}$, $v_\infty=25\ \mathrm{km}\,\mathrm{s}^{-1}$, and $F_\gamma=3.10\times10^{11}\ \mathrm{cm}^{-2}\,\mathrm{s}^{-1}$, after 0.1 Myr of evolution.
  The lines and symbols are the same as in Fig.~\ref{fig:M5V20E}.
  }
\label{fig:M4V25E}
\end{figure}

\begin{figure}
\centering
\includegraphics[width=0.9\hsize]{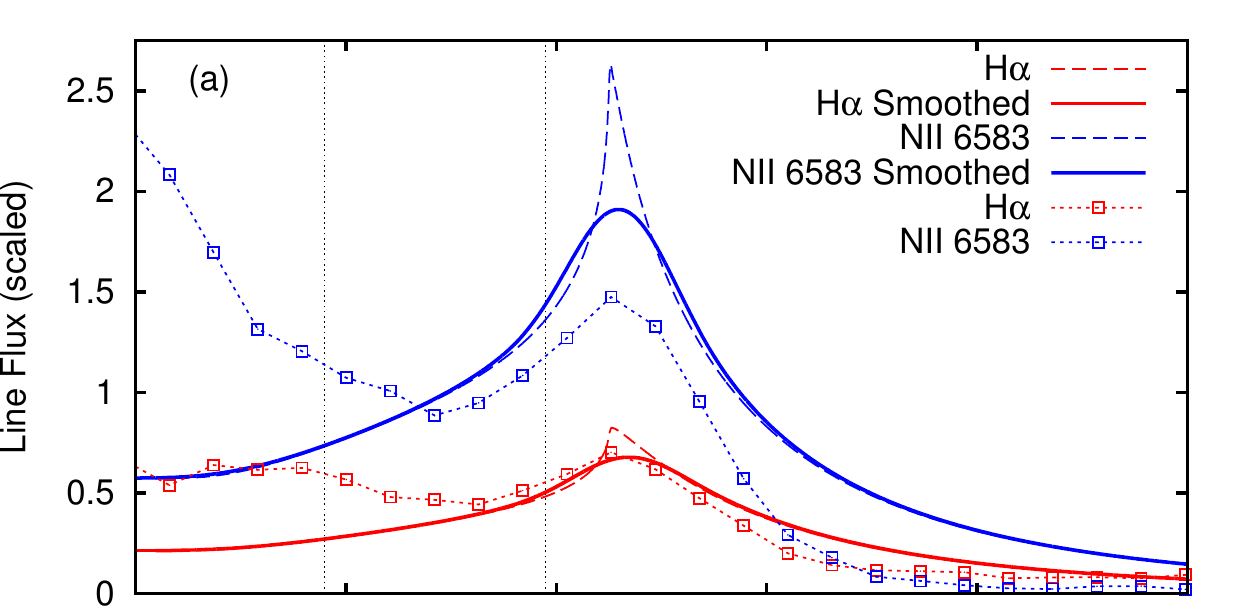}
\includegraphics[width=0.9\hsize]{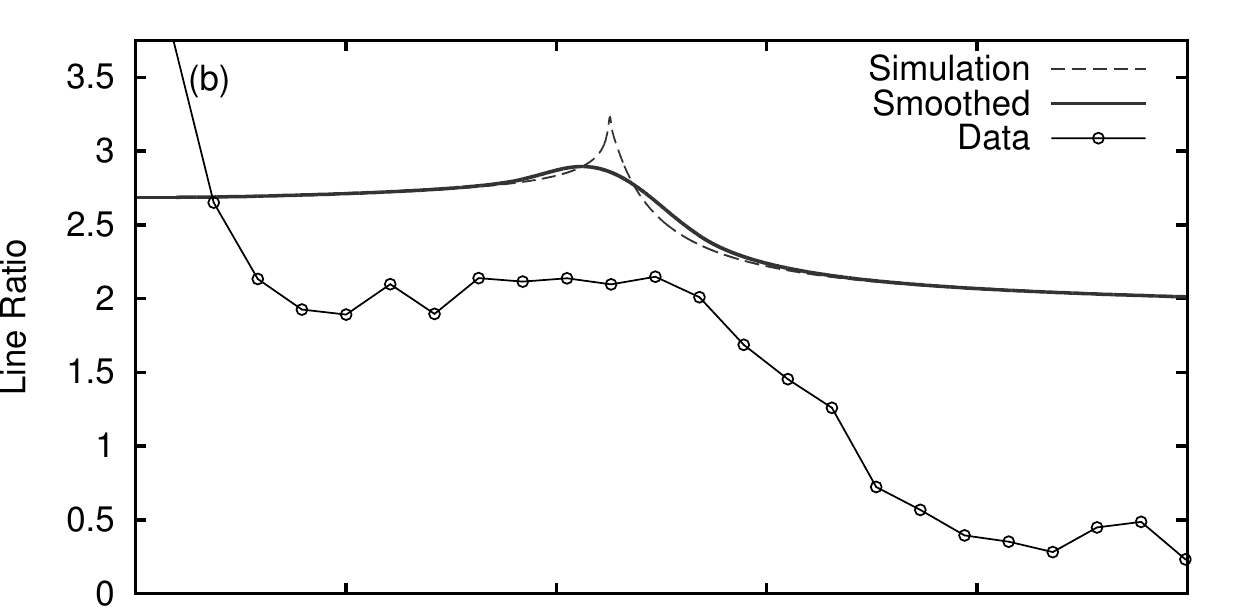}
\includegraphics[width=0.9\hsize]{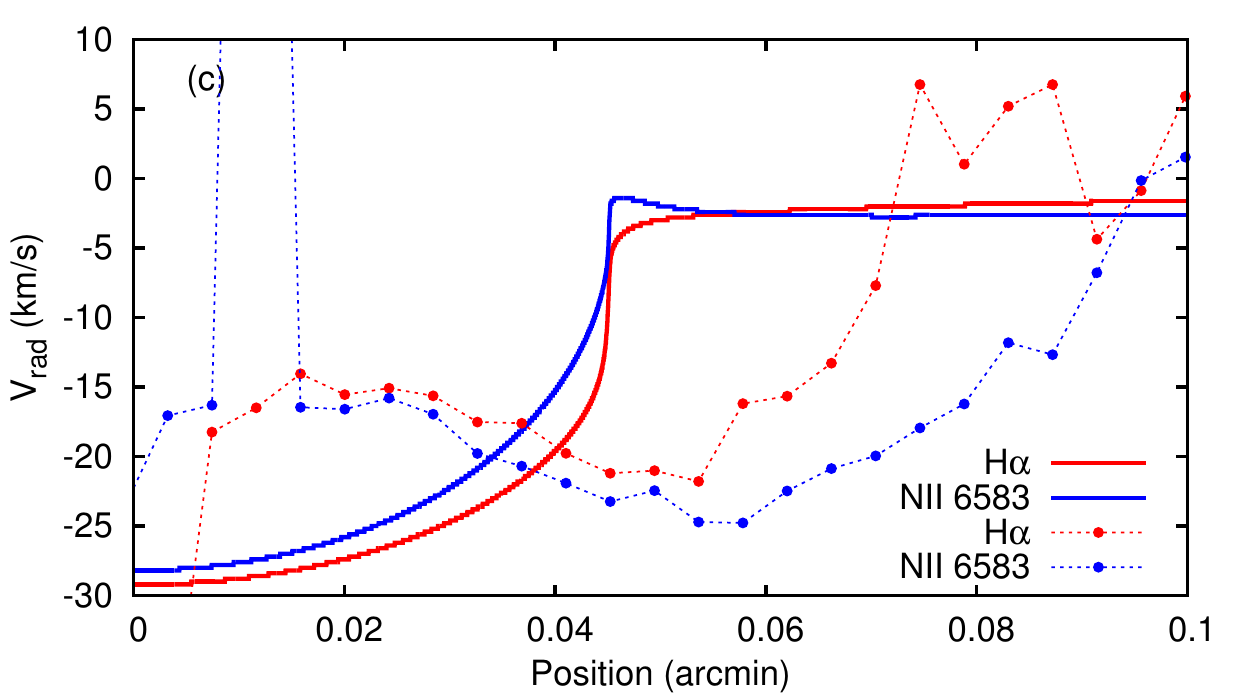}
\caption{
  The H$\alpha$ and [N\,II] spectral lines from the simulation M4V30E, with $\dot{M}=10^{-4}\ \mathrm{M}_{\odot}\,\mathrm{yr}^{-1}$, $v_\infty=30\ \mathrm{km}\,\mathrm{s}^{-1}$, and $F_\gamma=2.15\times10^{11}\ \mathrm{cm}^{-2}\,\mathrm{s}^{-1}$, after 0.1 Myr of evolution.
  The lines and symbols are the same as in Fig.~\ref{fig:M5V20E}.
  }
\label{fig:M4V30E}
\end{figure}

%%% ------------------------------------------------------------
%%% ------------------------------------------------------------
Here we plot the line intensity, radial velocity, and line ratio for the simulations that are not discussed in the text.
Figs.~\ref{fig:M5V20E} and \ref{fig:M5V30E} show M5V20E and M5V30E, respectively, and 
Figs.~\ref{fig:M4V15E}-\ref{fig:M4V30E} show results for M4V15E, M4V20E, M4V25E, and M4V30E.

%%%%%%%%%%%%%%%%%%%%%%%%%%%%%%%%%%%%%%%%%%%%%%%%%%%%%%%%%%%%%%%%%%%%%
%%%%%%%  Mdot=2\times10^{-5} Msun/yr, FAR 
%%%%%%%%%%%%%%%%%%%%%%%%%%%%%%%%%%%%%%%%%%%%%%%%%%%%%%%%%%%%%%%%%%%%%

%%%%%%%%%%%%%%%%%%%%%%%%%%%%%%%%%%%%%%%%%%%%%%%%%%%%%%%%%%%%%%%%%%%%%
%%%%%%%  Mdot=10^{-4} Msun/yr, FAR 
%%%%%%%%%%%%%%%%%%%%%%%%%%%%%%%%%%%%%%%%%%%%%%%%%%%%%%%%%%%%%%%%%%%%%

%%% ------------------------------------------------------------
%%% ------------------------------------------------------------
\section{Results for the WNW side of the nebula} \label{app:WNWside}
%%% ------------------------------------------------------------
%%% ------------------------------------------------------------

\begin{figure}
\centering
\includegraphics[width=0.9\hsize]{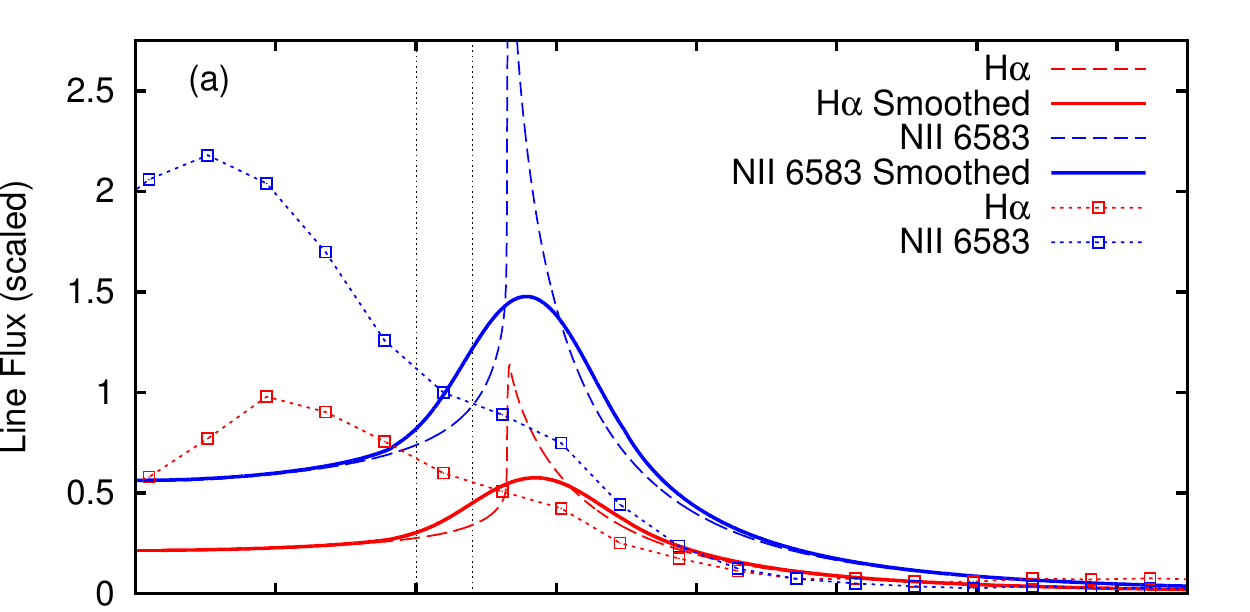}
\includegraphics[width=0.9\hsize]{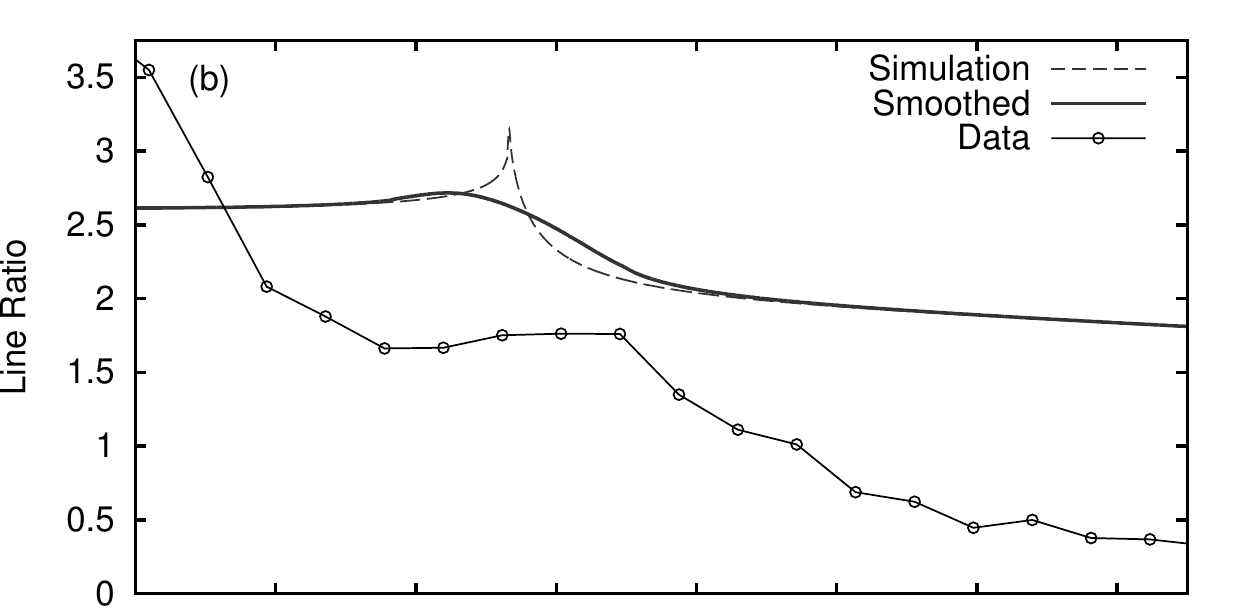}
\includegraphics[width=0.9\hsize]{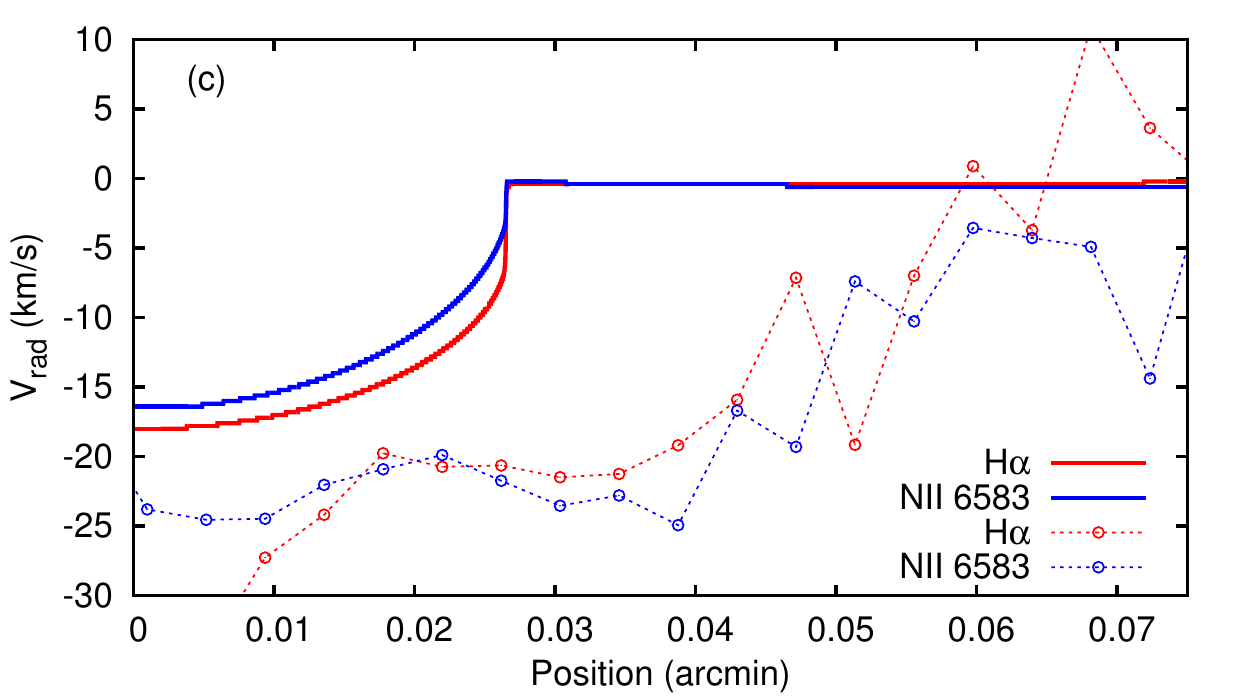}
\caption{
  The H$\alpha$ and [N\,II] spectral lines from the simulation M5V15W, after 0.1 Myr of evolution.
  The solid lines show the simulation, and the dotted lines with points show the observations to the WNW of W26.
  The panels show \textbf{(a)} line brightness, \textbf{(b)} line ratio [N\,II]/H$\alpha$, and \textbf{(c)} radial velocity of the peak emission, all as a function of distance from the star.
  Vertical dotted lines in panel (a) show the range of radii used to set the normalisation of the simulated and observed [N\,II] line brightness.
  }
\label{fig:M5V15W}
\end{figure}

\begin{figure}[h]
\centering
\includegraphics[width=0.9\hsize]{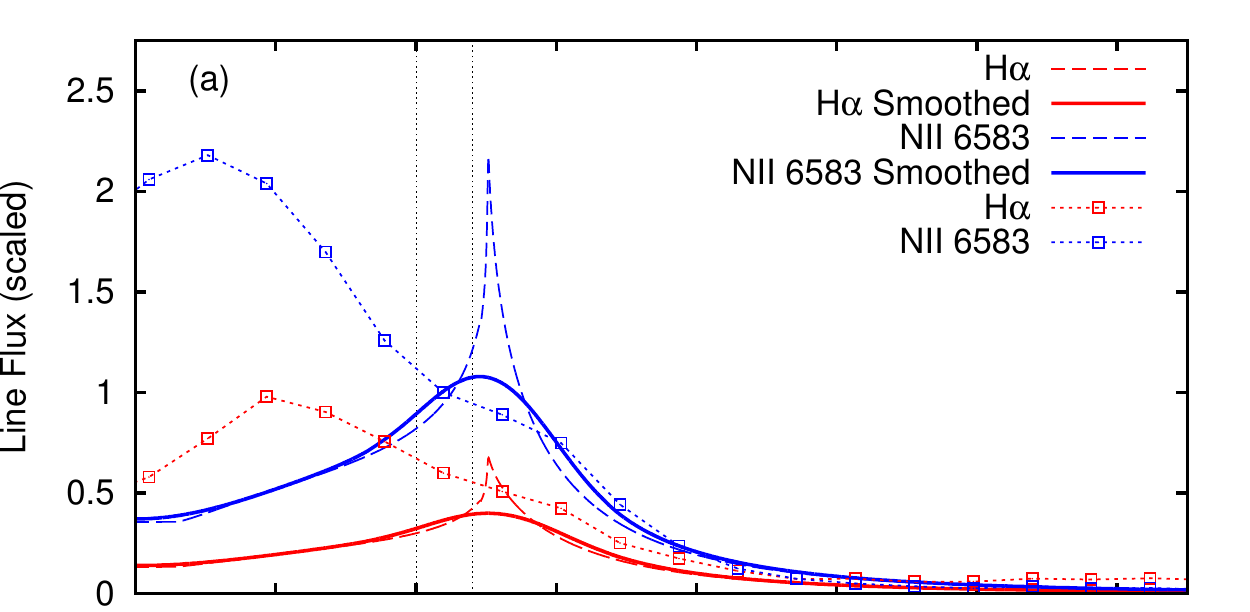}
\includegraphics[width=0.9\hsize]{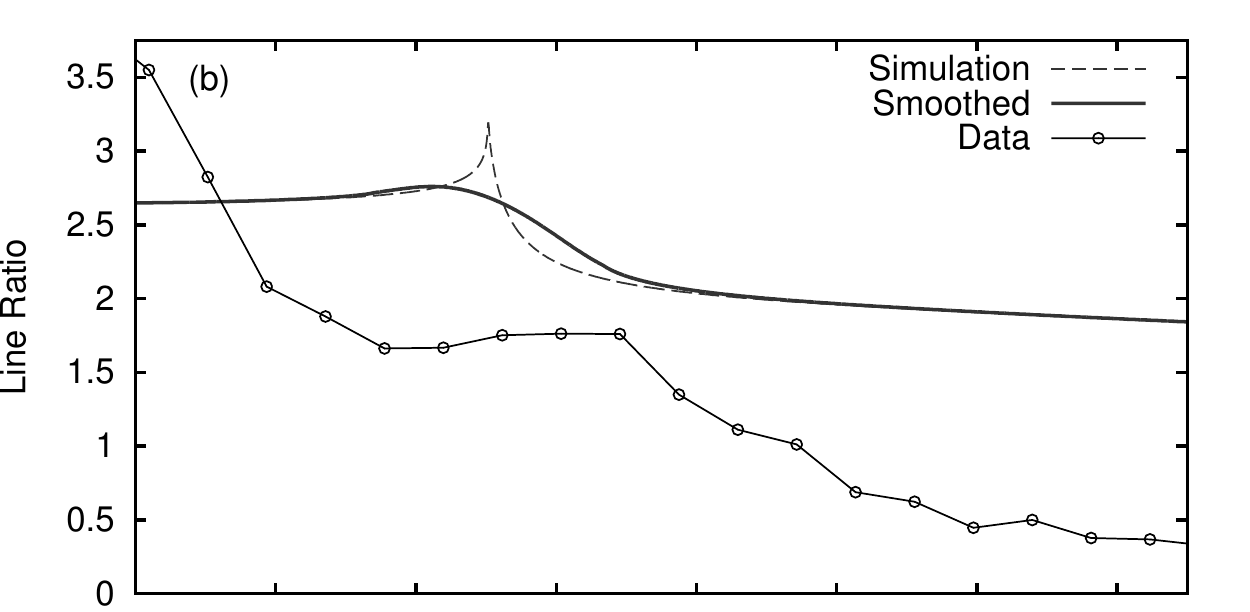}
\includegraphics[width=0.9\hsize]{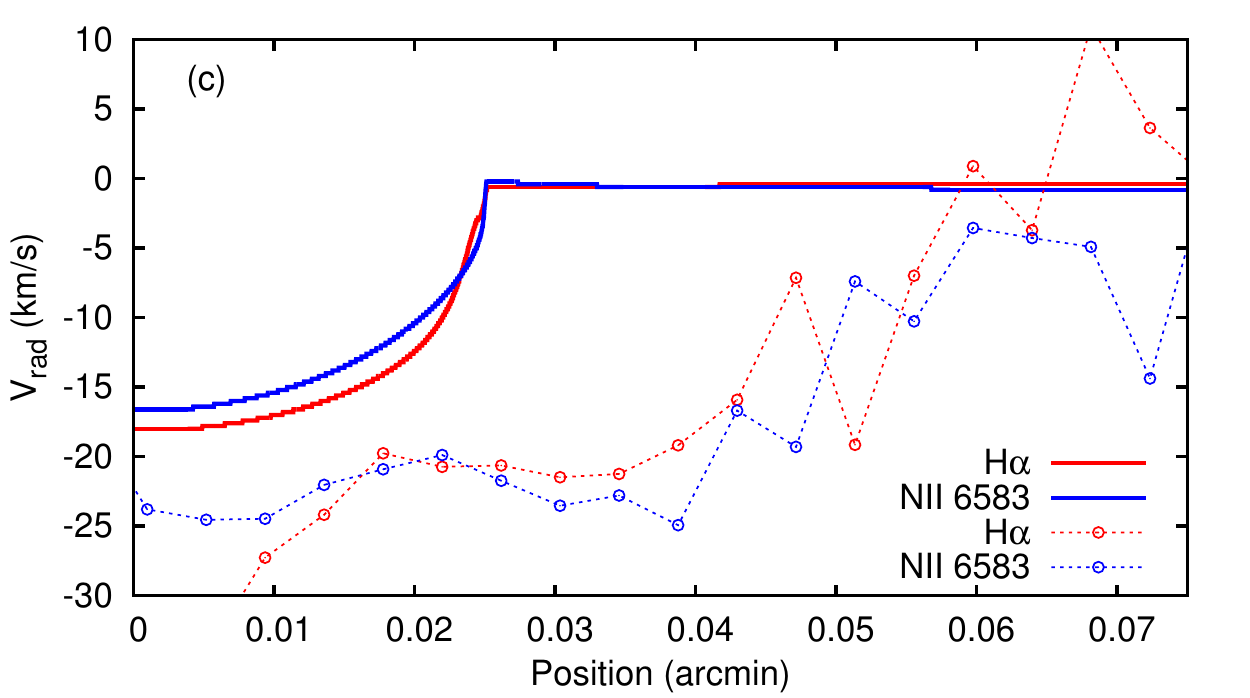}
\caption{
  The H$\alpha$ and [N\,II] spectral lines from the simulation M5V20W, with $\dot{M}=2\times10^{-5}\ \mathrm{M}_{\odot}\,\mathrm{yr}^{-1}$, $v_\infty=20\ \mathrm{km}\,\mathrm{s}^{-1}$, and $F_\gamma=1.29\times10^{11}\ \mathrm{cm}^{-2}\,\mathrm{s}^{-1}$, after 0.1 Myr of evolution.
  The lines and symbols are the same as in Fig.~\ref{fig:M5V15W}.
  }
\label{fig:M5V20W}
\end{figure}

\begin{figure}
\centering
\includegraphics[width=0.9\hsize]{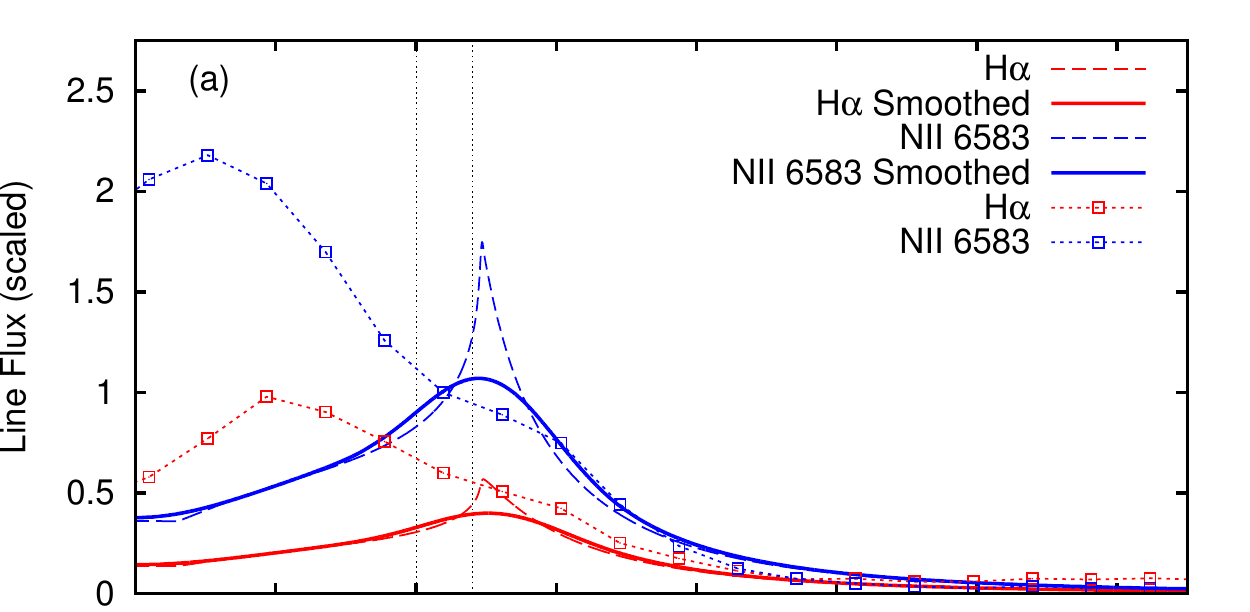}
\includegraphics[width=0.9\hsize]{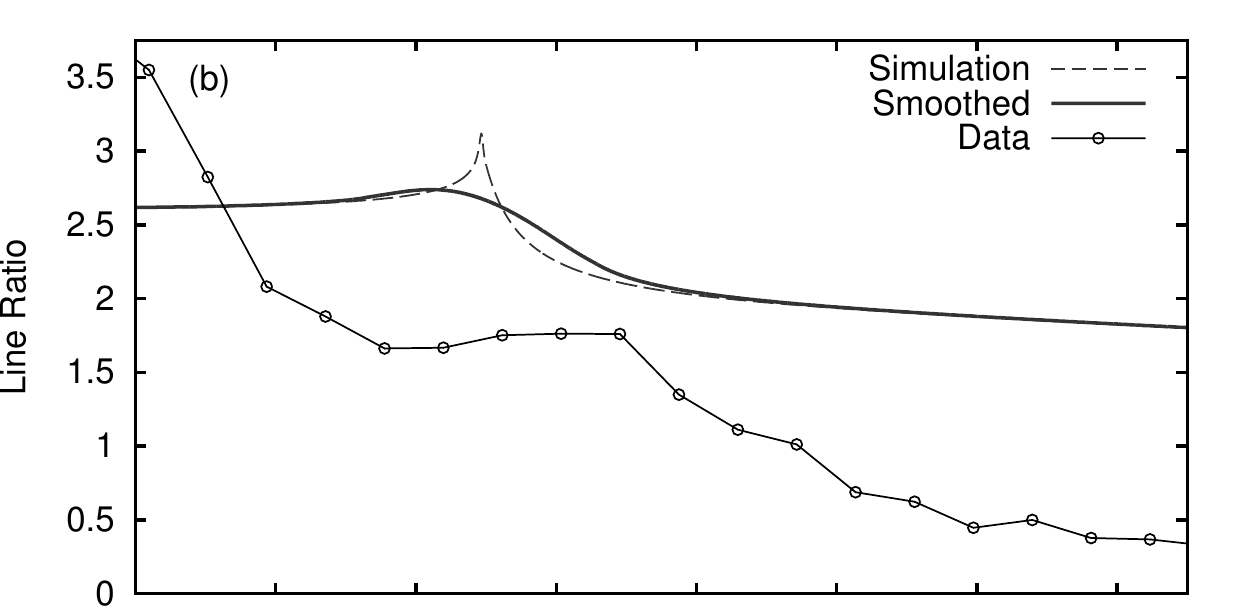}
\includegraphics[width=0.9\hsize]{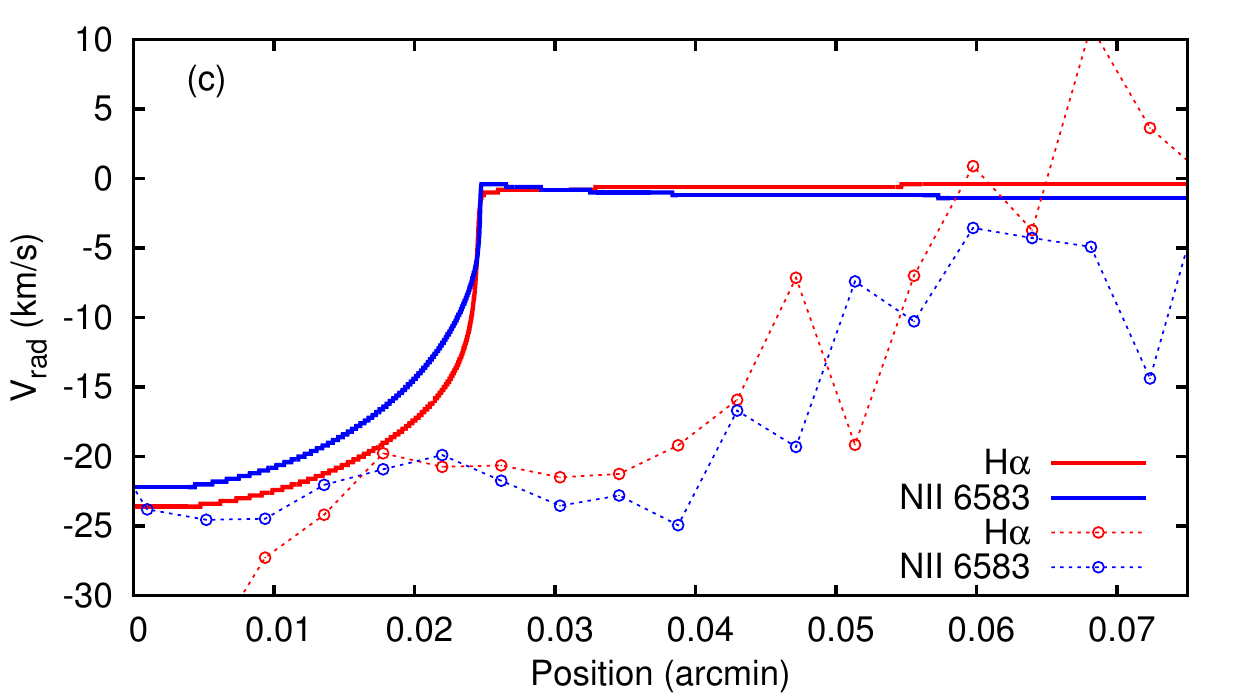}
\caption{
  The H$\alpha$ and [N\,II] spectral lines from the simulation M5V25W, with $\dot{M}=2\times10^{-5}\ \mathrm{M}_{\odot}\,\mathrm{yr}^{-1}$, $v_\infty=25\ \mathrm{km}\,\mathrm{s}^{-1}$, and $F_\gamma=8.29\times10^{10}\ \mathrm{cm}^{-2}\,\mathrm{s}^{-1}$, again after 0.1 Myr of evolution.
  The lines and symbols are the same as in Fig.~\ref{fig:M5V15W}.
  }
\label{fig:M5V25W}
\end{figure}

\begin{figure}
\centering
\includegraphics[width=0.9\hsize]{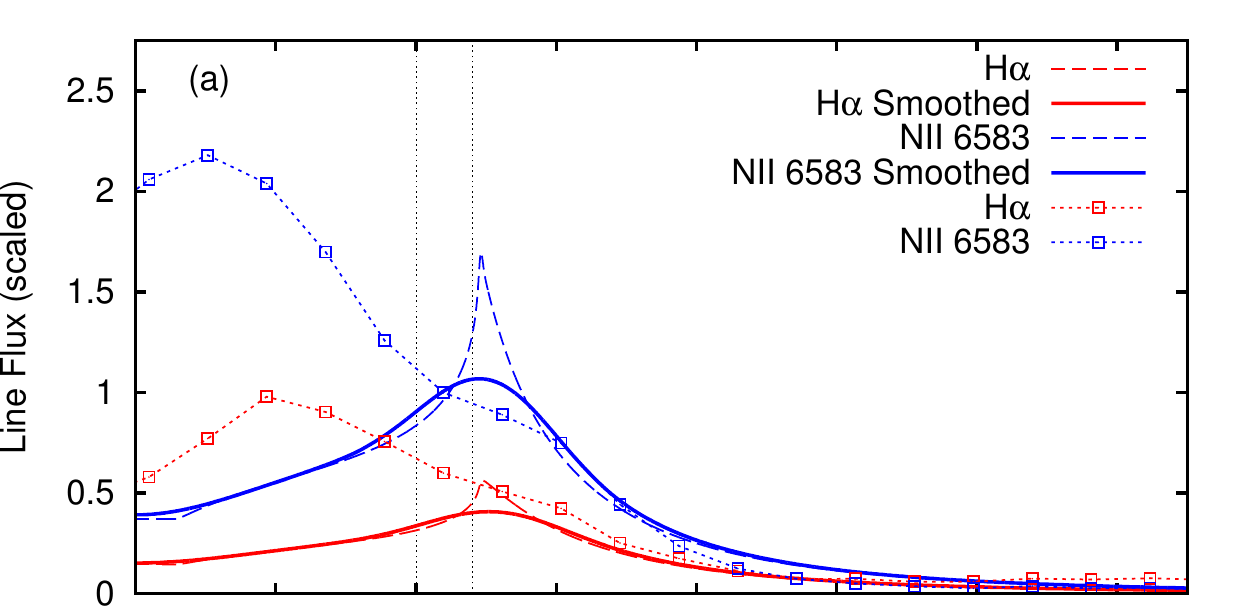}
\includegraphics[width=0.9\hsize]{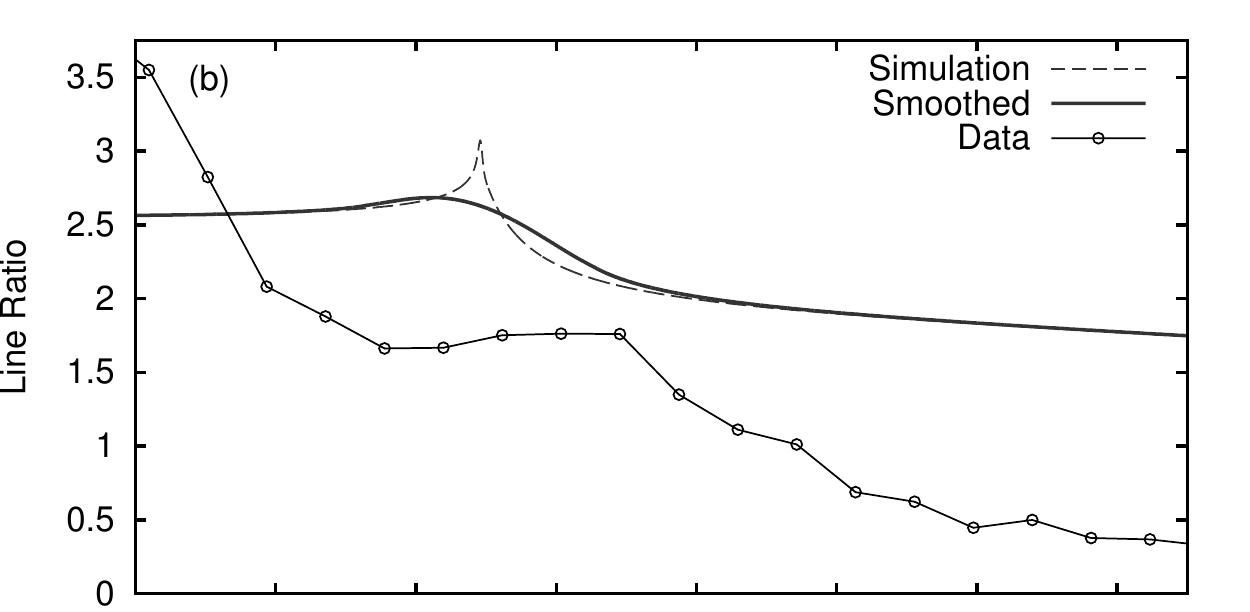}
\includegraphics[width=0.9\hsize]{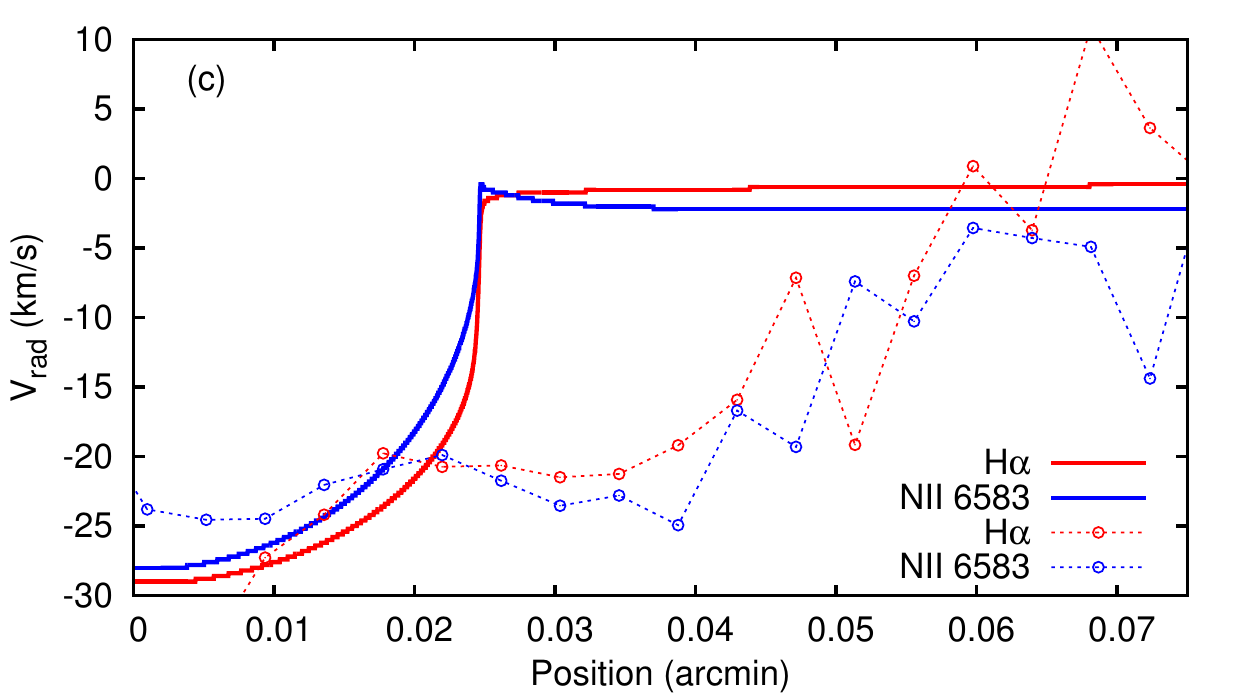}
\caption{
  The H$\alpha$ and [N\,II] spectral lines from the simulation M5V30W, with $\dot{M}=2\times10^{-5}\ \mathrm{M}_{\odot}\,\mathrm{yr}^{-1}$, $v_\infty=30\ \mathrm{km}\,\mathrm{s}^{-1}$, and $F_\gamma=5.83\times10^{10}\ \mathrm{cm}^{-2}\,\mathrm{s}^{-1}$, after 0.1 Myr of evolution.
  The lines and symbols are the same as in Fig.~\ref{fig:M5V15W}.
  }
\label{fig:M5V30W}
\end{figure}

\begin{figure}
\centering
\includegraphics[width=0.9\hsize]{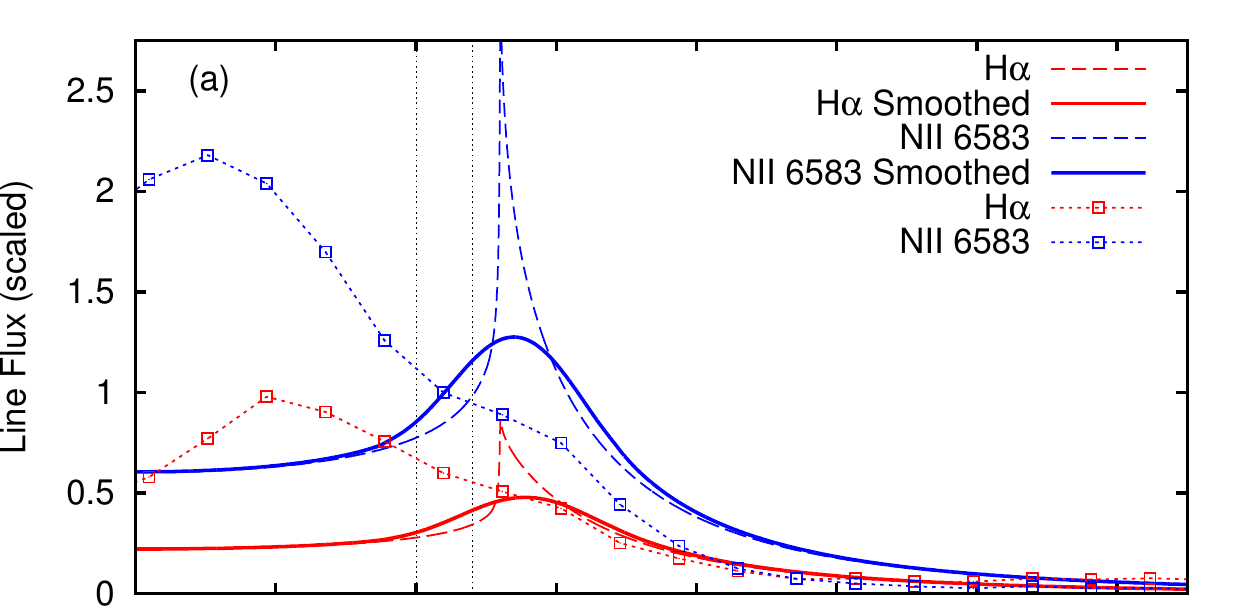}
\includegraphics[width=0.9\hsize]{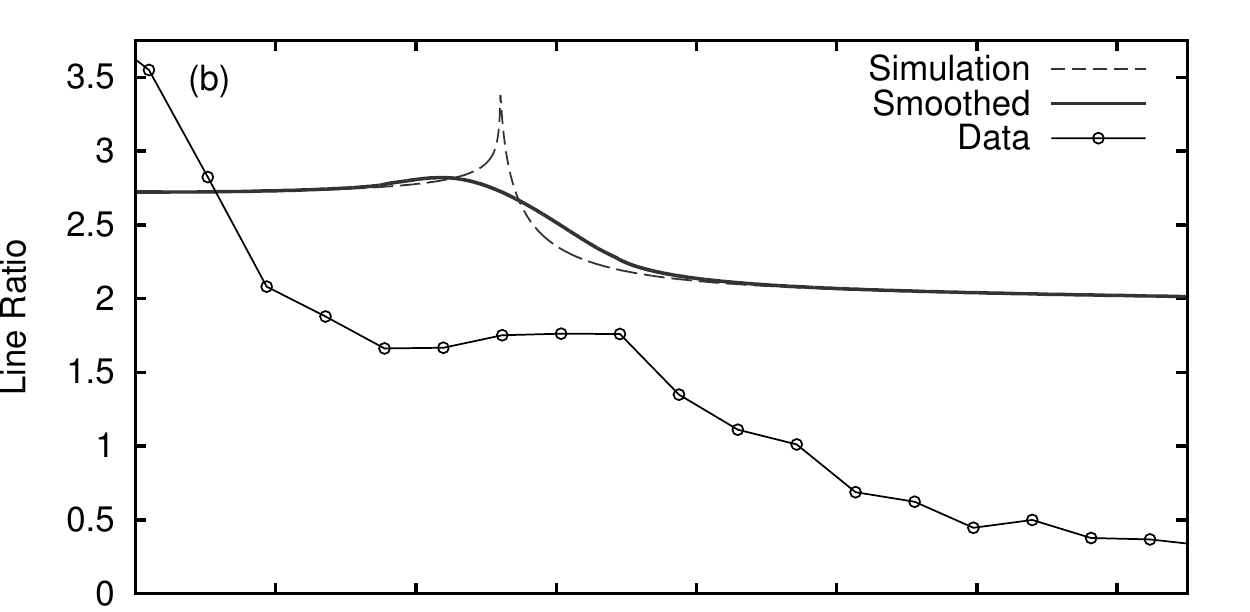}
\includegraphics[width=0.9\hsize]{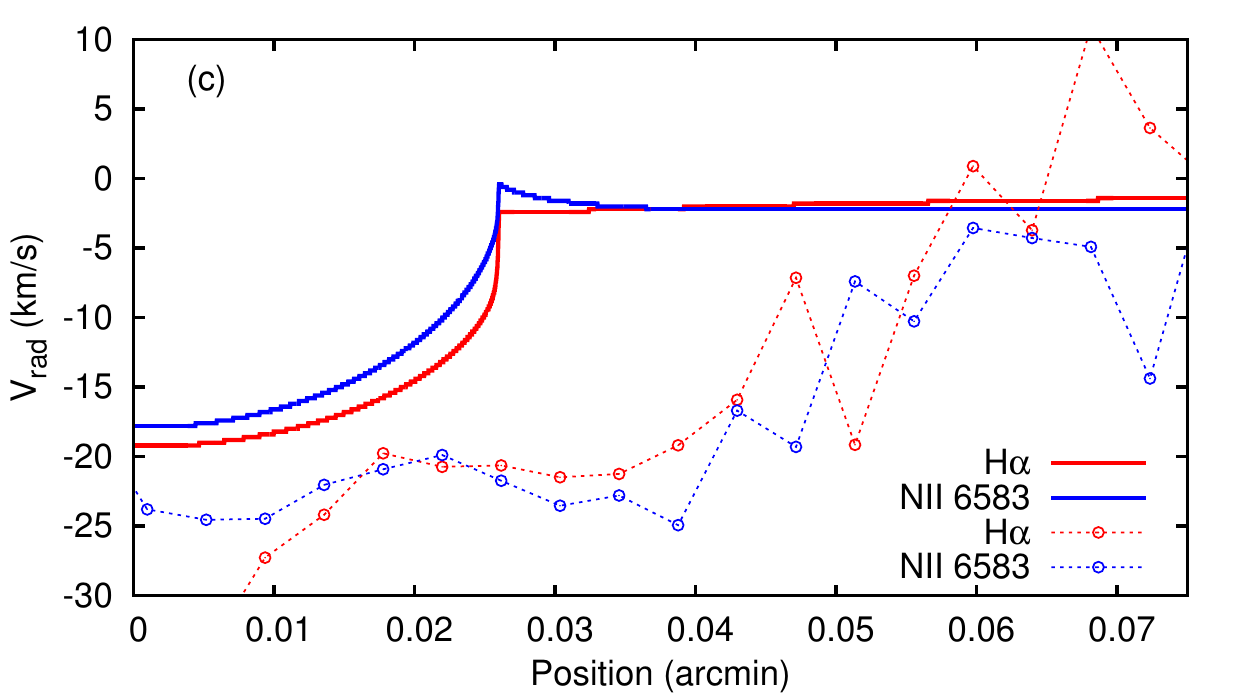}
\caption{
  The H$\alpha$ and [N\,II] spectral lines from the simulation M4V15W, with $\dot{M}=10^{-4}\ \mathrm{M}_{\odot}\,\mathrm{yr}^{-1}$, $v_\infty=15\ \mathrm{km}\,\mathrm{s}^{-1}$, and $F_\gamma=1.70\times10^{12}\ \mathrm{cm}^{-2}\,\mathrm{s}^{-1}$, after 0.1 Myr of evolution.
  The lines and symbols are the same as in Fig.~\ref{fig:M5V15W}.
  }
\label{fig:M4V15W}
\end{figure}

\begin{figure}
\centering
\includegraphics[width=0.9\hsize]{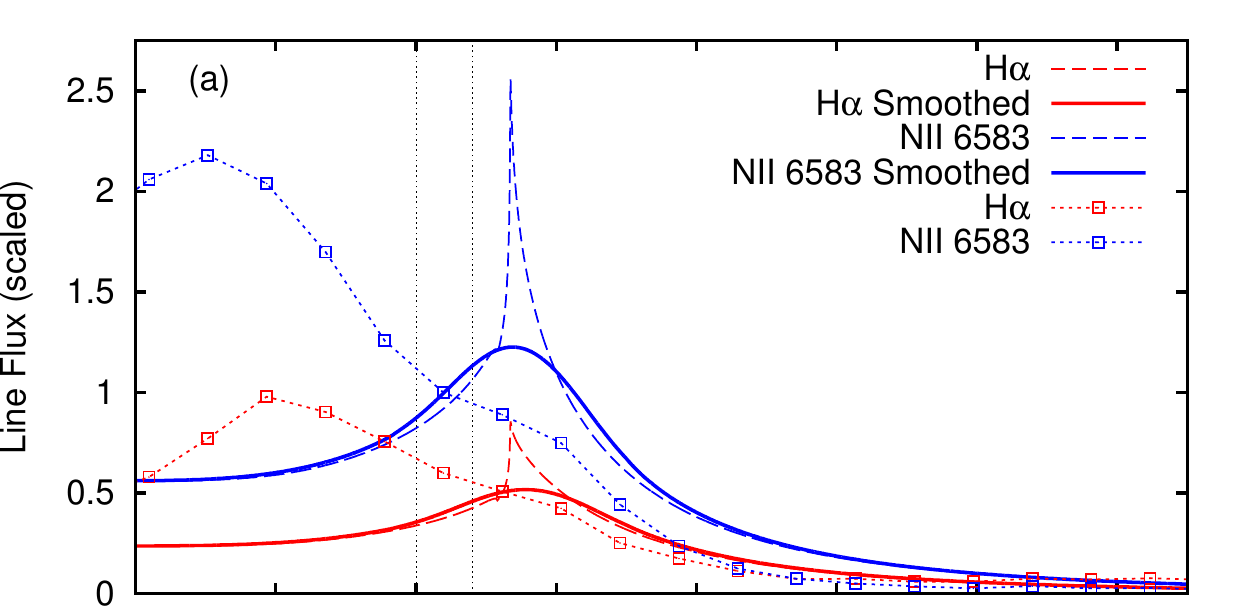}
\includegraphics[width=0.9\hsize]{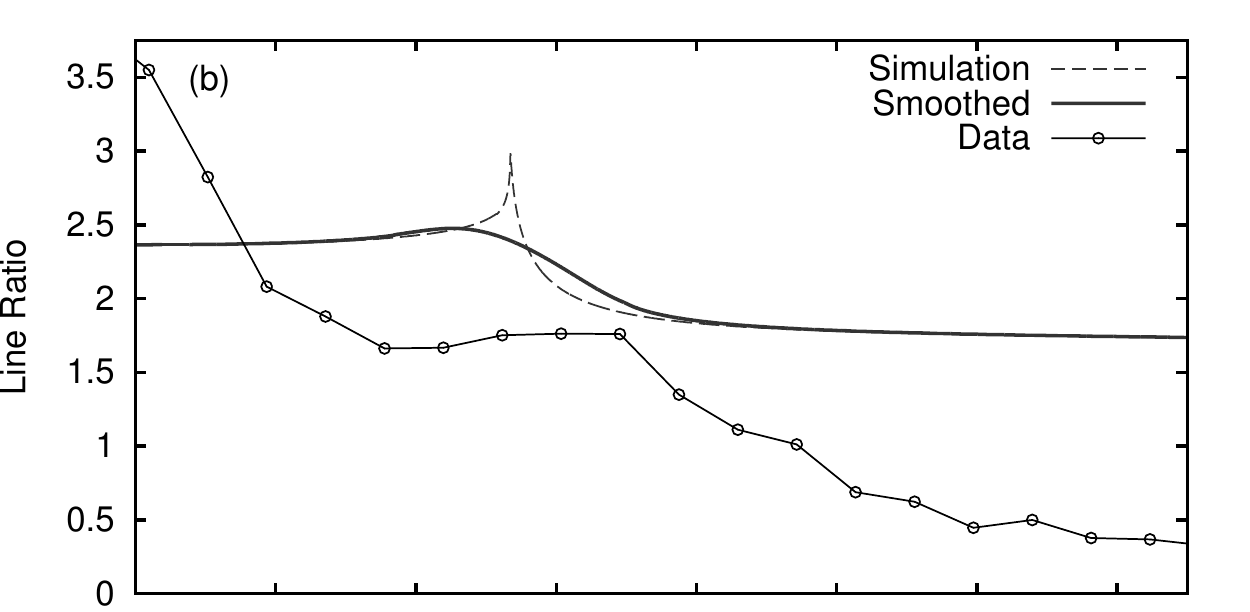}
\includegraphics[width=0.9\hsize]{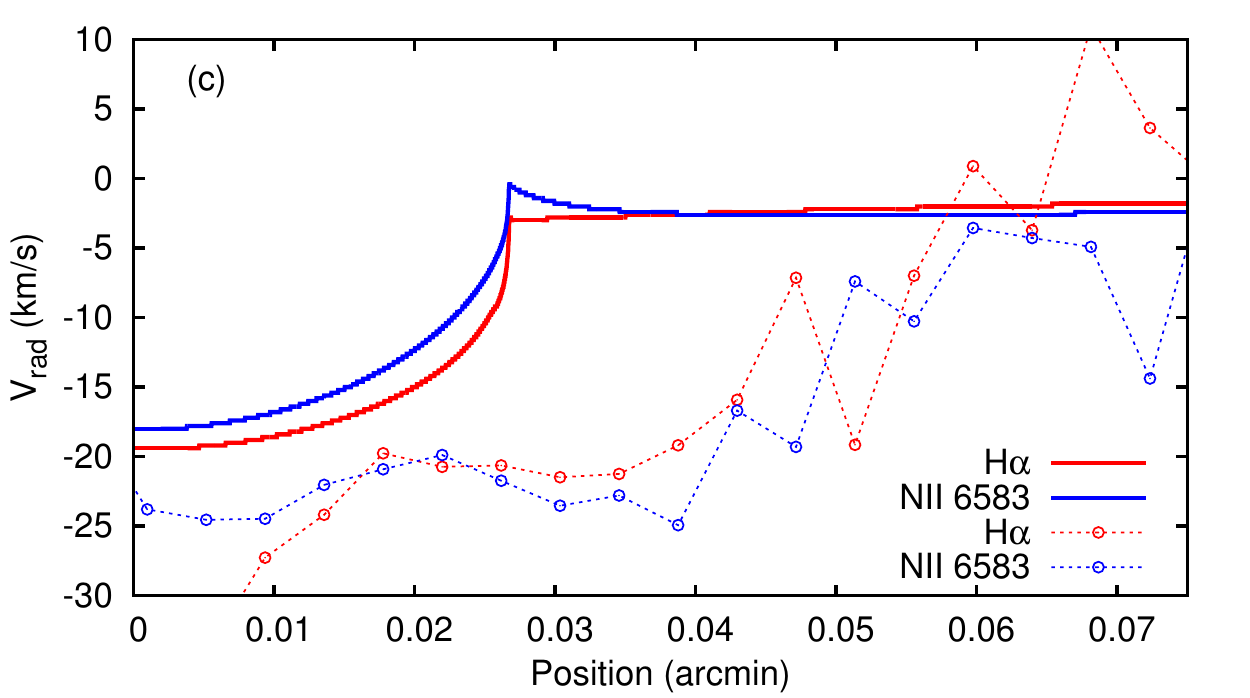}
\caption{
  The H$\alpha$ and [N\,II] spectral lines from the simulation M4V20W, with $\dot{M}=10^{-4}\ \mathrm{M}_{\odot}\,\mathrm{yr}^{-1}$, $v_\infty=20\ \mathrm{km}\,\mathrm{s}^{-1}$, and $F_\gamma=2.34\times10^{12}\ \mathrm{cm}^{-2}\,\mathrm{s}^{-1}$, after 0.1 Myr of evolution.
  The lines and symbols are the same as in Fig.~\ref{fig:M5V15W}.
  }
\label{fig:M4V20W}
\end{figure}

\begin{figure}
\centering
\includegraphics[width=0.9\hsize]{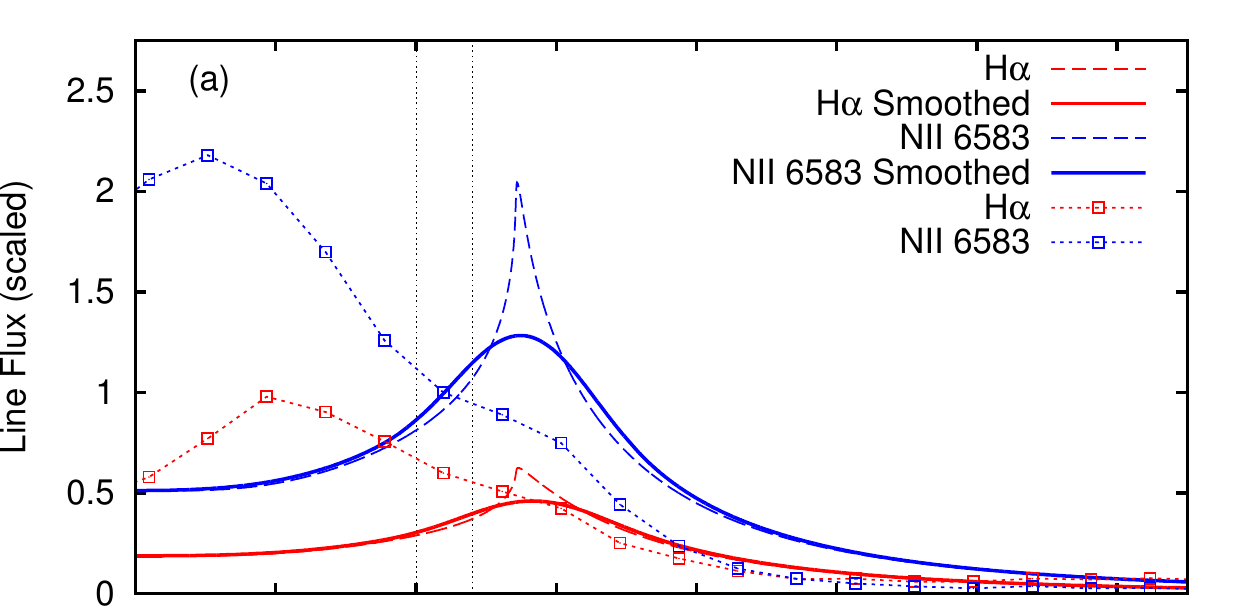}
\includegraphics[width=0.9\hsize]{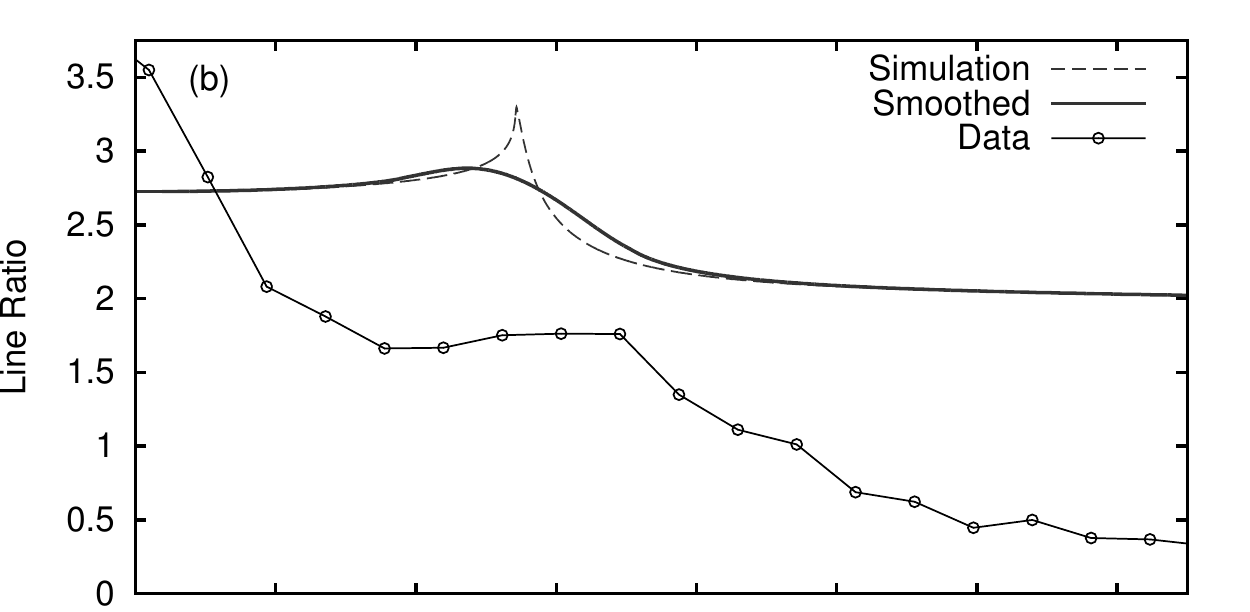}
\includegraphics[width=0.9\hsize]{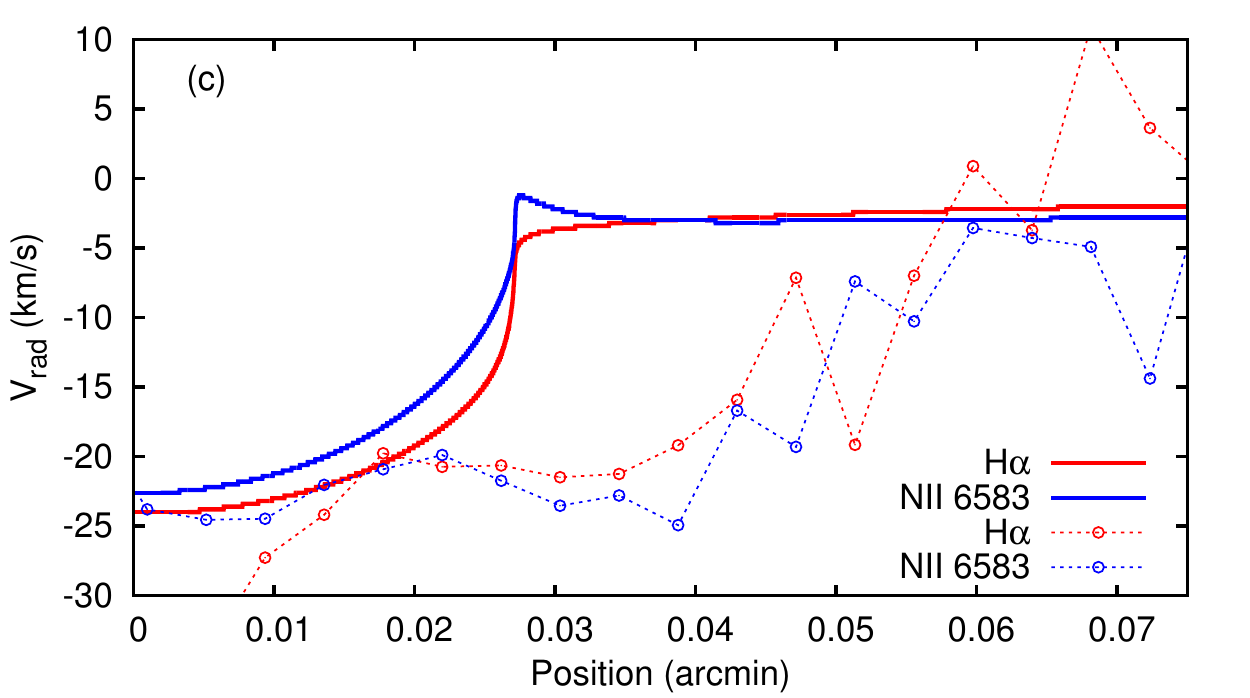}
\caption{
  The H$\alpha$ and [N\,II] spectral lines from the simulation M4V25W, with $\dot{M}=10^{-4}\ \mathrm{M}_{\odot}\,\mathrm{yr}^{-1}$, $v_\infty=25\ \mathrm{km}\,\mathrm{s}^{-1}$, and $F_\gamma=1.44\times10^{12}\ \mathrm{cm}^{-2}\,\mathrm{s}^{-1}$, after 0.1 Myr of evolution.
  The lines and symbols are the same as in Fig.~\ref{fig:M5V15W}.
  }
\label{fig:M4V25W}
\end{figure}

\begin{figure}
\centering
\includegraphics[width=0.9\hsize]{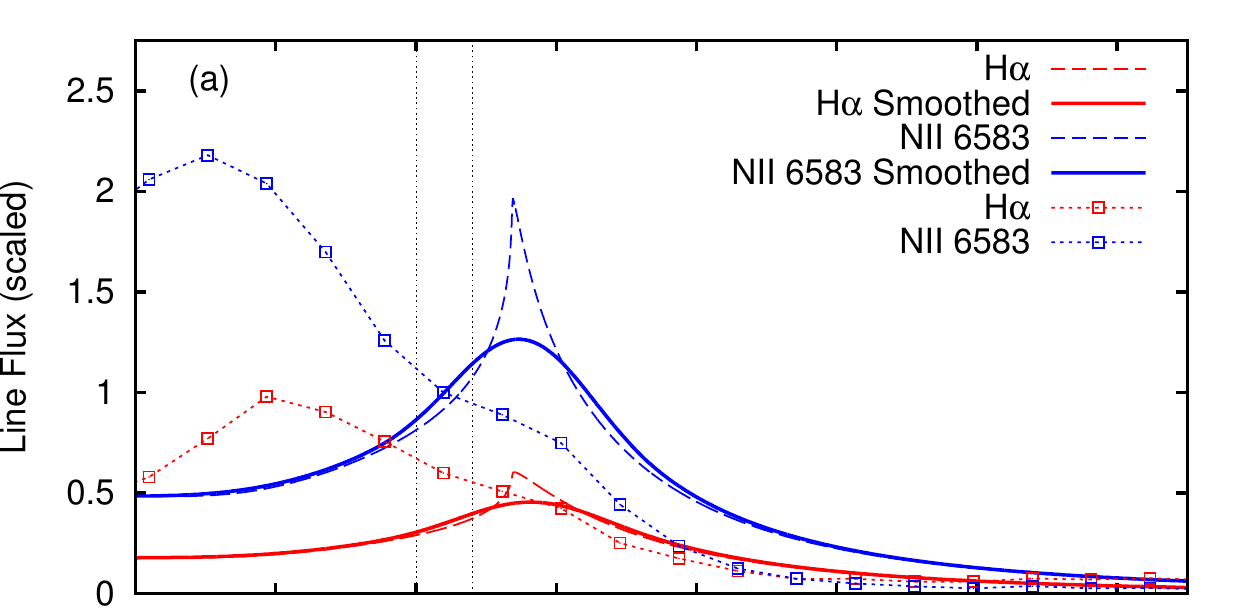}
\includegraphics[width=0.9\hsize]{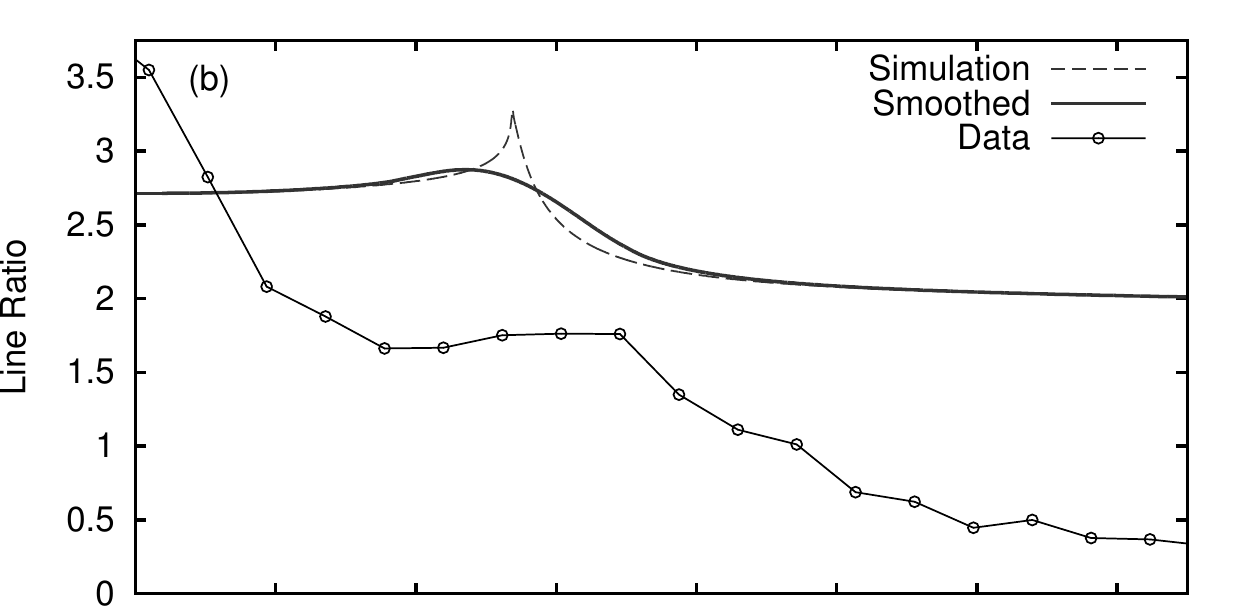}
\includegraphics[width=0.9\hsize]{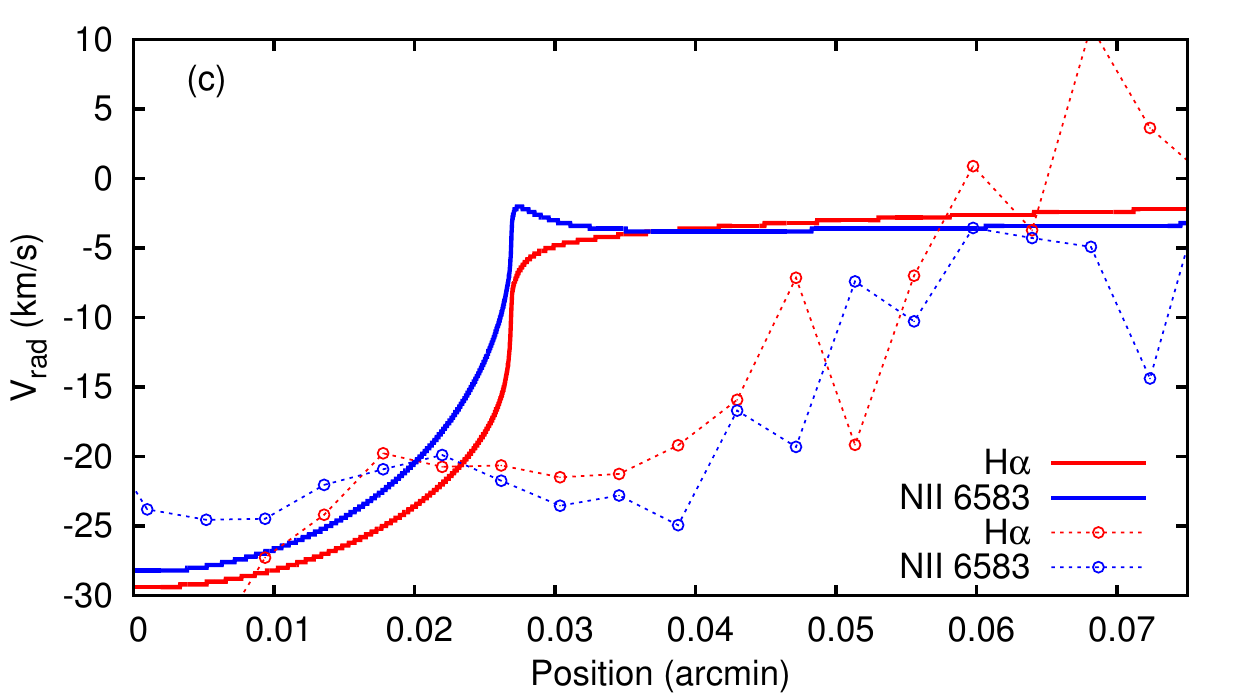}
\caption{
  The H$\alpha$ and [N\,II] spectral lines from the simulation M4V30W, with $\dot{M}=10^{-4}\ \mathrm{M}_{\odot}\,\mathrm{yr}^{-1}$, $v_\infty=30\ \mathrm{km}\,\mathrm{s}^{-1}$, and $F_\gamma=9.95\times10^{11}\ \mathrm{cm}^{-2}\,\mathrm{s}^{-1}$, after 0.1 Myr of evolution.
  The lines and symbols are the same as in Fig.~\ref{fig:M5V15W}.
  }
\label{fig:M4V30W}
\end{figure}

Here we plot the line intensity, radial velocity, and line ratio for the simulations labelled ``W'' in Table~\ref{tab:sims}.
These model the nebula Westnorthwest (WNW) from W26, in the direction of the extreme sgB[e] star W9.
As noted in the text, the nebula is confined closer to W26 on this side, possibly because of the proximity of W9.
The ``W'' simulations have larger photon fluxes than the corresponding ``E'' simulations, so that the ionized region penetrates deeper into the wind of the red supergiant.
The results of simulations are overplotted on the observed data in Fig.~\ref{fig:M5V15W} for simulation M5V15W, and similarly in Figs.~\ref{fig:M5V20W}--\ref{fig:M5V30W} for M5V20W, M5V25W, and M5V30W.
Figs.~\ref{fig:M4V15W}-\ref{fig:M4V30W} show results for the simulations M4V15W, M4V20W, M4V25W, and M4V30W, which have a denser wind because of a larger mass-loss rate.

The observed data shows no limb-brightened peak in the nebular emission, rather just a shoulder of emission between 0.02 and 0.03 arcmin.
If we discard the uncertain two points closest to zero (at W26) then the emission decreases monotonically with distance from W26, in strong constrast to the predictions of simulations.
This makes the discrepency with the limb-brightened emission from M5V15W (Fig.~\ref{fig:M5V15W}a) even larger for this side of W26.
The observed line ratio is slightly lower on this side of W26, but again is relatively constant across the nebula, decreasing only outside the bright part of the nebula ($\theta>0.04$ arcmin).
Additionally, the radial velocity of the lines (Fig.~\ref{fig:M5V15W}c) appears too low at all radii, but particularly for $\theta>0.03$ arcmin.

Simulation M5V25W (Fig.~\ref{fig:M5V25W}) has better agreement with observations, for the same reasons as M5V25E: the ionization front emission is less peaked, the radial velocity is more blueshifted in agreement with observations, and the line ratio has reasonable agreement with observations.
Again, however, the crucial problem is that the observed nebular emission is all blueshifted by about 25 $\mathrm{km}\,\mathrm{s}^{-1}$, whereas in the simulation the brightest emission is neither blue- nor redshifted.

%%%%%%%%%%%%%%%%%%%%%%%%%%%%%%%%%%%%%%%%%%%%%%%%%%%%%%%%%%%%%%%%%%%%%
%%%%%%%  Mdot=2\times10^{-5} Msun/yr, NEAR 
%%%%%%%%%%%%%%%%%%%%%%%%%%%%%%%%%%%%%%%%%%%%%%%%%%%%%%%%%%%%%%%%%%%%%

%%%%%%%%%%%%%%%%%%%%%%%%%%%%%%%%%%%%%%%%%%%%%%%%%%%%%%%%%%%%%%%%%%%%%
%%%%%%%  Mdot=10^{-4} Msun/yr, NEAR 
%%%%%%%%%%%%%%%%%%%%%%%%%%%%%%%%%%%%%%%%%%%%%%%%%%%%%%%%%%%%%%%%%%%%%

\end{document}